\documentclass[12pt]{article}
\usepackage{amsmath,amsfonts,amsbsy,amssymb,amsthm} 
\usepackage{natbib}
\usepackage{bbm}
\usepackage{booktabs}  % for nice tables
\usepackage{enumitem}  % resume enumerate
\usepackage[colorlinks=true,citecolor=blue,urlcolor=blue,linkcolor=blue]{hyperref}
\usepackage{adjustbox} % used to adjust widths 'n' heights of tables.
\usepackage{pifont}% http://ctan.org/pkg/pifont 
\newcommand{\cmark}{\ding{51}}%
\newcommand{\xmark}{\ding{55}}% 

\newcommand{\tightoverset}[2]{%
  \mathop{#2}\limits^{\vbox to -.5ex{\scriptsize\kern-1.3ex\hbox{$#1$}\vss}}}

\usepackage{comment}
\usepackage{color}

\newcommand \mbR{\mathbb{R}}

\newcommand \mcU{\mathcal{U}}
\newcommand \tu{\widetilde{u}}
\newcommand \bV{{\bf V}}
\newcommand \mcV{{\mathcal{V}}}
\newcommand \tX{\widetilde{X}}
\newcounter{example}[section]
\newenvironment{example}[1][]{\refstepcounter{example}\par\medskip
  \noindent \textbf{Example~\theexample. #1} \rmfamily}{\medskip}
\newcounter{remark}[section]
\newenvironment{remark}[1][]{\refstepcounter{remark}\par\medskip
  \noindent \textbf{Remark~\theremark. #1} \rmfamily}{\medskip}
\newcounter{algorithm}[section]
\newenvironment{algorithm}[1][]{\refstepcounter{algorithm}\par\medskip
   \noindent \textbf{Algorithm~\thealgorithm. #1} \rmfamily}{\medskip}
\newtheorem{assumption}{Assumption}[section]
\newtheorem{corollary}{Corollary}[section]
\newtheorem{theorem}{Theorem}[section]
\newtheorem{proposition}{Proposition}[section]
\newtheorem{lemma}{Lemma}[section]
\newtheorem{definition}{Definition}[section]

\DeclareMathOperator{\E}{\mathbb{E}}

\DeclareMathOperator{\V}{Var}
\DeclareMathOperator{\var}{Var}
\DeclareMathOperator{\cov}{Cov}

\newcommand{\blind}{0}

\begin{document}

\def\spacingset#1{\renewcommand{\baselinestretch}%
{#1}\small\normalsize} \spacingset{1}

%%%%%%%%%%%%%%%%%%%%%%%%%%%%%%%%%%%%%%%%%%%%%%%%%%%%%%%%%%%%%%%%%%%%%%%%%%%%%%

\if0\blind 
{ 
  \title{\bf
  Fast and Fair Simultaneous Confidence Bands for Functional Parameters
  %Fair Simultaneous Confidence Bands with Balanced False Positive Rate 
  }
  \author{
    Dominik Liebl\thanks{
    Dominik Liebl (dliebl@uni-bonn.de),
    Institute of Finance and Statistics,
    University of Bonn,
    Adenauerallee 24-42,
    53113 Bonn, Germany}
  \thanks{
    Hausdorff Center for Mathematics,
    Endenicher Allee 62,
    53115 Bonn, Germany}
  \and
    Matthew Reimherr\thanks{
    Department of Statistics,
    Penn State University,
    411 Thomas Building,
    University Park, PA 16802}
  }
  \date{}
  \maketitle
  } \fi

\if1\blind
{
  \bigskip
  \bigskip
  \bigskip
  \begin{center}
    {\LARGE\bf Simultaneous Confidence Bands with Balanced False Positive Rate}
\end{center}
  \medskip
} \fi

\bigskip

\begin{abstract} 
Quantifying uncertainty using confidence regions is a central goal of statistical inference.  Despite this, methodologies for confidence bands in Functional Data Analysis are still underdeveloped compared to estimation and hypothesis testing.  In this work, we present a new methodology for constructing simultaneous confidence bands for functional parameter estimates.  Our bands possess a number of positive qualities: (1) they are not based on resampling and thus are fast to compute, (2) they are constructed under the fairness constraint of balanced false positive rates across partitions of the bands' domain which facilitates the typical global, but also novel local interpretations, and (3) they do not require an estimate of the full covariance function and thus can be used in the case of fragmentary functional data. Simulations show the excellent finite-sample behavior of our bands in comparison to existing alternatives. The practical use of our bands is demonstrated in two case studies on sports biomechanics and fragmentary growth curves.
\end{abstract} 


\spacingset{1}

%%%%%%%%%%%%%%%%%%%%%%%%%%%%%%%%%%%%%%%%%%%%
\section{Introduction}
%%%%%%%%%%%%%%%%%%%%%%%%%%%%%%%%%%%%%%%%%%%%

As part of the big data revolution, statistical tools have made astonishing strides towards handling increasingly complex and structured data.  Data gathering technologies have allowed scientists, businesses, and government agencies to collect data on increasingly sophisticated phenomena, including high-dimensional measurements, functions, surfaces, and images.  A major tool set that has emerged for handling such highly complex and smooth structures is Functional Data Analysis, FDA \citep{RS_2005_book,FV2006_book,hsing2015theoretical,KR2017_book}. There, one has access to rich, dynamic, and smooth structures, while also having some degree of repetition, allowing researchers to fit complex and flexible nonparametric statistical models.

Despite the success and maturity of FDA, some foundational questions remain unanswered or underdeveloped. This work is concerned with a fundamental goal of statistical inference: quantifying estimation uncertainty with simultaneous confidence bands for functional parameters (mean functions, regression functions, etc.). Simultaneous confidence bands have received increasing attention in recent years, however, current solutions usually suffer from one of several major drawbacks: they are computationally expensive, they result in bands that are very conservative, they cannot be used in the case of fragmentary functional data, or they can only be interpreted globally---not locally. In this work we present a broad framework based on random process theory that solves the aforementioned issues. Our bands possess a number of desirable qualities, but two are especially noteworthy. 

First, they are constructed using a non-constant, adaptive critical value function which allows us to fill an important gap in the interpretability of existing simultaneous confidence bands: While classic $(1-\alpha)\times 100\%$ simultaneous confidence \emph{intervals} \citep{Dunn_1961} control each single confidence interval at a known false positive rate (typically $\tilde\alpha=\alpha/m$ in case of $m$ confidence intervals), existing $(1-\alpha)\times 100\%$ simultaneous confidence bands do not provide comparable information about their local control of the false positive rate $\tilde\alpha<\alpha$ over subsets of the bands' domain. Our bands solve this issue by using adaptive critical value functions balancing the false positive rate across (arbitrary) partitions of the domain of the functional parameter. This makes our bands both simple to interpret and applicable in contexts requiring statistical fairness guarantees.  
%%%
Balanced false positive rates are an established statistical fairness constraint in machine learning \citep{HaPrSr2016,MoRo_2021} required for statistical decision procedures which can cause harm/discrimination due to disproportionate concentrations of false positive events across subgroups. Existing simultaneous confidence bands are unable to balance such potentially harmful disproportionate concentrations of false positive events and thus require a fairness constraint---particularly, when used to inform high-stakes decisions such as in medicine \citep{Boschi_et_al_2021}, economics \citep{Liebl_2019}, or forensic gait analysis \citep{Kelly_2020}. If, for instance, the domain of the functional parameter denotes age, as in many biomedicine applications \citep{HS_2007,UF_2013}, then unbalanced false positive events can systematically affect certain age-subgroups resulting in unnecessary medical treatments with potential side effects for the affected age-subgroups. While there is a certain awareness of this fairness issue, for instance in biomechanics \citep{Pataky_2019}, our bands are the first to provide simultaneous inference under the statistical fairness guarantee of balanced false positive rates.

Second, our bands do not require estimation of the full covariance function of the estimator.  Instead, only a narrow band along the diagonal is needed and then only to ascertain the pointwise uncertainty of the estimate and its derivative.  No other FDA method in the literature, that we are aware of, for quantifying uncertainty globally posses such a property.  This has broad implications in FDA as many tools are being actively developed for data consisting of functional fragments, where estimation of the full covariance is generally impossible \citep{DH2016,DP2019,DHHK2019,LR2019,KL2020}.

% Literature:
The recent literature in FDA provides different approaches and methods to simultaneous inference for functional data, with simulation based methods being currently the most successful. \cite{D2011}, \cite{CYT12}, and \cite{WWWO2019} propose bands based on the parametric bootstrap. A further simulation based approach is the interval-wise testing (IWT) procedure proposed by \cite{PV2016_IWT} and \cite{PV2017_IWT} which uses permutations. Simulation based approaches are computationally intensive and, as shown in our simulation study can perform weakly in small samples. An exception are the multiplier bootstrap bands proposed by \cite{DKA_2020}, \cite{DK_2022} and \cite{TS2022}, which perform well also in small samples.
% The IWT procedure is able
% to keep the significance level, but has very low power in detecting local
%%%
% violations of the simultaneous null hypotheses.
Another typical approach is to apply a dimension reduction based on functional principal component analysis and to build multivariate confidence ellipses; see, for instance, \cite{YMW2005} and \cite{GGC2013}.  As shown by \cite{CR2018}, however, these procedures lead to ellipses with zero-coverage.  As a solution to this problem, \cite{CR2018} proposed an approach that transforms confidence hyper-ellipsoids into valid confidence bands. Also related is the work of \cite{LOPS2019} who adapt the Benjamini-Hochberg procedure to the case of functional data.
%%%
Early theoretical results on Kac-Rice formulas were developed by \cite{I1963}, \cite{CL1965}, and \cite{B1966}, among others. Fundamental introductions to random process/field theory, including different versions of the Kac-Rice formula, are given in the textbooks of \cite{AT2007_book}, \cite{AW2009_book}, and \cite{CL2013_book}. The random process/field theory based methods have seen a tremendous success in the applied neuroimaging and biomechanics literature \citep{PFAKN2007_book,WLNKS2009,PRV2016,WHY2018}, where these methods are subsumed under the term Statistical Parametric Mapping. Recently, \cite{TS2022} published random process/field theory based simultaneous confidence bands. The authors consider the case of $t$-processes and contribute new inference results, but for the well-known existing Gaussian kinematic formulas; i.e.~higher dimensional Kac-Rice formulas. By contrast, we consider the more general cases of elliptical processes, non-constant adaptive critical value functions, and fragmentary functional data.

% Paper-Outline
The paper and its contributions are structured as follows. Section \ref{sec:AwFair} motivates and introduces Definition \ref{Def:Fairness} of false positive rate balance for stochastic processes. In Section \ref{SEC:MT}, we collect our core methodological and theoretical contributions: Theorem \ref{TH:MAIN} contains our generalized Kac-Rice formula for variable critical value functions and general elliptical processes, Theorem \ref{TH:EST} contains our asymptotic results, Algorithm \ref{ALGO:FAIRNESS} shows how to select fair critical value functions, and Proposition \ref{PRO:FAIR_BAND} together with Lemma \ref{LE:FAIRNESS} shows that our fair bands fulfill the Definition \ref{Def:Fairness} of false positive rate balance. Simulations in Section \ref{SEC:SIMUL} show the excellent finite-sample behavior of our fair confidence bands for fully observed and fragmentary functional data.  Section \ref{SEC:APPL} contains applications to data from sports biomechanics and to fragmentary growth curves.  Lastly, Section \ref{SEC:DISS} contains our discussion.

%%%%%%%%%%%%%%%%%%%%%%%%%%%%
\section{Introducing Non-Constant Critical Value Functions}\label{sec:AwFair}
%%%%%%%%%%%%%%%%%%%%%%%%%%%%
We build upon a simple connection between the supremum of a stochastic process and the expected Euler characteristic of an excursion \citep{Piterbarg1982,Azais_et_al_2002}. Consider a real valued random functional parameter estimator over a closed interval, $X=\{X(t): t \in [0,1]\}$, where restricting the domain to $[0,1]$ is without loss of generality (for any closed finite interval) and common in FDA to ease notation.  Instead of finding the probability that $X$ ever crosses a critical value function, $u = \{u(t) : t \in [0,1]\}$, i.e., $P( \exists t\in[0,1]: X(t)\geq u(t))$, the expected Euler characteristic inequality suggests to consider the task of counting the events when $X$ crosses $u$ in an upward trajectory. While a seemingly small change, the latter ends up being far more tractable.  Thus far, the literature on random process/field theory has focused on constant critical values $u(t)\equiv u \in \mbR$; see, for instance, \cite{Worsley_et_al_2004}, \cite{AT2007_book}, and \cite{AW2009_book}. In this paper, we demonstrate how to handle the more general case of non-constant critical value functions, which, as we will see, opens up the exciting opportunity to produce tight and fair bands. The number of up-crossings of $X$ about $u$ on the interval $[0,1]$ is defined as
\begin{align*}
N_{u,X}([0,1])&:=\#\{0\leq t \leq 1:X(t)=u(t),X'(t)>u'(t)\}.
\end{align*}
If this quantity is zero, then the only way that $X(t)$ could have exceeded $u(t)$ was if $X$ started above of $u$ at $t=0$, since both functions, $X$ and $u$, are continuous.  This logic leads to the expected Euler characteristic inequality by applying Boole's and Markov's inequality
\begin{align}
P\big(\exists{t\in[0,1]} : X(t)\geq u(t)\big)
&=P\big(\{X(0)\geq u(0)\}\;
\text{or}
%\cup
\;\{N_{u,X}([0,1])\geq 1\}\big)\notag\\
&\leq P\big(X(0)\geq u(0)\big)+ P\big(N_{u,X}([0,1])\geq 1\big)\notag\\
&\leq P\big(X(0)\geq u(0)\big)+\E\big[N_{u,X}([0,1])\big]=\E\big[\varphi_{u,X}(0)\big],\label{EQ:EulerIneq}
\end{align}
where $\varphi_{u,X}(0):=\mathbbm{1}_{X(0)\geq u(0)}+N_{u,X}([0,1])$ denotes the Euler (or Euler-Poincar\'e) characteristic of the excursion set $\{t\in[0,1]:X(t)\geq u(t)\}$. Intuitively, this quantity counts exceedance events by starting at $t=0$, checking if $X(0)>u(0)$ and then moving through the domain to check for additional up-crossings. From a geometric point of view, the Euler characteristic over $[0,1]$-domains equals the number of connected components of the excursion set $\{t\in[0,1]:X(t)\geq u(t)\}$; see \cite{AT2007_book} for generalizations to more complex domains in higher dimensions. 

\vspace*{-3mm}

\paragraph{Sharpness of the expected Euler characteristic inequality.}
The sharpness of inequality \eqref{EQ:EulerIneq} was studied under different assumptions on the stochastic process $X$. \cite{Piterbarg1982} and \cite{Azais_et_al_2002} considered the case of smooth Gaussian processes with stationary covariance functions (i.e.~$\operatorname{Cov}(X(t),X(s))=\operatorname{Cov}(X(0),X(|t-s|))$ for all $t,s\in[0,1]$) and without global singularities (i.e.~$\operatorname{Cor}(X(t),X(s))=1$ only if $t=s$). These authors showed, under some further technical assumptions, that the approximation error for large $u\to\infty$ is $|P\big(\exists{t\in[0,1]} : X(t)\geq u(t)\big)-\E[\varphi_{u,X}(0)]|\leq c_1\exp\left(-c_2u^2/2\right)$ for some positive constants $c_1$ and $c_2$. \cite{TTA2005} showed that similar results hold for non-stationary covariance functions. These results generalize to our case of non-constant critical value functions $u(t)$ by considering $u\equiv\min_{0\leq t\leq 1}u(t)$ for large $u\to\infty$. Note that global singularities are allowed in \eqref{EQ:EulerIneq}, but make the approximation less sharp. Below in Proposition \ref{PRO:PRICE4FAIRNESS} (Price of fairness), we study the loss of accuracy of inequality \eqref{EQ:EulerIneq} when imposing more and more fairness constraints.  

\bigskip

To construct non-constant critical value functions $u(t)$ that are both tight and fair, it is necessary to generalize the expected Euler characteristic inequality \eqref{EQ:EulerIneq} and to allow starting the counting at some mid point, $t_0 \in [0,1]$ and then count crossings moving away from $t_0$ in both directions.  When moving from $t_0$ down to $0$, up-crossings as $t$ decreases are equivalent to down-crossings as $t$ increases (see Figure \ref{FIG:APP_CROSS} of the supplementary paper \cite{LR_Suppl}).  We therefore work with the following generalized expected Euler characteristic
\begin{align}\label{EQ:EulerIneq_General}
%P\big(\exists{t\in[0,1]} : X(t)\geq u(t) \big) \leq
  P\big(X(t_0) \geq u(t_0)\big)
+ \E[N_{u,X}^-([0,t_0])]
+ \E[N_{u,X}([t_0,1])]=\E[\varphi_{u,X}(t_0)],\quad t_0\in[0,1],
\end{align}
where $N_{u,X}^{-}[0,t_0]:=\#\{0\leq t \leq t_0:X(t)=u(t),X'(t)<u'(t)\}$ denotes the number of down-crossings over $[0,t_0]$.

In this work, we derive new expressions for $\E[\varphi_{u,X}(t_0)]$ allowing for varying critical value functions $u(t)$. Given such an expression and a significance level $\alpha\in(0,1)$, solving $\E[\varphi_{u,X}(t_0)]=\allowbreak\alpha/2$ allows one to derive powerful (two-sided) simultaneous confidence bands, where $t_0\in[0,1]$ will be determined by the imposed fairness constraint.  For instance, in many applications in FDA, one is interested in estimating a functional parameter, $\theta = \{\theta(t): t \in [0,1]\}$, for a particular FDA model. Suppose that $\hat{\theta}(t)$ is an estimator of $\theta(t)$, such that $\big(\hat{\theta}(t) - \theta(t)\big)$ is, possibly asymptotically, a mean zero Gaussian process.  Then we can take the standardized Gaussian process $X(t)=\big(\hat \theta(t) - \theta(t)\big)/(\V(\hat{\theta}(t)))^{1/2}\sim\mathcal{N}(0,1)$ for each $t\in[0,1]$, and compute the critical value function $u_{\alpha/2}(t)$ that solves $\E[\varphi_{u,X}(t_0)]=\alpha/2$.  Our Theorem \ref{TH:MAIN} ensures that 
\begin{equation}\label{EQ:SCB_Intro}
[\hat \theta_l(t),\hat \theta_u(t)]=\hat\theta(t) \pm u_{\alpha/2}(t) \sqrt{\V\big(\hat{\theta}(t)\big)},\quad t\in[0,1],
\end{equation}
is a valid $(1-\alpha) \times 100 \%$ simultaneous (two-sided) confidence band for the parameter $\theta(t)$ over $[0,1]$.  Our Theorem \ref{TH:EST} ensures that the simultaneous confidence band in \eqref{EQ:SCB_Intro} also holds asymptotically when the underlying covariance parameters (and therefore $u$) are estimated.  Moreover, our Corollary \ref{CO:STUDt} can be used for finite sample corrections using $t$-processes to take into account estimation errors in the covariance estimates.

The main hurdle in finding a $u_{\alpha/2}(t)$ that solves $\E[\varphi_{u,X}(t_0)] = \alpha/2$ is to find general expressions for $\E[N^-_{u,X}([0,t_0])]$ and $\E[N_{u,X}([0,t_0])]$, since a formula for the correction term, $P(X(t_0)\geq u(t_0))$, follows directly from the (possibly asymptotic) distribution of $X$. Focusing on symmetric distributions, a formula for the up-crossings can be translated into one for the down-crossings since $\E[N_{u,X}^-([0,t])] = \E[N_{-u,X}([0,t])]$.  Explicit formulas for $\E[N_{u,X}([0,1])]$ are grouped together under ``Kac-Rice formulas'' acknowledging the works of \cite{K1943} and \cite{R1944}. In this paper we generalize the existing Kac-Rice formulas by allowing for adaptive critical value functions $u(t)$. Another apparent challenge is that $\E[\varphi_{u,X}(t_0)] = \alpha/2$ will not, in general, have a unique solution when we allow for more general $u(t)$. However, far from being a liability, we show how this allows one to select bands adaptively, for instance, according to a given fairness constraint. While other criteria such as minimum width bands are possible too (see our discussion in Section \ref{SEC:DISS}), we focus on fairness, which has been a recent focus in machine learning, but so far was not discussed in the context of confidence bands.
%%%%%%%%%%%%%%
%%%%%%%%%%%%%%

\subsection{Fairness Constraint of False Positive Rate Balance}

\begin{figure}[t!]
  \centering
  \includegraphics[width=\textwidth]{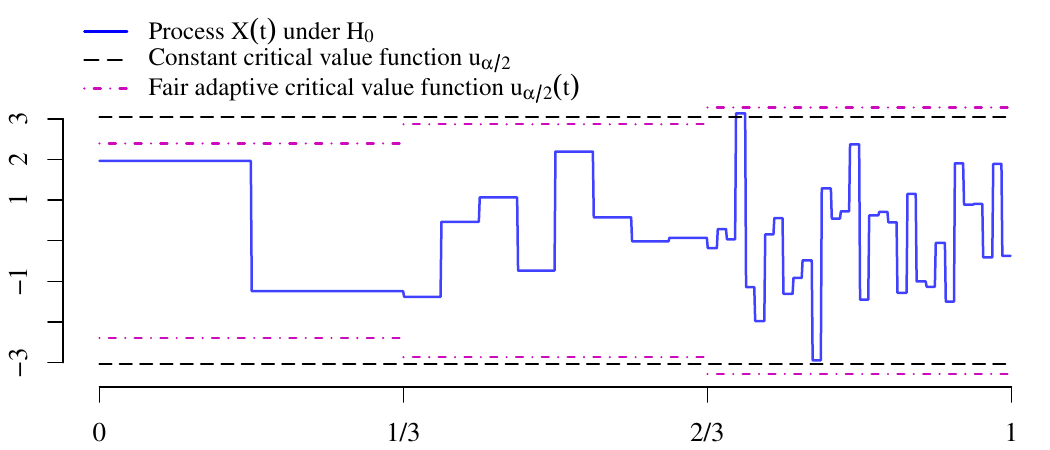}
  \caption{A realization of the piecewise constant $z$-test statistic process
    $X(t)$ as defined at the beginning of Section \ref{sec:AwFair}, the unfair
    constant critical value $u_{\alpha}=|z_{\alpha/24}|$ and the fair
    adaptive critical value function.} \label{FIG:Awareness4Fairness}
\end{figure}

To explain and motivate the definition of false positive rate balance, let us consider, for a moment, the trivial special case of a moment estimator
$\hat{\theta}_n(t)=n^{-1}\sum_{i=1}^nS_i(t)$, $t\in[0,1]$, that estimates the
mean function
$\theta(t)=\E(S_i(t))$ from an iid Gaussian random process sample
$S_i\overset{\textrm{iid}}{\sim}\mathcal{N}(\theta,C)$, $i=1,\dots,n$,
with independent, piecewise constant sample paths\footnote{The corresponding covariance
  function $C(s,t)=\cov(S_i(t),S_i(s)$ is non-stationary, bloc-diagonal
  with 42 blocs $C(t,s)=\kappa_j>0$ for $s,t\in[a_{j-1},a_j)$, $1\leq j\leq
  42$, formed by two equidistant intervals $[a_{j-1},a_j)$, $1\leq j\leq 2$, in
  $[1,1/3]$, eight equidistant intervals $[a_{j-1},a_j)$, $2\leq j\leq 10$, in
  $[1/3,2/3]$, and 32 equidistant intervals $[a_{j-1},a_j)$, $10\leq j\leq 42$, in
  $[2/3,1]$.}: two equidistant constant sections over
$[0,1/3]$, eight over $[1/3,2/3]$, and 32 over $[2/3,1]$;
see Figure \ref{FIG:Awareness4Fairness}.  Under the null hypothesis H$_0$:
$\theta(t)=\theta_0(t)$, the $z$-test statistic is a Gaussian process
$X(t)=\sqrt{n}\big(\hat{\theta}_n(t)-\theta_0(t)\big)/\sqrt{C(t,t)}\overset{\textrm{H}_0}{\sim}\mathcal{N}(0,1)$.

Testing simultaneous hypothesis
about the mean function $\theta$ requires
a critical value $u_{\alpha/2}>0$ that controls the false positive rate simultaneously for all $t\in[0,1]$ at a given significance
level $\alpha\in(0,1)$. For the here considered trivial case of a Gaussian
process $X$ with $42$
independent piecewise constant sections, the correct critical value is given by a Bonferroni correction
$u_{\alpha/2}=|z_{(\alpha/2)/42}|$ with $z_{(\alpha/2)/42}$ denoting the
$(\alpha/2)/42$-quantile of the standard normal distribution, since
\begin{equation*}
  P_{\text{H}_0}(\exists t\in[0,1]:|X(t)|\geq u_{\alpha/2})=42P(|Z|\geq|z_{(\alpha/2)/42}|)=42(\alpha/42)=\alpha,
\end{equation*}
where $Z\sim\mathcal{N}(0,1)$. However, since $C(s,t)$ is non-stationary, the
constant critical value $u_{\alpha/2}=|z_{(\alpha/2)/42}|$ leads to
unbalanced false positive rates. Only
$(2/42)\approx 5\%$
of all false positive events occur over $[0,1/3]$,
$(8/42)\approx 19\%$ over $[1/3,2/3]$, but
$(32/42)\approx 76\%$ over $[2/3,1]$,
\begin{equation*}
  \begin{array}{r@{\,:\,}lcccr}
    P_{\text{H}_0}(\exists t\in[0,1/3]&|X(t)|\geq u_{\alpha/2})&=&\phantom{3}2P(|Z|\geq|z_{(\alpha/2)/42}|)&=&\phantom{2}\left(2/42\right)\alpha\\
    P_{\text{H}_0}(\exists t\in[1/3,2/3]&|X(t)|\geq u_{\alpha/2})&=&\phantom{3}8P(|Z|\geq|z_{(\alpha/2)/42}|)&=&\phantom{2}\left(8/42\right)\alpha\\
    P_{\text{H}_0}(\exists t\in[2/3,1]&|X(t)|\geq u_{\alpha/2})&=&32P(|Z|\geq|z_{(\alpha/2)/42}|)&=&\left(32/42\right)\alpha.
  \end{array}
\end{equation*}
%%%%%%%%%%%%%%%%%%%%%%
% \spacingset{1}
% \spacingset{1.00001}
%%%%%%%%%%%%%%%%%%%%%%

To balance the false positive rates over the partition $0<1/3<2/3<1$
one needs an adaptive critical value function $u_{\alpha/2}(t)$ that allocates the
significance level $\alpha$ according to the Lebesgue measure (interval length)
of each partition, i.e.~$\lambda([0,1/3])\alpha=(1/3)\alpha$, $\lambda([1/3,2/3])=(1/3)\alpha$, and
$\lambda([2/3,1])=(1/3)\alpha$. In the trivial case of a Gaussian process with
independent, piecewise constant sample path sections, such a fair
critical value function is simply derived using local partition wise Bonferroni correction $u_{\alpha/2}(t)=|z_{(\alpha/2)/6}|$ for
$t\in[0,1/3)$, $u_{\alpha/2}(t)=|z_{\alpha/24}|$ for $t\in[1/3,2/3)$, and
$u_{\alpha/2}(t)=|z_{\alpha/96}|$ for $t\in[2/3,1)$, which yields balanced false
positive rates
\begin{equation*}
  \begin{array}{rcccr}
    P_{\text{H}_0}(\exists t\in[0,1/3]:X(t)\geq u_{\alpha/2}(t))&=&\phantom{3}2P(|Z|\geq|z_{(\alpha/2)/6}|)&=&\left(1/3\right)\alpha\\
    P_{\text{H}_0}(\exists t\in[1/3,2/3]:X(t)\geq u_{\alpha/2}(t))&=&\phantom{3}8P(|Z|\geq|z_{(\alpha/2)/24}|)&=&\left(1/3\right)\alpha\\
    P_{\text{H}_0}(\exists t\in[2/3,1]:X(t)\geq u_{\alpha/2}(t))&=&32P(|Z|\geq|z_{(\alpha/2)/96}|)&=&\left(1/3\right)\alpha
  \end{array}
\end{equation*}
while at the same time the fair critical value function $u_{\alpha/2}(t)$
correctly bounds the false positive rate over $[0,1]$, i.e.
$P_{\text{H}_0}(\exists t\in[0,1]:X(t)\geq u_{\alpha/2}(t))=\alpha$.

This trivial example (Fig.~\ref{FIG:Awareness4Fairness}) motivates our Definition \ref{Def:Fairness} of false positive rate balance for stochastic processes which adapts the fairness constraint of false positive rate balance from the machine learning literature \citep{HaPrSr2016,MoRo_2021}. However, unlike the above trivial case, we consider in this paper the non-trivial case of stochastic processes, $X_i$, with \emph{smooth} sample paths, $X_i\in C^1[0,1]$, and \emph{unknown} covariance function, $\cov(X_i(t),X_i(s))$ that is allowed to be non-stationary and non-zero for all $t,s\in[0,1]$. Thus our critical value functions are smooth adaptive functions allowing inference under the fairness constraint of false positive rate balance as defined in the following.

% do not consider the trivial case of piecewise constant
% processes, but the much more involved case of non-stationary smooth stochastic
% processes. However, the above example motivates our definition of \emph{false positive rate balance}
% for smooth stochastic processes with respect to any given partitions of the domain $[0,1]$.

\begin{definition}\label{Def:Fairness} {\normalfont\bf(False Positive Rate Balance)}
  Let $0=a_0<a_1<\dots<a_p=1$, $p\geq 2$, be a given partition of $[0,1]$, let $\alpha\in(0,1)$ denote the significance level and let
  H$_0$ denote the null hypothesis. A
  simultaneous statistical hypothesis test satisfies false positive rate balance
  if for each $j=1,\dots,p$
  \begin{align*}
    P(\textrm{Reject H$_0$ over $[a_{j-1},a_j]$}|\textrm{H$_0$ is true over $[a_{j-1},a_j]$})&\leq\lambda([a_{j-1},a_j])\,\alpha,\\
\text{and}\quad  P(\textrm{Reject H$_0$ over $[0,1]$}|\textrm{H$_0$ is true over $[0,1]$}) & \leq\lambda([0,1])\,\alpha,
  \end{align*}
  where $\lambda([a,b])=(b-a)$ denotes the Lebesgue measure of $[a,b]$.
  % For
  % exact tests, the weak inequalities are binding, for conservative tests they
  % are not binding.
\end{definition}
Simultaneous confidence bands that fulfill Definition \ref{Def:Fairness} are interpretable globally over the total domain $[0,1]$, as well as locally over  subintervals $[a_{j-1},a_j]\subset[0,1]$, $j=1,2,\dots,p$, or combinations of subintervals.
%subintervals $\bigcup_{j\in\mathcal{J}}[a_{j-1},a_j]$ with $\mathcal{J}\subseteq\{1,2,\dots,p\}$.

%%%%%%%%%%%%%%%%%%%%%%%%%%%%%%%%%%
\section{Theory and Methods}\label{SEC:MT}
%%%%%%%%%%%%%%%%%%%%%%%%%%%%%%%%%%

Our first main result is Theorem \ref{TH:MAIN}, which establishes the Kac-Rice formula for variable critical value functions, $u=\{u(t):t \in [0,1]\}$, and general elliptical processes, $X = \{X(t): t \in [0,1]\}$. The results are established under the very mild assumption that the considered process, $X$, is in $C^1[0,1]$ and that the critical value function, $u(t)$, is in $C[0,1]$ and continuously differentiable almost everywhere. In our second main result, Theorem \ref{TH:EST}, we consider the problem of estimating the confidence band for a process that is only asymptotically normal/elliptical.  A key point is that we only need consistent estimates of the covariance parameters about the diagonal. This interesting property makes our method applicable even for fragmentary functional data where an estimation of the full covariance function is impossible.
% We discuss how to
% produce such estimators in Section \ref{SSEC:CONSTRUCT}.  Since we allow for
% variable critical value functions $u(t)$, the equation
% $\E[\varphi_{u,X}(t_0)]=\alpha/2$ has, in general, no unique solution, but is
% solved by an entire class of critical value functions.  Therefore, we show that the
% entire solution class of estimated critical value functions converges in Hausdorff
% distance as the sample size tends to infinity, $n\to \infty$.
%  From this
% solution class, one
% can choose fair critical value functions (see Section \ref{SSEC:FAIR}).
Throughout this paper we will make the following assumption.
%%%%%%%%%%%%%%%%%%%%%%%%%%%%%%%%%
\begin{assumption} \label{A:ELIP}
We assume that $X=\{X(t): 0\leq t\leq 1\}$ is a centered elliptical process with $X\in C^1[0,1]$ almost surely.
\end{assumption}
%%%%%%%%%%%%%%%%%%%%%%%%%%%%%%%%%
The assumption that $X\in C^1[0,1]$ almost surely means that the realized paths of $X$ will, with probability one, be one time continuously differentiable, which excludes, for instance, Brownian motion.  As our methodology is based on counting up-crossings, some smoothness is required, otherwise the number of up-crossings may be infinite.

Of course, not every stochastic process is elliptical, but allowing $X$ to be elliptically distributed provides at least two major benefits. First, in applications where an elliptical distribution (e.g.~a $t$-distributed estimator) may be reasonably assumed, we can provide small sample inference. Second, in large samples, our framework of elliptical processes allows to make use of central limit theorems making restrictive distributional assumptions obsolete, and additionally provides the possibility to include a finite sample correction.

One key aspect of elliptical processes is that they can be expressed as scalar mixtures of Gaussians; see, for instance, \cite{BBT2014} for the case of functional data and \cite{FW2018_book} for a general overview in the case of multivariate data.
%%%%%%%%%%%%%
\begin{lemma}\label{LEM:EP}
  If $X=\{X(t): 0\leq t\leq 1\}$ is a centered elliptical process with a
  covariance operator that does not have finite rank,
  then there exists a strictly positive random variable $V > 0$ and a mean zero Gaussian process, $Z=\{Z(t):0\leq t \leq 1\}$, such that $V$ and $Z$ are independent and satisfy $\{X(t): 0\leq t \leq 1\} \overset{D}{=} \{V Z(t): 0 \leq t \leq 1\}$.
\end{lemma}
%%%%%%%%%%%%%

If $V$ is heavy tailed, then $X(t)$ need not even have a finite mean. We will thus refer to the covariance of $Z(t)$, $c(t,s)=\cov(Z(t),Z(s))$, as the \textit{dispersion function} of $X(t)$. If $V$ has a finite and non-zero variance, then the covariance function of $X$ is given by $\cov(X(t),X(s))=\var(V)c(t,s)$ for all $t,s\in[0,1]$. In the Gaussian case, where $V\equiv\sigma$, the covariance function of $X$ is given by $\cov(X(t),X(s))=\sigma^2c(t,s)$ for all $t,s\in[0,1]$. Since we are working with an arbitrary critical value function $u(t)$, we can always assume that it has been scaled by the standard deviation of $Z(t)$, $\sqrt{c(t,t)}$, so that, without loss of generality, we can assume $c(t,t) = 1$.

%%%%%%%%%%%%%%%%%%%%%%%%%%%%%%%%%%%
\subsection{Main Theoretical Results}\label{SSEC:TR}
%%%%%%%%%%%%%%%%%%%%%%%%%%%%%%%%%%%
%\subsubsection*{Generalized Kac-Rice Formula}
In this section we present our key theoretical results.  We begin with our first
theorem, which establishes a generalized Kac-Rice formula for varying critical
value functions $u(t)$ and elliptical processes $X(t)$.  We employ notation such as $\partial_{12} c(t,t)$ to denote partial derivatives of $c(t,s)$ in the first and second coordinates, but then evaluated at $t=s$.  Moreover, we write $C^1_{a.e.}[0,1]$ to denote the space of all continuous functions that are continuously differentiable almost everywhere on $[0,1]$. Note that $C^1[0,1]\subset C^1_{a.e.}[0,1]$, since $C^1_{a.e.}[0,1]$ also contains, for instance, piece-wise linear functions.
%%%%%%%%%%%%%%%%%%%%%%%%%%%%
\begin{theorem}\label{TH:MAIN}
Let Assumption \ref{A:ELIP} hold and assume that $u\in C^1_{a.e}[0,1]$.  Let $V>0$ be the mixing coefficient of $X$, such that $X(t)\overset{D}{=}VZ(t)$, and define $\mcV = V^{-2}$ along with its moment generating function, $M_{\mcV}(\cdot)$.
%%%%%%%%%
Let $\tau(t)^2 = \partial_{12} c(t,t) = \var(Z'(t))$, where $c(t,s)$ is the dispersion function of $X(t)$, equivalently the covariance function of $Z(t)$.  Assume $X(t)$ has a constant pointwise unit dispersion, $c(t,t) = \var(Z(t))= 1$, and $\tau(t)>0$ for all $t\in[0,1]$. Then we have for any fixed $t_0 \in [0,1]$
\begin{align}\label{EQ:GRF_Ell}
\begin{split}
\E[\varphi_{u,X}(t_0)]  = &
P\big(X(t_0) \geq u(t_0)\big) +
\int_{0}^1 \frac{\tau(t)}{2 \pi } M_\mcV \left(-\frac 1 2 \left[u(t)^2  + \frac{u'(t)^2}{ \tau(t)^2} \right]  \right) \ dt \\
& + \int_0^{t_0} \int_0^\infty  \frac{u'(t)}{2 \pi\tau(t)}
M_\mcV'\left( - \frac 1 2 \left[ u(t)^2+ \frac{(y-u'(t))^2}{\tau(t)^2} \right]  \right)  \ dy \ dt\\
& - \int_{t_0}^1 \int_0^\infty  \frac{u'(t)}{2 \pi\tau(t)}
M_\mcV'\left( - \frac 1 2 \left[ u(t)^2+ \frac{(y+u'(t))^2}{\tau(t)^2} \right]  \right)  \ dy \ dt.
\end{split}
\end{align}
\end{theorem}
%%%%%%%%%%%%%%%%%%%%%%%%%%%%

Note that since $\mcV$ is strictly positive the above integrals are finite regardless of the mixture distribution, meaning $X(t)$ can be heavy tailed and the expression still holds.  Thus $X(t)$ need not even have a finite mean or variance which allows us to model even extreme value processes.  The roughness parameter function $\tau$ allows to quantify the extent of the local multiple testing problem across the domain $[0,1]$.

The proof of Theorem \ref{TH:MAIN} proceeds in several steps.  First, we condition on the mixing parameter
$\E[\varphi_{u,X}(t_0)] =
 \E[\E[\varphi_{u,VZ}(t_0)|V]]=
 \E[\E[\varphi_{u/V,Z}(t_0)|V]]$
so that we can exploit the connection to Gaussian processes.  Conditioned on
$V$, we build up two key approximations, which are common in random process theory, though much
more complicated here since $u$ is not restricted to a constant function.  One
is based on approximating the number of up-crossings by integrating a particular
continuous kernel, and the second is based on a linear interpolation along
dyadics.  Both approximations make the expectation easier to calculate and then
we justify taking appropriate limits.

The following three corollaries consider important special cases that simplify the interpretation of our general formula \eqref{EQ:GRF_Ell} in Theorem \ref{TH:MAIN}.
%%%%%%%%%%%%%%%%%%%%%%%%%%%%%%%%%%%%%%%%%%%%%%%%%%%%
\begin{corollary}\label{CO:CONST} {\normalfont\bf(Constant band for elliptical processes)}
Let the conditions of Theorem \ref{TH:MAIN} hold. If $u(t) \equiv u$ for all $t\in[0,1]$, then one obtains the elliptical version of the classic Kac-Rice formula
$\E[\varphi_{u,X}(0)]=P\big(X(0) \geq u\big) +
\frac{\|\tau\|_{1}}{2 \pi} M_\mcV \left(- \frac{u^2}{2}  \right)$,
where $\|f\|_{1}=\int_0^1|f(t)|dt$ denotes the $L^1$-norm.
\end{corollary}
%%%%%%%%%%%%%%%%%%%%%%%%%%%%%%%%%%%%%%%%%%%%%%%%%%%%
Corollary \ref{CO:CONST} directly follows from Theorem \ref{TH:MAIN} since
$u'\equiv 0$ for constant critical values, $\tau(t) \geq 0$ for all $t\in[0,1]$
such that $\int_0^1\tau(t)dt=||\tau||_1$, and $X(t_0) \overset{D}{=} X(0)$ for
all $t_0\in[0,1]$.

%%%%%%%%%%%%%%%%%%%%%%%%%%%%%%%%%%%%%%%%%%%%%%%%%%%%
\begin{corollary}\label{CO:GAUSS} {\normalfont\bf(Gaussian processes)} Let the conditions of Theorem \ref{TH:MAIN} hold.\\
%%%%%%%%%%%%%%%%%%%%%%
a) Variable band:
%%%%%%%%%%%%%%%%%%%%%%
When $X(t)$ is a Gaussian process, we set $V\equiv\sigma$ such that $\var(X(t))=\allowbreak\sigma^2\var(Z(t))=\allowbreak\sigma^2$ for all $t\in[0,1]$. Formula \eqref{EQ:GRF_Ell} in Theorem \ref{TH:MAIN} then leads to
%%%%%%%%%
\begin{align}\label{EQ:GRF_Gauss}
\begin{split}
\E[\varphi_{u,X}(t_0)]  =
\Phi\left(\frac{-u(t_0)}{\sigma}\right) & +
\int_0^1 \frac{\tau(t)}{2 \pi}
\exp\left\{- \frac{1}{2 \sigma^2}\left[ {u(t)^2} + \frac{u'(t)^2}{\tau(t)^2} \right]\right\} \ dt  \\
& +\int_0^{t_0} \frac{u'(t)}{\sqrt{2 \pi \sigma^2}}
\exp\left\{- \frac{u(t)^2}{2 \sigma^2}\right\} \Phi\left(\frac{u'(t)}{ \sigma \tau(t)}\right) \ dt \\
& -\int_{t_0}^1 \frac{u'(t)}{\sqrt{2 \pi \sigma^2}}
\exp\left\{- \frac{u(t)^2}{2 \sigma^2}\right\} \Phi\left(\frac{-u'(t)}{ \sigma \tau(t)}\right) \ dt.
\end{split}
\end{align}
\begin{comment}
or equivalently
\begin{align*}
\E[\varphi_{u,X}(t_0)]  =
	 \Phi( -u(t_0) / \sigma) & +
\int_0^1 \frac{\tau(t)}{2 \pi}
\exp\left\{- \frac{1}{2 \sigma^2}\left[ {u(t)^2} + \frac{u'(t)^2}{\tau(t)^2} \right]\right\} \ dt  \\
& +\int_0^{t_0} \frac{u'(t)}{\sqrt{2 \pi \sigma^2}}
\exp\left\{- \frac{u(t)^2}{2 \sigma^2}\right\} \ dt \\
& -\int_{0}^1 \frac{u'(t)}{\sqrt{2 \pi \sigma^2}}
\exp\left\{- \frac{u(t)^2}{2 \sigma^2}\right\} \Phi\left(\frac{-u'(t)}{ \sigma \tau(t)}\right) \ dt.
\end{align*}
\end{comment}
%%%%%%%%%%%%%%%%%%%%
b) Constant band:
%%%%%%%%%%%%%%%%%%%%
If additionally $u(t) \equiv u$, one obtains the classic Kac-Rice formula
\begin{equation}\label{EQ:KacRice_Gauss}
 \E[\varphi_{u,X}(0)]  =
	 \Phi\left(\frac{-u}{\sigma}\right)+
\frac{ \|\tau\|_{1}}{2 \pi}\exp\left\{-\frac{u^2}{2 \sigma^2}\right\}.
\end{equation}
\end{corollary}
%%%%%%%%%%%%%%%%%%%%%%%%%%%%%%%%%%%%%%%%%%%%%%%%%%%%

%%%%%%%%%%%%%%%%%%%%%%%%%%%%%%%%%%%%%%%%%%%%%%%%%%%%
\begin{corollary}\label{CO:STUDt} {\normalfont\textbf{(}}\textbf{t}{\normalfont\textbf{-processes)}} Let the
  conditions of Theorem \ref{TH:MAIN} hold. \\
%%%%%%%%%%%%%%%%%%%%%%
a) Variable band:
%%%%%%%%%%%%%%%%%%%%%%
When $X(t)$ is a $t$-process with $\nu$ degrees of freedom, meaning $\mcV \sim \chi^2_\nu/ \nu$, formula \eqref{EQ:GRF_Ell} in Theorem \ref{TH:MAIN} leads to
\begin{align}\label{EQ:GRF_t}
\begin{split}
&\E[\varphi_{u,X}(t_0)] =
	 F_t( -u(t_0);\nu) +\int_{0}^1 \frac{\tau(t)}{2 \pi }
\left(
1 +  \frac{u(t)^2}{ \nu }  + \frac{u'(t)^2}{  \nu \tau(t)^2}
\right)^{-\nu/2} \ dt \\
& +\int_{0}^{t_0} \frac{u'(t)}{2 \pi \tau(t) }\left(1+ \frac{u(t)^2}{\nu }\right)^{-\nu/2 -1} \frac{\Gamma((\nu+1)/2) \sqrt{(\nu+1) \pi }a(t)}{\Gamma((\nu+2)/2)} F_t\left(\frac{u'(t)}{a(t)};\nu+1\right)\ dt \\
& -\int_{t_0}^1 \frac{u'(t)}{2 \pi \tau(t) }\left(1+ \frac{u(t)^2}{\nu }\right)^{-\nu/2 -1} \frac{\Gamma((\nu+1)/2) \sqrt{(\nu+1) \pi }a(t)}{\Gamma((\nu+2)/2)} F_t\left(\frac{-u'(t)}{a(t)};\nu+1\right)\ dt,
\end{split}
\end{align}
where $a(t)^2 := \nu \tau(t)^2 (1+ u(t)^2/\nu )/(\nu+1)$, $\Gamma$ denotes the gamma function, and $F_t(.;\nu)$ denotes the cumulative distribution function of a $t$-distribution with $\nu$ degrees of freedom.\\[1.5ex]
\begin{comment}
or equivalently
\begin{align*}
\E[\varphi_{u,X}(t_0)]  & =
	 F_\nu( -u(t_0)) \\
	 & +
\int_{0}^1 \frac{\tau(t)}{2 \pi }
\left(
1 +  \frac{u(t)^2}{ \nu }  + \frac{u'(t)^2}{  \nu \tau(t)^2}
\right)^{-\nu/2} \ dt \\
& +\int_{0}^{t_0} \frac{u'(t)}{2 \pi \tau(t) }\left(1+ \frac{u(t)^2}{\nu }\right)^{-\nu/2 -1} \frac{\Gamma((\nu+1)/2) \sqrt{(\nu+1) \pi }a(t)}{\Gamma((\nu+2)/2)} \ dt, \\
& -\int_{0}^1 \frac{u'(t)}{2 \pi \tau(t) }\left(1+ \frac{u(t)^2}{\nu }\right)^{-\nu/2 -1} \frac{\Gamma((\nu+1)/2) \sqrt{(\nu+1) \pi }a(t)}{\Gamma((\nu+2)/2)} F_t(-u'(t)/a(t);\nu+1)\ dt.
\end{align*}
\end{comment}
%%%%%%%%%%%%%%%%%%%%%%
b) Constant band:
%%%%%%%%%%%%%%%%%%%%%%
If additionally $u(t) \equiv u$, one obtains the t-version of the classic Kac-Rice formula
\begin{equation}\label{EQ:KacRice_t}
\E[\varphi_{u,X}(0)] =
F_t(-u;\nu) +
\frac{||\tau||_1}{2 \pi } \left( 1 +  \frac{u^2}{ \nu } \right)^{-\nu/2}.
\end{equation}
\end{corollary}
%%%%%%%%%%%%%%%%%%%%%%%%%%%%%%%%%%%%%%%%%%%%%%%%%%%%

As with the Gaussian distribution, the t-expression does not have a
closed form in general.  However, in terms of numerically finding $u(t)$, the
above formulations can be readily employed. Equivalently for more general
elliptical distributions the critical value $u$ can be found numerically, as long as one has a convenient form for $M_\mcV$.

Our second main result concerns statistical estimation of the bands in practice.
Let $X_n=\hat{\theta}_n-\theta$ be the correctly (under the
null hypothesis) centered estimator $\hat{\theta}_n$ of some functional
parameter $\theta$. Here, we assume that $X_n$ is only
asymptotically (large $n$)
Gaussian/elliptical in $C^1[0,1]$.  In practice, we usually do not know the
dispersion function $c(t,s)$ and thus cannot immediately normalize to assume
$\cov(Z(t),Z(t))=c(t,t)=1$ as assumed in Theorem \ref{TH:MAIN} which is without loss of
generality due to the variable critical value functions. Instead, we
assume that we have a sequence of uniformly consistent estimates
$\hat{c}_n(t,t)$ of the dispersion $c(t,t)$ and its partial derivatives, where
$c(t,t)>0$.

%%%%%%%%%%%%%%%%%%%%%%%%%%%%%%%%%%
\begin{assumption}\label{a:est}
Let $c(t,s) = \cov(Z(t),Z(s))$ be the dispersion function of $X$ and assume the
limiting distribution (for large $n$) of $V$ in $X=VZ$ is known. Assume that we have a sequence of estimators, $\hat{c}_n(t,s)$, for $n=1,2,\dots$ satisfying,
% \begin{align*}
%   \sup_{0 \leq t \leq 1}&|\hat{c}_n(t,t) - c(t,t)| = o_P(1)\\
%   \sup_{0 \leq t \leq 1}&|\partial_1 \hat{c}_n(t,t) - \partial_1 c(t,t)| = o_P(1)\\
%   \sup_{0 \leq t \leq 1}&|\partial_{12} \hat{c}_n(t,t) - \partial_{12} c(t,t)| = o_P(1)
% \end{align*}
$\sup_{0 \leq t \leq 1}|\hat{c}_n(t,t) - c(t,t)| = o_P(1)$,
$\sup_{0 \leq t \leq 1}|\partial_1 \hat{c}_n(t,t) - \partial_1 c(t,t)| =
o_P(1)$, and
$\sup_{0 \leq t \leq 1}|\partial_{12} \hat{c}_n(t,t) - \partial_{12} c(t,t)| = o_P(1)$
and that $c(t,t) > 0$ for all $0 \leq t \leq 1$.
\end{assumption}
%%%%%%%%%%%%%%%%%%%%%%%%%%%%%%%%%%
In cases where
$c$ can be estimated by averaging a random sample of stochastic
processes, Theorem 1 in \cite{DK_2022} can be used to establish these convergence results.

\begin{remark}\label{RE:ASSest}
If $V$ has a finite and non-zero variance, then $c(t,s)$ can be estimated by
$\hat{c}_n(t,s)=\var(V)^{-1}\widehat{\cov}_n(X(t),X(s))$. Since $c(t,s)$
is symmetric, $\partial_1 \hat{c}_n(t,t)=\partial_2 \hat{c}_n(t,t)$ and
$\partial_1 c(t,t)=\partial_2 c(t,t)$. Moreover, note that we do not require estimation of the full
dispersion, $c(t,s)$, only its diagonal, which is needed to normalize the
process, and the derivatives along the diagonal, which are needed to estimate
$\tau(t)$. Practically, this would
usually require that $\hat{c}_n$ converges to $c$ in say $C^2$. However, notice that
%%%
$\partial_1 c(t,t) = \cov(Z(t),Z'(t))$ and that $\partial_{12}c(t,t) = \var(Z'(t))=\tau(t)^2$.
%%%
Therefore, in many settings, it is possible to estimate these covariance
quantities more directly, as for instance in the arguably most relevant case,
where $X(t)$ is formed by averaging a random sample of stochastic processes.
Equation \eqref{EQ:tau_2} in Example \ref{EX:FFSCB_t} gives an example of such a
more direct estimation of $\tau$ by using that $\tau(t)^2=\var(Z'(t))$.
\end{remark}

\begin{remark}
Given an estimator $\hat{c}_n$ we can normalize $X_n$ such
that, by Slutsky's lemma, it follows that
$\tilde X_n(t) := c_n^{-1/2}(t,t) X_n(t) \overset{D}{\to} c^{-1/2}(t,t) X(t) = \tilde X(t)\quad\text{as}\quad n\to\infty$.
The process $\tilde X(t)$ fulfills then the assumptions of Theorem \ref{TH:MAIN}, so we
need only plug in its corresponding $\tilde \tau(t)$ into Theorem \ref{TH:MAIN}
to find corresponding bands.  By Assumption \ref{a:est}, we can construct a
sequence of uniformly consistent estimators $\tilde \tau_n(t) \to \tilde
\tau(t)$.  While this presents a minor notational annoyance, practically, it
simply amounts to constructing the band from standardized data or using
$\hat{c}_n(t,s)/\sqrt{\hat{c}_n(t,t),\hat{c}_n(s,s)}$ to find $u(t)$; see Examples
\ref{EX:FFSCB_Gauss} and \ref{EX:FFSCB_t} below.
%%%%%%%%%%%%%%%
%This is because
 %\[\tilde c_n(t,s) := \frac{c_n(t,s)}{\sqrt{c_n(t,t) c_n(s,s)}} \Longrightarrow
%\tilde \tau_n(t) = \partial_x \partial_y \tilde c_n(x,y) |_{x=y=t}.\]
 %And we have
 %\begin{align*}
 %\partial_x \partial_y \tilde c_n(x,y)
 %& = \partial_x \left[\frac{c_{n,2}(x,y)}{\sigma_n(x) \sigma_n(y)}  - \frac{c_n(x,y) \sigma_n'(y)}{\sigma_n(x) \sigma_n^2(y)} \right] \\
 %& = \frac{c_{n,1,2}(x,y)}{\sigma_n(x) \sigma_n(y)}
% - \frac{c_{n,2}(x,y) \sigma_n'(x)}{\sigma_n^2(x) \sigma_n(y)}
%- \frac{c_{n,1}(x,y) \sigma_n'(y)}{\sigma_n(x) \sigma_n^2(y)}
%+  \frac{c_{n}(x,y) \sigma_n'(x) \sigma_n'(y)}{\sigma_n^2(x) \sigma_n^2(y)}
% \end{align*}
%Clearly, as long as the dispersion is continuous and $C(t,t) > 0$ for all $t$, we can rescale $X_n(t)$ so that $X(t)$ has constant dispersion $\sigma = 1$.  Thus, as long as we have a uniformly consistent estimate, we can assume without loss of generality that $\sigma =1$.  However, note that we assume that the distribution of $V$ is known, which is equivalent to assuming one knows what stochastic family $X$ is part of (e.g. Gaussian, t, etc).
%%%%%%%%%%%
\end{remark}

Our other key assumption is that the space of potential bands, $\mcU \subseteq C^1_{a.e}[0,1]$, is convex, compact, and contains the constant functions (up to an appropriate bound).   Compactness is commonly needed in estimation theory to ensure that the estimators are well behaved.  The convexity combines with the constant functions to eliminate some pathological  limiting problems. In particular, it ensures that if $u \in \mcU$ isn't on the boundary, then $u(t) + c$ will also be in $\mcU$ for $c$ small.  This helps ensure that no asymptotic band is ``isolated'', in the sense that one cannot construct a corresponding sequence of estimators for which it is the corresponding limit.

\begin{theorem}\label{TH:EST}
Let Assumptions \ref{A:ELIP} and \ref{a:est} hold and fix $\alpha \in (0,1)$ and $t_0 \in [0,1]$.  %Assume $\tau \in C[0,1]$ and that $\tau_n \in C[0,1]$ is a sequence of estimators satisfying $\sup_{0\leq t\leq 1}|\tau_n(t) - \tau(t)| = o_P(1)$.
Define $\tilde \tau_n(t) = \partial_{12} \tilde c_n(t,s)$ where $\tilde c_n(t,s):=c_n(t,s)/\sqrt{c_n(t,t)c_n(s,s)}$ and define $\tilde \tau(t)$ analogously.
 %are such that $u(t) \geq c> 0$ for all $t$ and some fixed constant $c \in \mbR$.  Assume that
Assume that $\{u \in \mcU\} \subseteq C^1_{a.e.}[0,1]$, where
$\mcU$ is convex, compact, and contains the constant functions (up to an
appropriate threshold to maintain compactness).  For $\E[\varphi_{u,X}(t_0)]$ as
in \eqref{EQ:GRF_Ell}, for a general $\eta \in C[0,1]$ with $\eta(t) \geq 0$ for all
$t\in[0,1]$, and for a non-negative real-valued slackness function
$\varsigma:u\in\mathcal{U}\mapsto \varsigma(u)\in\mathbb{R}_{\geq 0}$ that is continuous for all $u\in \mcU$, define the function\\[-10mm]
%%%%
\begin{align*}
  f(u, \eta)&:= \E[\varphi_{u,X}(t_0)]+\varsigma(u)\\
%{\color{black}P\big(X(t_0)\geq u(t_0)\big)} & +
%\int_{0}^1 \frac{\eta(t)}{2 \pi } M_\mcV \left(-\frac 1 2 \left[u(t)^2  + \frac{u'(t)^2}{ \eta(t)^2} \right]  \right) \ dt \\
%& + \int_0^{t_0} \int_0^\infty  \frac{u'(t)}{2 \pi \eta(t)}
%M_\mcV'\left( - \frac 1 2 \left[ u(t)^2+ \frac{(y-u'(t))^2}{\eta(t)^2} \right]  \right)  \ dy \ dt \\
%& - \int_{t_0}^1 \int_0^\infty  \frac{u'(t)}{2 \pi \eta(t)}
%M_\mcV'\left( - \frac 1 2 \left[ u(t)^2+ \frac{(y+u'(t))^2}{\eta(t)^2} \right]  \right)  \ dy \ dt,
\text{and the sets}\quad S_n&:= f_{\tilde \tau_n}^{-1}(\alpha) = \{u \in \mcU:
f(u,\tilde\tau_n) = \alpha/2\}\\
\text{and }\quad S &:= f_{\tilde\tau}^{-1}(\alpha)= \{u \in \mcU: f(u,\tilde \tau) = \alpha/2\}.\\[-10mm]
\end{align*}
Then we have the following:\\[-10mm]
\begin{enumerate}[wide, labelwidth=!, labelindent=0pt, itemsep=-1ex]
\item The sets $\{S_n\}$ and $S$ are nonempty and closed with probability 1.
\item If $\{u_n:n=1,\dots\}$ is any sequence with $u_n \in S_n$, then $f(u_n, \tilde \tau) \overset{P}{\to} \alpha/2$ as $n \to \infty.$
\item $S_n \to S$ in Hausdorff distance\footnote{The Hausdorff distance is given by
$d(S_n,S) = \max\left\{ \sup_{u_n \in S_n} \inf_{u \in S} \|u_n - u\|,
\sup_{u \in S} \inf_{u_n \in S_n} \|u_n - u\|
 \right\}$.} with probability one.
%%%%%%%%%%
%%%%%%%%
% \item Let $G:\mcU \to [0,\infty)$ be a continuous function used to partially order $\mcU$, in the sense that $u_1 \preceq u_2$ if $G(u_1) \leq G(u_2)$.  Then $G(S_n) \overset{P}{\to} G(S)$ in Hausdorff distance.  Furthermore, if there is a unique $u \in S$ such that $G(u) = \min_{u' \in S}G(u'),$ then any sequence $u_n$ that minimizes $G(u')$ for $u'\in S_n$ will converge in probability to $u$, and such a sequence exists with probability 1.
\end{enumerate}
\end{theorem}
%%%%

The slackness function, $0\leq\varsigma(u)<\infty$, facilitates the consideration of
additional constraints like fairness constraints. Theorem \ref{TH:EST} provides
several important asymptotic results that tie consistency of the band to
consistency of the functional parameter estimates. The overall message is that
if one finds a $\hat u$ such that $f(\hat u, \tilde \tau_n) = \alpha/2$, then,
with probability tending to one, $\hat u$ will be close to a $u$ that satisfies
$f(u,\tilde \tau) = \alpha/2$, which would be the target, but $\tilde \tau$ must
be estimated. However, there is a certain awkwardness in
stating and establishing these results since each $\alpha\in(0,1)$ leads to entire set of
critical value functions one could use.  The first result simply says, with probability one, there
exists a non-empty set of candidate critical value functions using either the true, $\tilde \tau$, or estimate,
$\tilde \tau_n$.  The second result states that any sequence of critical value functions selected
using the estimate, $\tilde \tau_n$ will asymptotically give the correct
coverage. The third result states that, as sets, the set of critical value
functions using
the estimate, $\tilde \tau_n$, converges to the set one would obtain using the
true $\tilde \tau$.
% The fourth and last result states that if one uses some specific criteria to order the bands, then the `best' band using the estimate will converge to the one using the true $\tau$, so long as its unique.

The set $\mcU$ is used to define the class of bands that one wants to consider.
In the case of the classic Kac-Rice formula, one would take $\mcU$ to be the set
of constant functions (up to some bound to ensure compactness).  However, our
theory allows for much more general compact sets containing non-constant,
possibly infinite dimensional critical value functions.
For example, if
$\{e_i:i=1,\dots,m\} \subset C^1[0,1]$ is a finite collection of linearly
independent functions, then we can take $\mcU = \mbox{span}\{e_1,\dots,e_m\}
\cap B_c(\delta)$, with $m$ not necessarily finite and $B_c(\delta)$ being the closed ball of radius $\delta < \infty$ around constant critical value functions $c$.

%This gives a convenient form for producing adaptive bands since they
%will take the form $u(t) = \sum a_i e_i(t)$ and then one need only search over
%the coefficients.

% For example, the $e_j$ could be taken to be b-splines or
% Fourier bases.  Note, however, that $\mcU$ can still be genuinely infinite
% dimensional, with $m=\infty$;
% % , as long as the coefficients are forced to zero in
% % a systematic way,
% for example, $\mcU$ may be an appropriate Sobolev or Reproducing Kernel Hilbert Space (RKHS) ball.

\subsection[]{Fair Critical Value Function
  $\boldsymbol{u^\star_{\alpha/2}}$}\label{SSEC:FAIR}

Any critical value function $u_{\alpha/2}\in\mcU_{\alpha/2}(t_0)$ with
$\mcU_{\alpha/2}(t_0)=\{u\in\mcU:\E[\varphi_{u,X}(t_0)] + \varsigma(u)=\alpha/2\}$, for
given $\alpha\in(0,1)$ and $t_0\in[0,1]$, leads to a valid $(1-\alpha)\times 100\%$ simultaneous confidence band $[\hat{\theta}_l,\hat{\theta}_u]$
as in \eqref{EQ:SCB_Intro}. This generates a whole family of valid simultaneous
confidence bands
$\mathcal{F}_{\alpha,t_0}=\{[\hat{\theta}_l,\hat{\theta}_u]:u_{\alpha/2}\in\mcU_{\alpha/2}(t_0)\}$
and one can develop procedures for selecting specific simultaneous
confidence bands
$[\hat{\theta}_l^\star,\hat{\theta}_u^\star]\in\mathcal{F}_{\alpha,t_0}$
according to some constraint like minimal squared or absolute average
band width or, as consided in the following, according to a fairness constraint.

Algorithm \ref{ALGO:FAIRNESS} below selects a fair critical value function
$u^\star_{\alpha/2}$ that enables inference under the
fairness constraint of false positive rate balance (Definition \ref{Def:Fairness}) over any given
partition $0=a_0<a_1<a_2<\dots<a_p=1$.
% The idea is to determine
% $u^\star_{\alpha/2}$ by solving the generalized Kac-Rice equations
% $\E[\varphi_{u^\star_{\alpha/2},X}(t_0)]=\lambda([a_{j-1},a_j])\alpha/2$
% separately for each sub-process $\{X(t):t\in[a_{j-1},a_j]\}$, where
% $\lambda([a_{j-1},a_j])=a_j-a_{j-1}$ denotes the Lebesgue measure of
% $[a_{j-1},a_j]$.
The following theorem states that the fair critical value
function $u^\star_{\alpha/2}$, selected by Algorithm \ref{ALGO:FAIRNESS}, allocates the fair proportional shares
$(\alpha/2)\lambda([a_{j-1},a_j])=(\alpha/2)(a_{j}-a_{j-1})$ of the nominal (two-sided) significance level $\alpha/2$ to
each subinterval $[a_{j-1},a_j]$, $j=1,2,\dots,p$, and combinations
thereof.
\begin{lemma}\label{LE:FAIRNESS} {\normalfont\bf(Fairness of $\boldsymbol{u^\star_{\alpha/2}}$)} Let the
  conditions of Theorem \ref{TH:MAIN} hold. Choose a significance level
  $\alpha\in(0,1)$ and consider a partition $0=a_0<a_1<a_2<\dots<a_p=1$ with $1\leq p<\infty$. The critical value function, $u^\star_{\alpha/2}$, selected by
  Algorithm \ref{ALGO:FAIRNESS}, allocates the fair proportional shares
  $(\alpha/2)(a_{j}-a_{j-1})$ of the nominal (one-sided) significance level $\alpha/2$ to
  each local sub-process $\{X(t):t\in[a_{j-1},a_j]\}$, $j=1,2,\dots,p$, such
  that for any $\mathcal{J}\subseteq\{1,2,\dots,p\}$
\begin{align}\label{EQ:PROB_FAIR_01}
  P\left(\exists t\in\bigcup_{j\in\mathcal{J}}[a_{j-1},a_{j}]: X(t)\geq u^\star_{\alpha/2}(t)\right)&\leq\sum_{j\in\mathcal{J}}\frac{\alpha}{2}(a_{j}-a_{j-1}).
\end{align}
\end{lemma}

Inverting Lemma \ref{LE:FAIRNESS} to the case of simultaneous confidence bands allows us to show that our simultaneous confidence bands fulfill the fairness constraint of false positive rate balance (Definition \ref{Def:Fairness}); see Proposition \ref{PRO:FAIR_BAND} below.
\smallskip

% The most important special cases of Theorem \ref{LE:FAIRNESS} are the fair
% inference results over single subintervals, i.e.~$P(\exists
% t\in[a_{j-1},a_j]:X(t)\geq u^\star_{\alpha/2}(t))\leq\frac \alpha 2
% (a_j-a_{j-1})$ for each $j=1,\dots,p$, and over the total domain $[0,1]$,
% i.e.~$P\left(\exists t\in[0,1]:X(t)\geq u^\star_{\alpha/2}(t)\right)\leq\frac{\alpha}{2}$.
% \begin{align*}
%   P(\exists
%   t\in[a_{j-1},a_j]:X(t)\geq u^\star_{\alpha/2}(t))&\leq
%   \frac \alpha 2 (a_j-a_{j-1})\quad\text{for each}\quad j=1,\dots,p,\\
%  \text{and}\quad P\left(\exists t\in[0,1]:X(t)\geq u^\star_{\alpha/2}(t)\right)&\leq\frac{\alpha}{2}.
% \end{align*}

%\spacingset{1}
\begin{algorithm}\label{ALGO:FAIRNESS} {\normalfont\bf(Selecting the fair
    critical value function $\boldsymbol{u^\star_{\alpha/2}}$)}
  \begin{description}[wide, labelwidth=!, labelindent=0pt, itemsep=-1ex]
  \item[Initialization:]
    Choose a significance level $\alpha\in(0,1)$ and a partition
    $0=a_0<a_1<a_2<\dots<a_p=1$ with $1\leq p<\infty$. Let $u_{p}(t)$ denote the
    following initially constant (over $[a_0,a_1]$) piecewise linear function:
    \[
      u_{p}(t):=c_1+c_2(t-a_{1})_{+}+\dots+c_{p}(t-a_{p-1})_{+}, \quad\text{where}\quad(x)_+=\max\{x,0\}.
    \]
    %\begin{quote}
      {\sc Comment:}
    We use $u_{p}(t)$ and $u_{p}'(t)$
    for modelling $u(t)$ and $u'(t)$ in \eqref{EQ:GRF_Ell} of Theorem
    \ref{TH:MAIN}. By partitioning the integrals in \eqref{EQ:GRF_Ell} into separate integrals over
    subintervals $[a_{j-1},a_{j}]$, $j=1,\dots,p$, we determine the
    coefficients $c_j$ of $u_{p}$ consecutively for each $j=1,\dots,q$.\\[.5ex]
  %\end{quote}
\item[For $j=1$:]
    Since $u_p(t)=c_1$ and
    $u_{p}'(t)=0$ for all $t\in[a_0,a_1]$, the integrals in
    \eqref{EQ:GRF_Ell}, restricted to $[a_0,a_1]$, simplify such that $c_{1}>0$ can be determined by solving
    %%%%%%%%%
    \begin{equation}\label{EQ:FAIRNESS_ALGO_1}
      P\big(X(a_1) \geq u_{p}(a_1)\big) + \int_{a_0}^{a_1} \frac{\tau(t)}{2 \pi}
      M_\mcV\left(- \frac{c_1^2}{2}\right) \ dt = \frac{\alpha}{2}\big(a_1-a_0\big).
    \end{equation}
    %%%%%%%%%
    %%%%%%%%%%%%%%
  \item[For $j=2,3,\dots,p$ do:]\phantom{TEXT}\\
    Given $c_{1},\dots,c_{j-1}$, $u_p(t)$ and $u'_p(t)$ for
    $t\in[a_{j-1},a_j]$ only depend on $c_j$ such that
    \begin{align*}
      u_p(t)\equiv u_p(t;c_j)=c_1+\sum_{\ell=2}^jc_\ell(t-a_\ell)_{+}\quad\text{and}\quad
      u_p'(t)\equiv u_p'(t;c_j)=\sum_{\ell=2}^jc_\ell.
    \end{align*}

    \begin{description}[labelwidth=!, labelindent=1cm, itemsep=-1ex]
    \item[If $j$ is even,] determine the parameter $-\infty<c_{j}<\infty$ that solves
      %%%%%%%%%%%%%%%
      \begin{align}\label{EQ:FAIRNESS_ALGO_2}
        \begin{split}
        & P\big(X(a_{j-1}) \geq u_{p}(a_{j-1};c_j)\big) + \int_{a_{j-1}}^{a_{j}} \frac{\tau(t)}{2 \pi } M_\mcV \left(-\frac 1 2 \left[u_{p}(t;c_j)^2  + \frac{u_{p}'(t;c_j)^2}{ \tau(t)^2} \right]  \right) \ dt \\
        -& \int_{a_{j-1}}^{a_{j}} \int_0^\infty  \frac{u_{p}'(t;c_j)}{2 \pi\tau(t)} M_\mcV'\left( - \frac 1 2 \left[u_{p}(t;c_j)^2+ \frac{(y+u_{p}'(t;c_j))^2}{\tau(t)^2} \right]  \right)  \ dy \ dt = \frac{\alpha}{2}\big(a_j-a_{j-1}\big).
      \end{split}
      \end{align}
        %%%%%%%%%%%%%%%
    \item[If $j$ is odd,] determine the parameter $-\infty <c_{j}<\infty$ that
      solves
      \begin{align}\label{EQ:FAIRNESS_ALGO_3}
        \begin{split}
        & P\big(X(a_{j}) \geq u_{p}(a_{j};c_j)\big) + \int_{a_{j-1}}^{a_{j}} \frac{\tau(t)}{2 \pi } M_\mcV \left(-\frac 1 2 \left[u_{p}(t;c_j)^2  + \frac{u_{p}'(t;c_j)^2}{ \tau(t)^2} \right]  \right) \ dt \\
        +& \int_{a_{j-1}}^{a_{j}} \int_0^\infty  \frac{u_{p}'(t;c_j)}{2 \pi\tau(t)} M_\mcV'\left( - \frac 1 2 \left[u_{p}(t;c_j)^2+ \frac{(y-u_{p}'(t;c_j))^2}{\tau(t)^2} \right]  \right)  \ dy \ dt =\frac{\alpha}{2}\big(a_j-a_{j-1}\big).
        \end{split}
      \end{align}
      %%%%%%%%%%%%%%%%%%%%%%%
    \end{description}
  \item[End do]
  \item[Return:] $u^\star_{\alpha/2}:=u_p$%$\{u^\star_{\alpha/2}(t):t\in[0,1]\}=\{u_p(t):t\in[0,1]\}$
  \end{description}
\end{algorithm}
%\spacingset{1.00001}

% An important feature of Algorithm \ref{ALGO:FAIRNESS} is the
% alternating construction for even and odd values of $j=1,\dots,p$. This way the point-probability
% components $P(X(a_j)\geq u(a_j))$ are only needed for odd $j=1,3,\dots$ since
% they are used by two adjacent intervals $[a_{j-1},a_j]$ and
% $[a_{j},a_{j+1}]$; except in the case of an odd $p$, where the final
% point-probability component $P(X(a_p)\geq u(a_p))$ is only used once, namely, by the final
% interval $[a_{p-1},a_p]$. This strategy only requires $\lceil
% p/2\rceil=\#\{1\leq j\leq p: j \textrm{ odd}\}$ many point-probability
% components for $p$ subintervals, which leads to a more efficient (i.e.~less
% conservative) inference than when using a naive version of Algorithm
% \ref{ALGO:FAIRNESS} requiring $p$ point-probability components for $p$ subintervals.

Imposing fairness constraints can make
inference conservative: in tendency, the larger the number of fairness
partitions, $p$, the more
conservative the inference procedure becomes. Such costs of imposing fairness constraints are well
known in the literature \citep{CDPFGH2017}. The expensive part in the
construction of the fair critical value function $u^\star_{\alpha/2}$ are
the additional correction terms, $P(X(a_j)\geq u(a_j))$, needed for odd $j=3,5,\dots$ with $j\leq p$. While the expected Euler
characteristic inequality in \eqref{EQ:EulerIneq} with at most two, $1\leq p\leq
2$, partitioning
intervals $[0,a_1]$ and $[a_1,1]$, with $a_1\in[0,1]$, requires only one correction
term, $P(X(a_1)\geq u(a_1))$, larger numbers of fairness partitions, $p\geq 3$, require
additional correction terms. Algorithm
\ref{ALGO:FAIRNESS} is designed to minimize the fairness costs by using as few
as possible corrections terms; namely, only one for each pair of partitioning
intervals $[0,a_j]$ and $[a_j,1]$, for odd $j=1,3,\dots$ with $j\leq p$. This feature of Algorithm
\ref{ALGO:FAIRNESS} allows us to quantify the price of fairness.

\begin{proposition}\label{PRO:PRICE4FAIRNESS} {\normalfont\bf (Price of
    fairness)}
  The expected Euler characteristic inequality when using the fair critical
  value function $u^\star_{\alpha/2}$ determined by Algorithm
  \ref{ALGO:FAIRNESS} is
  \begin{align}\label{EQ:PRICE4FAIRNESS}
  P\big(\exists{t\in[0,1]} : X(t)\geq u(t)\big)\leq
  \E[\varphi_{u^\star_{\alpha/2},X}(a_1)] + \varsigma_p(u^\star_{\alpha/2})=\frac{\alpha}{2},
\end{align}
\begin{equation*}
  \text{where}\quad\varsigma_p(u^\star_{\alpha/2})=
  \left\{\begin{array}{ll}
           0&\text{if }1\leq p\leq 2\\
           \sum_{\substack{3\leq j\leq
           p\\j\,:\,\textrm{odd}}}P\big(X(a_j) \geq u^\star_{\alpha/2}(a_j)\big)>0&\text{if }p\geq 3.\\
         \end{array}\right.
\end{equation*}
\end{proposition}

That is, for $1\leq p\leq 2$ partitioning intervals, the expected Euler characteristic inequality \eqref{EQ:PRICE4FAIRNESS} based on the fair critical value function, $u^\star_{\alpha/2}$, does not require more slackness than the basic expected Euler characteristic inequality \eqref{EQ:EulerIneq} based on a non-fair critical value function with $a_1=t_0$. For $p\geq 3$ partitioning intervals, the fair critical value function, $u^\star_{\alpha/2}$, becomes costly as we need to consider $\#\{3\leq j\leq p: j \textrm{ odd}\}$ additional correction terms.

%%%%%%%%%%%%%%%%%%%%%%%%%%%%%%%%%%%%%%%%%%%%%%%%%%%
\subsection[]{Constructing Fair Confidence Bands}\label{SSEC:CONSTRUCT}
%%%%%%%%%%%%%%%%%%%%%%%%%%%%%%%%%%%%%%%%%%%%%%%%%%%
In this section we first consider two prototypical examples for constructing simultaneous confidence bands using the above described results and methods.  Example \ref{EX:FFSCB_Gauss} is a generic example which applies to any asymptotically Gaussian functional parameter estimator $\hat{\theta}$.  Example \ref{EX:FFSCB_t} considers the practically relevant special case of $t$-distributed estimators, $\hat\theta$, of the mean function $\theta$.  Second, we introduce the formal fairness property of our simultaneous confidence bands.

%% %%%%%%%%%%%%%%%%%%%%%%%%%%%
%% Examples
%% %%%%%%%%%%%%%%%%%%%%%%%%%%%
\begin{example}\label{EX:FFSCB_Gauss} {\normalfont\bf(Asymptotically Gaussian Estimators)}
Let $\{\hat\theta(t),0\leq t\leq 1\}$ be an asymptotically Gaussian (parametric or nonparametric) estimator of a functional parameter $\{\theta(t),0\leq t\leq 1\}$ such that
\begin{equation}\label{EQ:ANORM}
  \sqrt{r_n}\left(\hat\theta(t)-\operatorname{Bias}\big(\hat\theta(t)\big)-\theta(t)\right)\overset{D}{\to}\mathcal{N}(0,C_\theta(t,t))\quad\text{for all}\quad t\in[0,1],
\end{equation}
as $n\to\infty$, where $\sqrt{r_n}$ denotes the parametric, $\sqrt{r_n}=\sqrt{n}$, or nonparametric, $\sqrt{r_n}=o(\sqrt{n})$, convergence rate of $\hat\theta(t)$, $\operatorname{Bias}(\hat\theta(t))$ denotes the known (or consistently estimable) possibly non-zero bias, and $C_\theta(t,t)$ denotes the known (or consistently estimable) covariance function.  This situation is fairly general and arises, for instance, when estimating the mean or covariance function from sparse or dense functional data \citep{LH2010,D2011,ZW2016}, in eigenfunction/value estimation \citep{KR2013,K2019}, in function-on-scalar regressions \citep{CGO2016}, in concurrent function-on-function regressions \citep{MCH2018}, etc. Generally, as discussed in \cite{CR2018}, the asymptotic normality in \eqref{EQ:ANORM} requires \textit{tightness} of the estimate to ensure that the convergence in distribution occurs in the strong topology.  Typically, this rules out estimates from ill-posed inverse problems such as, for instance, the classic scalar-on-function regression estimators \citep[cf.][]{CMS2007}.
% %%%%%%%%%%%%%%%%%%%%%%%%%%%%%%%%%%%%%

From \eqref{EQ:ANORM} we have that
\begin{equation*}
X_n(t)=\frac{\left(\hat\theta(t)-\operatorname{Bias}\big(\hat\theta(t)\big)-\theta(t)\right)}{\sqrt{C_\theta(t,t)/r_n}}\overset{D}{\to}\mathcal{N}(0,1)\quad\text{for all}\quad t\in[0,1],
\end{equation*}
such that $X_n\overset{D}{\to}X$, where $X(t)\overset{D}{=}V Z(t)$ with $V\equiv\sigma=1$ for all $t\in[0,1]$. That is, the covariance function of $X$ equals its dispersion function $\cov(X(t),X(s))=c(t,s)$ with $c(t,t)=1$ for all $t,s\in[0,1]$ and the only missing parameter required to compute the formula in Corollary \ref{CO:GAUSS} is the roughness parameter function $\tau(t)=(\partial_{12}c(t,t))^{1/2}$ which quantifies the multiple testing problem locally for each $t\in[0,1]$.  Since $c(t,s)=\cov(X(t),X(s))$ and $\cov(X(t),X(s))=\tilde{C}_\theta(t,s)$ with $\tilde{C}_\theta(t,s)=C_\theta(t,s)/\sqrt{C_\theta(t,t)C_\theta(s,s)}$ is follows that $\tau(t)=(\partial_{12}\tilde{C}_\theta(t,t))^{1/2}$ which is directly computable from the covariance function $C_\theta$.

Plugging in the parameters $\sigma=1$ and $\tau(t)=(\partial_{12}\tilde{C}_\theta(t,t))^{-1/2}$ into the Gaussian formula of Corollary \ref{CO:GAUSS} allows us to compute the fair critical value function, $u_{\alpha/2}^\star$, as described in Algorithm \ref{ALGO:FAIRNESS}.  The fair simultaneous confidence band for $\theta$ is then given by
\begin{equation}\label{EQ:FairBand_z}
\operatorname{FF}_z(t)=[\hat\theta_l^\star(t),\hat\theta_u^\star(t)]:=\hat\theta(t)\pm u_{\alpha/2}^\star(t)\sqrt{C_\theta(t,t)/r_n}\quad\text{for all}\quad t\in[0,1].
\end{equation}
We denote this simultaneous confidence bands as Fast and Fair (FF) simultaneous confidence bands as it is considerably faster to compute than the popular simulation/resampling based alternatives.  One-sided confidence bands, $[\hat\theta_l^\star,\infty]$ or $[-\infty,\hat\theta_u^\star]$, can be constructed from the one-sided version of $u_\alpha^\star$ which can be computed using Algorithm \ref{ALGO:FAIRNESS}, but with substituting $\alpha/2$ by $\alpha$. Usually, we do not know the covariance function $C_\theta$ or roughness parameter function $\tau$ and the have to plug in consistent estimators $\hat{C}_\theta$ and $\hat\tau$. The next example shows such estimators for the case where $\theta$ is the mean function.
\end{example}

\begin{example}\label{EX:FFSCB_t} {\normalfont\bf(Estimators of the Mean Function (Unknown Covariance))}
%%%%%%%%%%%%%%%%%%%%%%%%%%%%%%%%%%%%%%%%%%%%%%%%%%%
Let us consider an iid sample $\{S_i\}_{i=1}^n$ from a Gaussian process
$\mathcal{N}(\theta,C_\theta)$ with unknown mean function $\theta(t)=\E[S_i(t)]$ and unknown covariance function $C_\theta(t,s)=\cov(S_i(t),S_i(s))$. We estimate the mean and covariance functions using the sample estimators 
$\hat\theta(t)=n^{-1}\sum_{i=1}^nS_i(t)$ 
and
$\hat{C}_{\theta,n}(t,s)=(n-1)^{-1}\sum_{i=1}^n(S_i(t)-\hat\theta(t))(S_i(s)-\hat\theta(s))$, for $t,s\in[0,1]$. This setup yields that
\begin{equation*}
  X_n(t)=\frac{\hat{\theta}(t)-\theta(t)}{\sqrt{\hat{C}_\theta(t,t)/n}}\sim t_{\nu}\quad\text{with}\quad\nu=n-1\quad\text{for all}\quad t\in[0,1],
\end{equation*}
such that $X_n(t)\overset{D}{=}V_n Z(t)$ with $V_n=\sqrt{\nu/\chi^2_\nu}$ and $Z(t)\sim\mathcal{N}(0,1)$ for all $t\in[0,1]$, where
$\cov(X_n(t),X_n(s))=\var(V_n)\cov(Z(t),Z(s))=(\nu/(\nu-2))c(t,s)$ for all $\nu>2$ with $c(t,t)=1$ for all $t,s\in[0,1]$. Moreover, straight forward derivations show that $\cov(X_n(t),X_n(s))=C_\theta(t,s)(C_\theta(t,t)C_\theta(s,s))^{-1/2}$. Thus, $c(t,s)=((\nu-2)/\nu)\cov(X_n(t),X_n(s))$ can be estimated consistently by $\hat{c}_n(t,s)=\big((\nu-2)/\nu\big)\hat{C}_{\theta,n}(t,s)\big(\hat{C}_{\theta,n}(t,t)\hat{C}_{\theta,n}(s,s)\big)^{-1/2}\;\text{ for }\;t,s\in[0,1]$, and the roughness parameter function $\tau(t)$ can be estimated consistently by
% Since $\cov(X_n(t),X_n(s))=C_\theta(t,s)\big(C_\theta(t,t)C_\theta(s,s)\big)^{-1/2}\to c(t,s)$ as $n\to\infty$ for all $t,s\in[0,1]$, the roughness parameter function $\tau(t)$ can be estimated consistently by
\begin{equation}\label{EQ:tau_1}
  \hat{\tau}_1(t)=\big(\partial_{12}\hat{c}_n(t,t)\big)^{1/2}.
\end{equation}
Alternatively, as explained in Remark \ref{RE:ASSest}, we can use the following equivalent estimator:
\begin{equation}\label{EQ:tau_2}
\hat{\tau}_2(t)=\left(\widehat{\V}\big(\tilde{S}_1'(t),\dots,\tilde{S}_n'(t)\big)\right)^{1/2}\quad\text{for}\quad t\in[0,1],
\end{equation}
where $\widehat{\V}\big(\tilde{S}_1'(t),\dots,\tilde{S}_n'(t)\big)=(n-1)^{-1}\sum_{i=1}^n(\tilde{S}'(t)-n^{-1}\sum_{i=1}^n\tilde{S}'(t))^2$ denotes the empirical variance of the standardized and differentiated sample functions $\tilde{S}_i(t)=(S_i(t)-\hat\theta(t))/(\hat{C}_\theta(t,t))^{1/2}$. Plugging in the degrees of freedom $\nu=n-1$ and an estimate, $\hat{\tau}_1(t)$ or $\hat{\tau}_2(t)$, of $\tau(t)$ into the generalized Kac-Rice formula \eqref{EQ:GRF_t} in Corollary \ref{CO:STUDt} allows us to compute the fair critical value function, $u_{\alpha/2}^\star$, as described in Algorithm \ref{ALGO:FAIRNESS}. This then leads to the fair simultaneous confidence band
\begin{equation}\label{EQ:FairBand_t}
\operatorname{FF}_t(t)=[\hat{\theta}_l^\star(t),\hat{\theta}_u^\star(t)]:=\hat{\theta}(t)\pm\hat{u}_{\alpha/2}^\star(t)\sqrt{\hat{C}_\theta(t,t)/n},
\end{equation}
where the hat in $\hat{u}_{\alpha/2}^\star(t)$ indicates that the critical value function is based on an estimate, $\hat\tau_1$ or $\hat\tau_2$, of the roughness parameter function $\tau$.
%%%%%%%%
\begin{comment}
\todo[color=red!30,inline]{Alternatively:
$$\hat{c}(t,s)=\left(\frac{\nu-2}{\nu}\right)\hat{C}_\theta(t,s)\left(\hat{C}_\theta(t,t)\hat{C}_\theta(s,s)\right)^{-1/2}$$
and
$$\tilde{S}_i(t)=\left(\frac{\nu-2}{\nu}\right)^{1/2}(S_i(t)-\hat\theta(t))\left(\hat{C}_\theta(t,t)\right)^{-1/2}$$
respectively. But the factor $\left(\frac{\nu-2}{\nu}\right)$ is not implemented. Maybe it would be better to do so?!\\
\textbf{Answer:} No! Implementing this scaling factor makes things worse. The estimation errors of $\hat{C}_\theta(t,s)\left(\hat{C}_\theta(t,t)\hat{C}_\theta(s,s)\right)^{-1/2}$ are harming and the scaling leads then to a size inflation in small samples. (This is what I saw via simulations.)}
\end{comment}
%%%%%%%
\end{example}

\begin{remark} In the case where the random sample functions $\{S_i\}_{i=1}^n$ are fully observed, the estimators $\hat\tau_1$ and $\hat\tau_2$ in \eqref{EQ:tau_1} and \eqref{EQ:tau_2} are equivalent. However, if the sample functions are only sparsely observed $\hat\tau_2$ becomes infeasible since $S'_i$ cannot be computed from sparse functional data. In this case, an estimator similar to $\hat\tau_1$ can be used, where the sample estimator $\hat{C}_\theta$ has to be substituted by a nonparametric covariance estimator. 
\end{remark}

%%%%%%%%%%%%%%%%%%%%%%%%%%%%%%%%%%%%%%%%%%%%%%%%%%%%%%%%%%%%%%%%%%%%%%%%%%%%%%
% Fair Confidence Bands
%%%%%%%%%%%%%%%%%%%%%%%%%%%%%%%%%%%%%%%%%%%%%%%%%%%%%%%%%%%%%%%%%%%%%%%%%%%%%%

The following proposition formally describes the fairness property of the fair simultaneous confidence bands $\operatorname{FF}_z$ and $\operatorname{FF}_t$ in \eqref{EQ:FairBand_z} and \eqref{EQ:FairBand_t}.
%%%%%%%%
\begin{proposition}\label{PRO:FAIR_BAND} {\normalfont\bf(Fair confidence bands)}
Let the conditions of Theorems \ref{TH:MAIN} and \ref{TH:EST} hold and let $u_{\alpha/2}^\star$ be selected by Algorithm \ref{ALGO:FAIRNESS} with respect to a given partition $0=a_0<a_1<a_2<\dots<a_p=1$ and a given $\alpha\in(0,1)$. Let $[\hat\theta_l^\star,\hat\theta_u^\star]$ denote the simultaneous confidence band $\operatorname{FF}_z$ in \eqref{EQ:FairBand_z} or $\operatorname{FF}_t$ in \eqref{EQ:FairBand_t}. Then we have that
\begin{align}\label{EQ:VALID_SCB_ab_1}
P\left(\theta(t)\in[\hat\theta_l^\star(t),\hat\theta_u^\star(t)],\,\forall t\in\bigcup_{j\in\mathcal{J}}[a_{j-1},a_{j}]\right)
&\geq 1- \sum_{j\in\mathcal{J}}\alpha(a_{j}-a_{j-1})\geq 1-\alpha
\end{align}
for any subset $\mathcal{J}\subseteq\{1,2,\dots,p\}$. Thus, under H$_0$: $\theta=\theta_0$, the band $[\hat\theta_l^\star,\hat\theta_u^\star]$ fulfills the false positive rate balance Definition \ref{Def:Fairness}.
\end{proposition}

Proposition \ref{PRO:FAIR_BAND} simply inverts Lemma \ref{LE:FAIRNESS} to the case of simultaneous confidence bands. It states that the bands $[\hat\theta_l^\star,\hat\theta_u^\star]$ can be used as fair screening tools that allow to detect deviations from a null hypotheses under a balanced false positive rate across given subintervals $[a_{j-1},a_j]$, $j=1,2,\dots,p$, and combinations thereof. This facilitates both global and local interpretations: While $[\hat\theta_l^\star,\hat\theta_u^\star]$ is a valid $(1-\alpha)\times 100\%$ simultaneous confidence band over $[0,1]$, it is also a valid $(1-\sum_{j\in\mathcal{J}}\alpha(a_{j}-a_{j-1}))\times 100\%$ simultaneous confidence band over $\cup_{j\in\mathcal{J}}[a_{j-1},a_{j}]$; see also our application in Section \ref{SSEC:APPL_BIOM}.

%%%%%%%%%%%%%%%%%%%%%%%%%%%%%%%
\section{Simulations}\label{SEC:SIMUL}
%%%%%%%%%%%%%%%%%%%%%%%%%%%%%%%
This section contains our simulation study to assess and illustrate the proposed fair simultaneous confidence bands. We focus on the special case of simultaneous inference for the mean function since this allows us to compare our bands with many alternative bands form the literature. Section \ref{SEC:SIMUL_FULLFDA} considers the case of fully observed random functions and Section \ref{SEC:SIMUL_PARTFDA} considers the case of fragmentary functions for which it is impossible to estimate the total covariance operator. We focus on the practically relevant case of \textit{unknown} covariances, $C_\theta$.  All bands are computed based on the usual sample estimators $\hat{C}_\theta$ and $\hat{\theta}$ as defined in Example \ref{EX:FFSCB_t}.

%%%%%%%%%%%%%%%%%%%%%%%%%%%%%%%%%%%%%%%%
\subsection{Fully Observed Functions}\label{SEC:SIMUL_FULLFDA}
%%%%%%%%%%%%%%%%%%%%%%%%%%%%%%%%%%%%%%%%
We generate samples of random functions $\{S_i\}_{i=1}^n\overset{\operatorname{iid}}{\sim}\mathcal{N}(\theta,C_\theta)$ with mean function $\theta(t)$, $t\in[0,1]$, and covariance function $C_\theta(t,s)$, $t,s\in[0,1]$.  The sample functions, $S_i$, are evaluated at $101$ equidistant grid points $t=j/100$, $j=0,\dots,100$, to emulate continuous functional data.  We consider the following mean function scenarios based on shifting and scaling the polynomial mean function $\theta_0(t)=10t^3-15t^4+6t^6$ by some $\Delta\geq 0$:
\begin{description}[wide, labelwidth=!, labelindent=0pt, itemsep=-1ex]
\item[Mean1] $\theta(t)=\theta_0(t) + \Delta$ (`shift')
\item[Mean2] $\theta(t)=\theta_0(t) (1+\Delta)$ (`scale')
\item[Mean3] $\theta(t)=\theta_0(t) + \Delta\mathbbm{1}\{(0\leq t\leq 1/8)\}$ (`local')
\end{description}
The three mean function scenarios, Mean1-3, are shown in the upper row of Figure \ref{FIG:SIM_2}.

\begin{figure}[h!tb]
\centering
\includegraphics[height=.77\textheight, width=\textwidth]{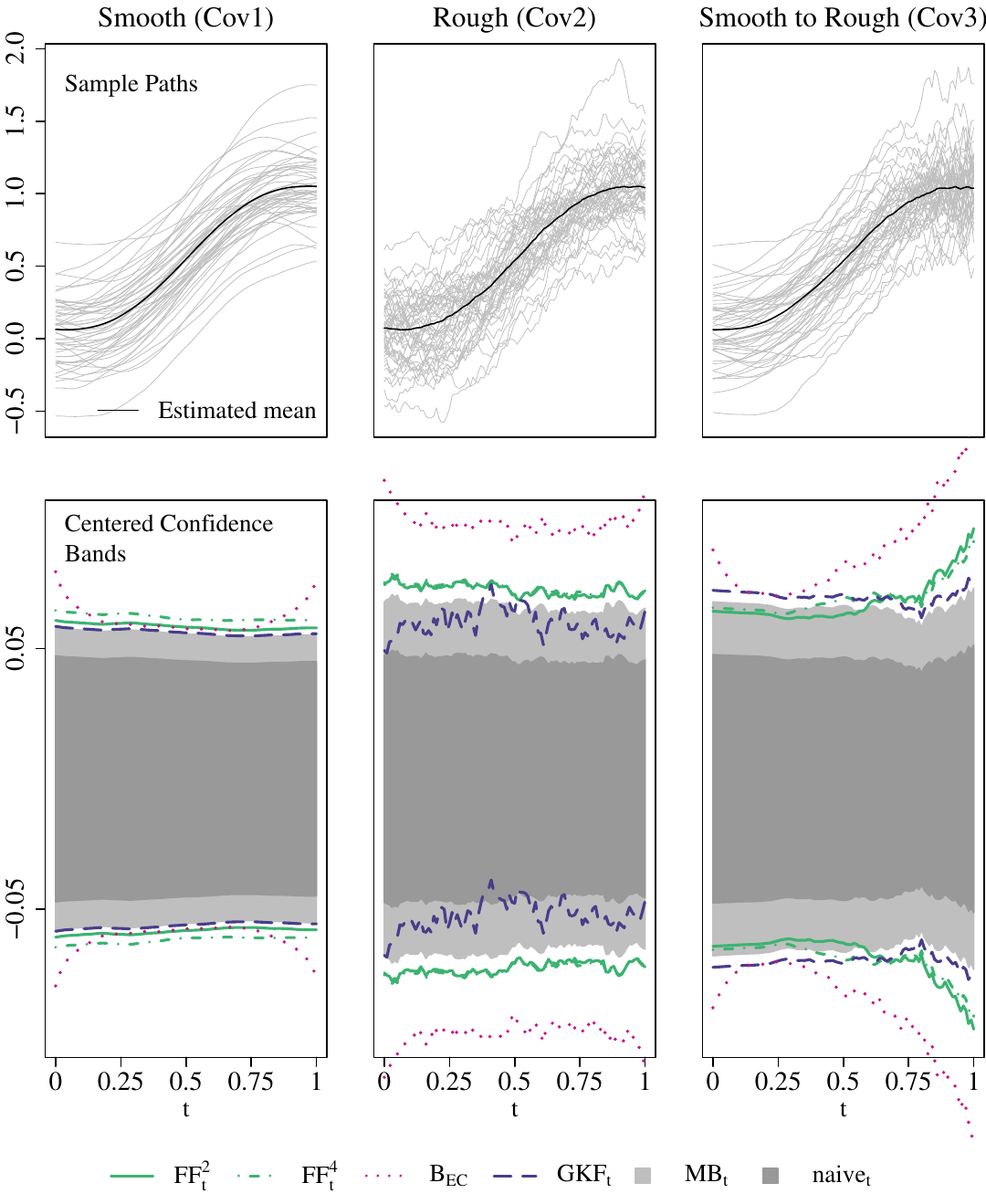}
\caption{{\sc Upper Row:} Sample paths and mean estimates.  {\sc Lower
    Row:} Centered 95\% confidence bands, where centering by $\hat\theta$
  is applied to improve the visual exposition.}
\label{FIG:SIM_1}
\end{figure}

For the covariance operator, we take the Mat\'ern covariance  $C_\theta(t,s)=\allowbreak 0.25^2\big(2^{1-\upsilon}/\Gamma(\upsilon)\big)\allowbreak \big(\sqrt{2\upsilon}|t-s|\big)^\upsilon K_\upsilon\big(\sqrt{2\upsilon}|t-s|\big)$, where $\Gamma$ is the gamma function, $K_\upsilon$ is the modified Bessel function of the second kind, and where $\upsilon\geq 0$ controls the roughness of the sample paths $\{S_i(t),0\leq t\leq1\}$.  We consider three different covariance scenarios:
\begin{description}[wide, labelwidth=!, labelindent=0pt, itemsep=-1ex]
\item[Cov1] Stationary Mat\'ern covariance $C_\theta$ with $\upsilon=3/2$ (`smooth').
\item[Cov2] Stationary Mat\'ern covariance $C_\theta$ with $\upsilon=1/2$ (`rough').
\item[Cov3] Non-stationary Mat\'ern-type covariance $C_\theta(t,s)=\allowbreak 0.25^2\allowbreak \big(2^{1-\upsilon_{ts}}/\Gamma(\upsilon_{ts})\big)\allowbreak \big(\sqrt{2\upsilon_{ts}}|t-s|\big)^{\upsilon_{ts}} K_{\upsilon_{ts}}\big(\sqrt{2\upsilon_{ts}}|t-s|\big)$, where $\upsilon_{ts}=2+\sqrt{\max(t,s)}(1/4 - 2)$ (`smooth to rough').
\end{description}
Cov1 results in smooth continuously differentiable sample functions as typical, for instance, for functional chemometric/spectrometric data \citep[see][Ch.~2]{FV2006_book}. Cov2 leads to rough non-differentiable sample functions as considered, for instance, in the literature on functional impact points \citep{McKS2010,Poss_et_al_2020} and represents a violation of our smoothness Assumption \ref{A:ELIP}.  The non-stationary covariance Cov3 leads to sample functions with inhomogeneous roughness (`smooth to rough').  Sample paths, $S_i$, for each of the covariance scenario, Cov1-3, are shown in the upper row of Figure \ref{FIG:SIM_1}. Each of the covariance scenarios contains both local ($t\approx s$) and global ($t\ll s$ or $t\gg s$) dependencies. The lower dependency bounds are 
$\operatorname{corr}(X(t),X(s))\geq
\operatorname{corr}(X(0),X(1))=0.48$ for all $t,s\in[0,1]$ in case of Cov1, $\operatorname{corr}(X(t),X(s))\geq
\operatorname{corr}(X(0),X(1))=0.37$  for all $t,s\in[0,1]$ in case of Cov2, and $\operatorname{corr}(X(t),X(s))\geq 
\operatorname{corr}(X(0),X(1))=0.29$  for all $t,s\in[0,1]$ in case of Cov3.
%%%%%%%%%%%%%%%%%%%%%%
%\spacingset{1}
%\spacingset{1.00001}
%%%%%%%%%%%%%%%%%%%%%%

\vspace*{-3mm}

\paragraph{Fair confidence bands and benchmark bands.} The following list describes the specifications of our fair confidence bands and introduces the considered benchmark bands.
%%%%%%%%%%%%
 % We compare our FF bands with the following benchmarks:
%%%%%%%%%%%%
\begin{description}[wide, labelwidth=!, labelindent=0pt, itemsep=-1ex]
\item[FF$\boldsymbol{^p_z}$ and FF$\boldsymbol{^p_t}$] Our Fast and Fair (FF) simultaneous confidence bands
\begin{equation*}
\operatorname{FF}^p_{\text{distr}}(t)=[\hat\theta^\star_l(t),\hat\theta^\star_u(t)]=\hat\theta(t)\pm \hat{u}_{\alpha/2}^\star(t) \big(\hat{C}_\theta(t,t)/n\big)^{1/2},\quad t\in[0,1],\quad
\end{equation*}
with $\text{distr}\in\{z,t\}$ denoting the distribution and $p\in\{1,2,4\}$ the number of equidistant domain partitions such as $a_0=0$, $a_1=1/4$, $a_2=2/4$, $a_3=3/4$, and $a_4=4/4$ in case of $p=4$.  The critical value function, $\hat{u}_{\alpha/2}^\star(t)$, is computed as described in Algorithm \ref{ALGO:FAIRNESS}. For estimating the roughness parameter function, $\tau$, we use $\hat\tau_2$ in \eqref{EQ:tau_2}.  When $\hat{u}_{\alpha/2}^\star$ is based on the Gaussian formula \eqref{EQ:GRF_Gauss} we write FF$_z^p$, and when $\hat{u}_{\alpha/2}^\star$ is based on the $t$-distribution formula \eqref{EQ:GRF_t} we write FF$_t^p$. The bands FF$^1_z$ and FF$^1_t$ use constant critical values, $u_1^\star(t)\equiv u_1^\star$, found by solving the equations \eqref{EQ:KacRice_Gauss} and \eqref{EQ:KacRice_t}, respectively.
%%%%%%%%%%%%%%%%%%%%%%
\item[B$\boldsymbol{_{\text{S}}}$] The simulation based bootstrap band proposed
  by \cite{D2011} uses a
  constant critical value determined by a parametric bootstrap, where we use 10,000 bootstrap replications.
\item[B$\boldsymbol{_{\text{EC}}}$] A modification of Scheff\'e's method
  to transform ellipsoid confidence regions into confidence bands proposed by
  \cite{CR2018}. %The B$_{\text{EC}}$ band is chosen adaptively to minimize the
  % average squared width of the band.
\item[GKF$\boldsymbol{_t}$] The simultaneous confidence band using the Gaussian kinematic formula
  of $t$-processes by \cite{TS2022}.
\item[MB$\boldsymbol{_t}$] The simultaneous confidence band based on a
  Rademacher multiplier-$t$ bootstrap as proposed in \cite{TS2022}.
\item[IWT] The interval-wise testing procedure proposed by \cite{PV2017_IWT} and further extended in \cite{AHPSV2018_IWT}. %This method uses permutation techniques to compute a simultaneous $p$-value function, $\{p_{\operatorname{IWT}}(t),0\leq t\leq 1\}$.
\item[naive$\boldsymbol{_t}$] The naive pointwise confidence interval based on the $t$-distribution without adjusting for multiple testing.  (Only used as a visual reference.)
\end{description}
The IWT procedure is implemented in the \textsf{R}-package \texttt{fdatest} \citep{fdatest_Rpgk}. The GKF$_t$ band and the MB$_t$ band are implemented in the \textsf{R}-package \texttt{SIRF} \citep{SIRF_Rpgk}. Our FF$_t^p$ and FF$_t^P$ bands and all remaining bands are implemented in our \textsf{R}-package \texttt{ffscb} \citep{ffscb_Rpgk}. In the following, we report the results for the $t$-distribution based bands FF$_t^p$ which take into account the estimation errors in $\hat{C}_\theta$.  The results of the Gaussian versions, FF$_z^p$ are summarized in the tables of Appendix \ref{APP:ADD_MATERIAL} of the supplementary paper \cite{LR_Suppl}.  In the case of small samples, the FF$_z^p$ bands are typically too narrow; however, for large samples, the FF$_z^p$ bands perform similarly to, and partially better than the FF$_t^p$ bands.

\vspace*{-3mm}

\paragraph{Band shapes.}  The lower row of Figure \ref{FIG:SIM_1} shows exemplary results of the (centered) confidence bands FF$_t^2$, FF$_t^4$, B$_{EC}$, GKF$_t$, MB$_t$, and naive$_t$ for each of the covariance scenarios Cov1-3. The bands FF$_t^1$ and B$_{\text{S}}$ are omitted to improve the visual exposition: the omitted B$_{\text{S}}$ band performs worse (too tight) than the alternative bootstrap based band MB$_t$, and the omitted FF$_t^1$ band performs similar to, but more stable than the GKF$_t$ band.  As expected, the adaptive FF$_t^2$ and FF$_t^4$ bands do not show any adaptive shapes in the stationary covariance scenarios, Cov1 and Cov2, but they show adaptive shapes in the non-stationary covariance scenario, Cov3, where they are tight over the initial area with high positive correlations and wide over the final area with low correlations. The B$_{\text{EC}}$ band shows a seemingly adaptive behavior, but (a) its shape cannot be interpreted, and (b) the band becomes very wide (conservative) in the case of rough processes. The rough covariance scenario Cov2 violates the smoothness Assumptions of FF$_t^2$, FF$_t^4$, and GKF$_t$, but while our fair bands remain stable, the GKF$_t$ band gets unstable in this scenario. The FF$_t^4$ band is wider (more conservative) than the FF$_t^2$ band as expected due to the price of fairness property (Proposition \ref{PRO:PRICE4FAIRNESS}).

\vspace*{-3mm}

\paragraph{Computation times.} Table \ref{TAB:CompTimes} in the
Appendix \ref{APP:ADD_MATERIAL} of the supplementary paper \cite{LR_Suppl} shows summary statistics of the computation times for all bands. The bands that do not use resampling (FF$_z^p$, FF$_t^p$, GKF$_t$, and B$_{\text{EC}}$) have all similar computation times ($0.025$ sec to $0.05$ sec) and are all considerably faster (factor $1/5$ to $1/33$) to compute than the resampling based alternatives (B$_{\text{S}}$, IWT, and MB$_t$).

%%%%%%%%%%%%%%%%%%%%%%%%%%%%%%%%%
\subsubsection*{Verifying Type-I Error Rate}
%%%%%%%%%%%%%%%%%%%%%%%%%%%%%%%%%
To evaluate our fair confidence bands and to compare them with the above
introduced alternative approaches, we use the duality of confidence bands with
simultaneous hypothesis tests and test the hypothesis
\begin{center}
	H$_0$: $\theta(t)=\theta_0(t)$ $\forall$ $t\in[0,1]$\quad vs.\quad H$_1$: $\exists t\in[0,1]$ s.t.~$\theta(t)\neq\theta_0(t)$.\\
\end{center}
We reject H$_0$ if $\theta_0$ is not covered by a band for at least one of grid
points $t$.  For the IWT procedure we reject H$_0$ if
$p_{\operatorname{IWT}}(t)<\alpha$ for at least one of the grid points. As the significance level we choose $\alpha=0.05$ which means that we consider 95\% simultaneous confidence bands.

To verify the type-I error rates, we draw 50,000 Monte Carlo samples
$\{S_i\}_{i=1}^n\overset{\operatorname{iid}}{\sim}\mathcal{N}(\theta_0,C_\theta)$
for each covariance function scenario Cov1-3.  We examine a challenging small
sample size of $n=15$ and a large sample size $n=100$.  Since the IWT procedure
is computationally very expensive we had to reduce the number of Monte Carlo
replications for this method to 5,000.\footnote{\cite{AHPSV2018_IWT} use only
  1,000 Monte Carlo replications.} Table \ref{TAB:SIM_1} contains the empirical type-I error rates and we can summarize the results as following:
\begin{itemize}[wide, labelwidth=!, labelindent=0pt, itemsep=-1ex]
\item[(a)] All bands are able to keep the nominal $\alpha$-level in all
  scenarios and sample sizes, except for the B$_{\text{S}}$ band
  which is too tight leading to over-rejections in all scenarios.
\item[(b)] Among the methods that are able to keep the $\alpha$-level, the
  type-I error rates of the FF$_t^1$ and MB$_t$ bands are closest to
  $\alpha=0.05$.
  % FF$_t^1$ is particularly good in small sample sizes and MB$_t$
  % in larger sample sizes.
\item[(c)] The B$_{\text{EC}}$ band is very conservative for large sample sizes
  and rough processes (Cov2-3).
\item [(d)] The larger $p$ the more conservative the FF$_t^p$ bands
  become due to the price of fairness property (Proposition \ref{PRO:PRICE4FAIRNESS}).
  %These costs are well known in the machine learning literature \citep{CDPFGH2017}.
\end{itemize}
%%%%%%%%%%%%%%%%%%%%%%%%%%%%%
%\spacingset{1}
\begin{table}[h!]
\centering
\caption{Type-I error rates.}
\label{TAB:SIM_1}
\begin{tabular}{l ccc c ccc}
\toprule
&\multicolumn{3}{c}{$n=15$}&&\multicolumn{3}{c}{$n=100$}\\
\cline{2-4}\cline{6-8}\\[-2.1ex]
  Band          & Cov1  & Cov2  & Cov3  && Cov1  & Cov2  & Cov3 \\
  \midrule
  FF$_t^1$      & 0.051 & 0.038 & 0.044 && 0.048 & 0.033 & 0.038 \\
  FF$_t^2$      & 0.037 & 0.036 & 0.038 && 0.036 & 0.029 & 0.034 \\
  FF$_t^4$      & 0.025 & 0.032 & 0.031 && 0.025 & 0.025 & 0.029 \\
  B$_{\text{EC}}$ & 0.051 & 0.036 & 0.035 && 0.027 & 0.001 & 0.006 \\
  B$_{\text{S}}$  & 0.088 & 0.122 & 0.120 && 0.055 & 0.061 & 0.059 \\
  MB$_t$        & 0.039 & 0.036 & 0.036 && 0.048 & 0.050 & 0.048 \\
  GKF$_t$       & 0.037 & 0.018 & 0.023 && 0.046 & 0.024 & 0.029 \\
  IWT           & 0.036 & 0.028 & 0.027 && 0.036 & 0.029 & 0.028 \\
\bottomrule
\multicolumn{8}{l}{Nominal type-I error rate: $\alpha=0.05$}
\end{tabular}
\end{table}
%\spacingset{1.00001}
%%%%%%%%%%%%%%%%%%%%%%%

% The reason for the weak performance of the bootstrap band, B$_{\text{S}}$, is the sensibility of the parametric bootstrap against the estimation errors in $\hat{C}_\theta$.  The bootstrap band performs well under the unrealistic scenario of a known covariance, $C_\theta$ \citep[see][]{CR2018}.

%%%%%%%%%%%%%%%%%%%%%%%%%%%%%%%%%%%%%%%%%%%%%%%%%
\subsubsection*{Verifying False Positive Rate Balance}\label{SSSEC:SIM_FAIR}
%%%%%%%%%%%%%%%%%%%%%%%%%%%%%%%%%%%%%%%%%%%%%%%%%
In this section, we assess the interpretable fairness property (Proposition \ref{PRO:FAIR_BAND}) of our FF$_t^p$ bands. To evaluate the fairness constraints of our bands, we use again the duality of confidence bands with hypothesis tests ($\alpha=0.05$), but this time we consider interval specific hypotheses:
\begin{center}
  H$^j_0$: $\theta(t)=\theta_0(t)$ $\forall$ $t\in[a_{j-1},a_{j}]$\quad vs.\quad
  H$^j_1$: $\exists t\in[a_{j-1},a_{j}]$ s.t.~$\theta(t)\neq\theta_0(t)$,\quad$j=1,\dots,p$\\
\end{center}
where we reject H$^j_0$ if $\theta_0$ is not covered by a band for at least one grid point $t$ in $[a_{j-1},a_{j}]$. The following interval scenarios are considered:
\begin{description}[wide, labelwidth=!, labelindent=0pt, itemsep=-1ex]
\item[4 Intervals:] $[0,1/4]$, $[1/4,2/4]$, $[2/4,3/4]$, and $[3/4,1]$ with fair nominal significance level for each interval: $\alpha(a_{j}-a_{j-1})=\alpha/4$, $j\in\{1,2,3,4\}$.
\item[2 Intervals:] $[0,1/2]$ and $[1/2,1]$ with fair nominal significance level for each interval: $\alpha(a_{j}-a_{j-1})=\alpha/2$, $j\in\{1,2\}$.
\end{description}

Table \ref{TAB:FAIR} shows the empirical type-I error rates of our FF$_t^p$ bands for $p\in\{1,2,4\}$ for the case of for $n=100$ and the challenging non-stationary covariance scenario Cov3, where adaptivity matters. The results for all other bands and all other covariance scenarios, Cov1 and Cov2, and sample sizes are reported in the tables of Appendix \ref{APP:ADD_MATERIAL} of the supplementary paper \cite{LR_Suppl}; however, the results for all scenarios can be summarized as following:
\begin{itemize}[wide, labelwidth=!, labelindent=0pt, itemsep=-1ex]
\item[(a)]
  The FF$_t^2$ and the FF$_t^4$ band keep the nominal
  interval specific type-I error rates in the 2 interval and the 4 interval
  scenario, respectively. 
\item[(b)] The FF$_t^4$ band is able to keep the nominal interval specific type-I
  error rates in \emph{both} interval scenarios. This is in
  accordance with Proposition
  \ref{PRO:FAIR_BAND} since the scenarios are nested.
\item[(c)] The non-adaptive FF$_t^1$ band is only able to keep the nominal type-I error rates over $[0,1]$ since Cov3 is a non-stationary covariance
  scenario. (The same applies to all benchmark bands as shown in Appendix \ref{APP:ADD_MATERIAL}  of the supplementary paper \cite{LR_Suppl}.)
\end{itemize}
%%%%%%%%%%%%%%%%%%%%%%%%%%%%%%%
%\spacingset{1}
\begin{table}[h!tb]
\centering
\caption{Checking false positive rate balance for $n=100$ and covariance scenario Cov3.}
\label{TAB:FAIR}
\begin{tabular}{l ccccc ccc cc}
  \toprule
  &\multicolumn{5}{c}{4 Intervals}&\multicolumn{3}{c}{2 Intervals}&\multicolumn{2}{c}{1 Interval}\\
  &\multicolumn{5}{c}{$\alpha/4=0.0125$}&\multicolumn{3}{c}{$\alpha/2=0.025$}&\multicolumn{2}{c}{$\alpha=0.05$}\\
  \cmidrule(lr){2-6} \cmidrule(lr){7-9} \cmidrule(lr){10-11}\\[-1ex]
  Band&$\left[0,\frac{1}{4}\right]$ & $\left[\frac{1}{4},\frac{2}{4}\right]$ & $\left[\frac{2}{4},\frac{3}{4}\right]$ & $\left[\frac{3}{4},1\right]$ &Check& $\left[0,\frac{1}{2}\right]$ & $\left[\frac{1}{2},1\right]$&Check& $\left[0,1\right]$&Check\\
  \midrule
  % FF$_t^4$& 0.011 & 0.017 & 0.012 & 0.014 &       & 0.020 & 0.023 &       & 0.034 &    \\
  % FF$_t^2$& 0.014 & 0.018 & 0.018 & 0.016 &       & 0.023 & 0.029 &       & 0.041 &    \\
  % FF$_t^1$& 0.008 & 0.010 & 0.015 & 0.030 &       & 0.013 & 0.037 &       & 0.044 &    \\
  %%%%
  FF$_t^1$& 0.007 & 0.009 & 0.013 & 0.025 &\xmark& 0.012 & 0.031&\xmark&0.038&\cmark\\
  FF$_t^2$& 0.015 & 0.018 & 0.016 & 0.009 &\xmark & 0.024 & 0.021&\cmark&0.034&\cmark\\
  FF$_t^4$& 0.012 & 0.013 & 0.011 & 0.011 &\cmark & 0.017 & 0.018&\cmark&0.029&\cmark\\
  \bottomrule
\end{tabular}
\end{table}
%\spacingset{1.00001}
%%%%%%%%%%%%%%%%%%%%%%%%%%%%%%%%%

%%%%%%%%%%%%%%%%%%%%%%%%%%%%%%%%%
\subsubsection*{Comparing Power}
\begin{figure}[htb]
\centering
\includegraphics[height=.5\textheight]{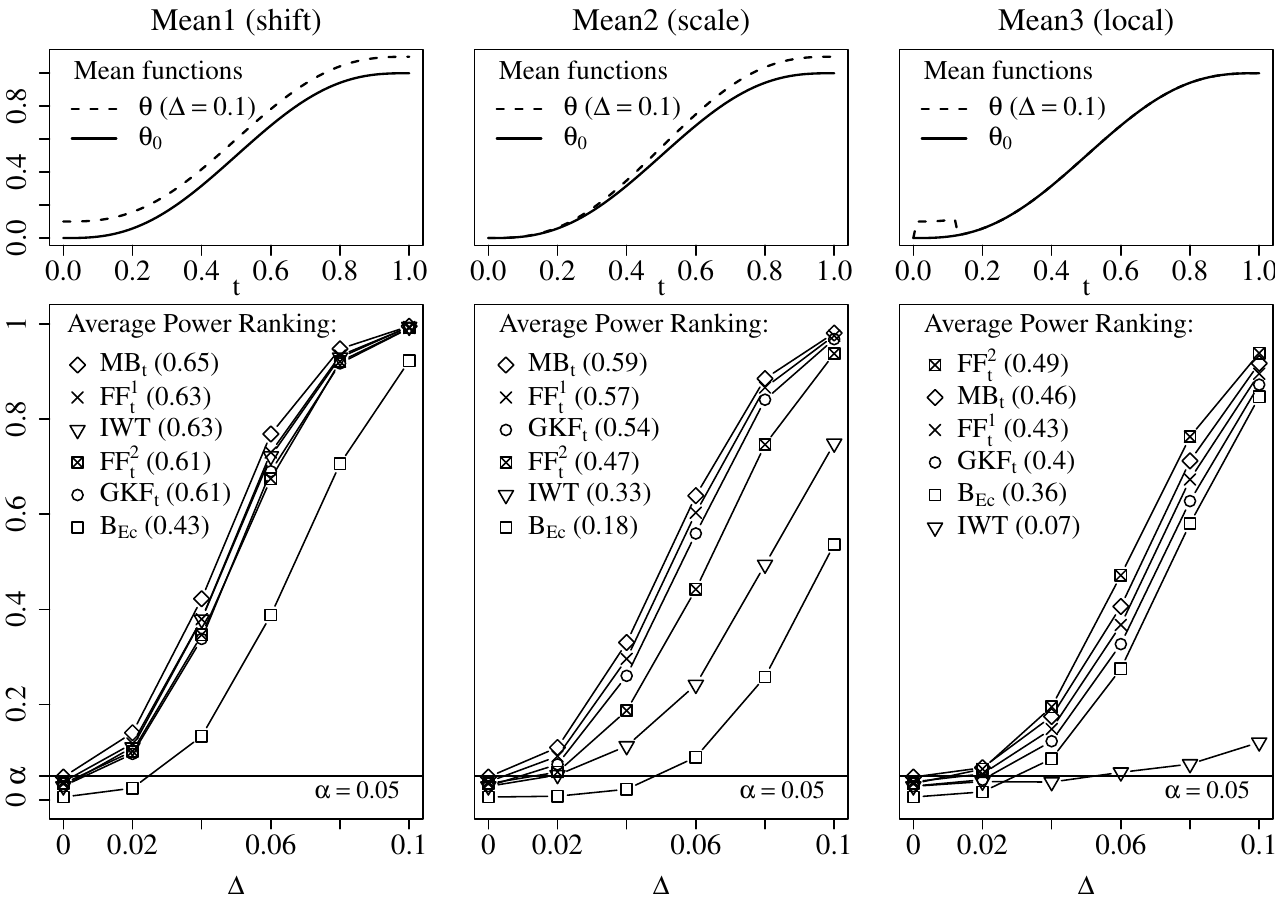}
\caption{{\sc Upper Row:} Different mean function scenarios Mean 1-3. {\sc Lower Row:}  Power comparisons for the `smooth to rough' covariance scenario Cov3 and sample size $n=100$. The average power values are averages over all mean perturbations $\Delta>0$.}
\label{FIG:SIM_2}
\end{figure}

%%%%%%%%%%%%%%%%%%%%%%%%%%%%%%%%%
For comparing the power of the hypothesis tests, we generate data using the mean
function scenarios Mean1-3 with increasingly large perturbations $\Delta>0$.
The hypothetical mean function, $\theta_0$, and the different true mean
functions, $\theta$, are shown in the upper row of Figure \ref{FIG:SIM_2}.  For
the small sample szenario, $n=15$, we consider the perturbations
$\Delta\in\{0.05, 0.15, 0.25, 0.35, 0.45\}$ and for the large sample szenario,
$n=100$, we consider the smaller perturbations
$\Delta\in\{0.02,0.04,0.06,0.08,0.1\}$.  Each of these mean function
perturbations is considered for each of the covariance scenarios Cov1-3.  We
reduce the number of Monte Carlo repetitions from 50,000 to 10,000, except for
the computationally costly IWT procedure where we keep the 5,000 repetitions as
used above.

%%%%%%%%%%%%%%%%%%%%%%
%\spacingset{1}
%\spacingset{1.00001}
%%%%%%%%%%%%%%%%%%%%%%
The lower row of Figure \ref{FIG:SIM_2} shows the power plots for the large
sample, $n=100$, and the challenging non-stationary covariance scenario Cov3.
The legends are ranked according to the methods' average powers over $\Delta>0$.
The results of the B$_{\text{S}}$ band and the FF$_t^4$ band are omitted to improve the visual exposition,
but all results for all scenarios can be found in Appendix
\ref{APP:ADD_MATERIAL} of the supplementary paper \cite{LR_Suppl}. The common
conclusion is as following:
\begin{itemize}[wide, labelwidth=!, labelindent=0pt, itemsep=-1ex]
\item[(a)] In scenario Mean1, all bands show similar power curves, except for
  the B$_{\text{EC}}$ band which has a relatively low power.
\item[(b)] In scenario Mean2, the bands MB$_t$, FF$_t^1$, GKF$_t$, and FF$_t^2$
  show similar power curves, except for the IWT procedure and the
  B$_{\text{EC}}$ band which have relatively low powers.
\item[(c)] In scenario Mean3, the band FF$_t^2$ has highest power since the fairness
  adaptation facilitates the detecting the local violation of the null
  hypothesis which is located in the initial area with high correlations.
  The IWT method of \cite{PV2017_IWT} is essentially not able to detect
  the considered local violation of the null hypothesis (Mean3).
  \item[(d)] In all scenarios, the power of the bootstrap band, B$_{\text{S}}$,
    (not shown in Figure \ref{FIG:SIM_2})
    is inflated as this band cannot keep the nominal $\alpha$-level (see Table \ref{TAB:SIM_1}).
\end{itemize}

%%%%%%%%%%%%%%%%%%%%%%%%%%%%%%%%%%%%%%%%%%%%%%%%%%%%%%%%%%%%%
\subsection{Fragmentary Functions}\label{SEC:SIMUL_PARTFDA}
\begin{figure}[h!tb]
\centering
\includegraphics[height=.5\textheight]{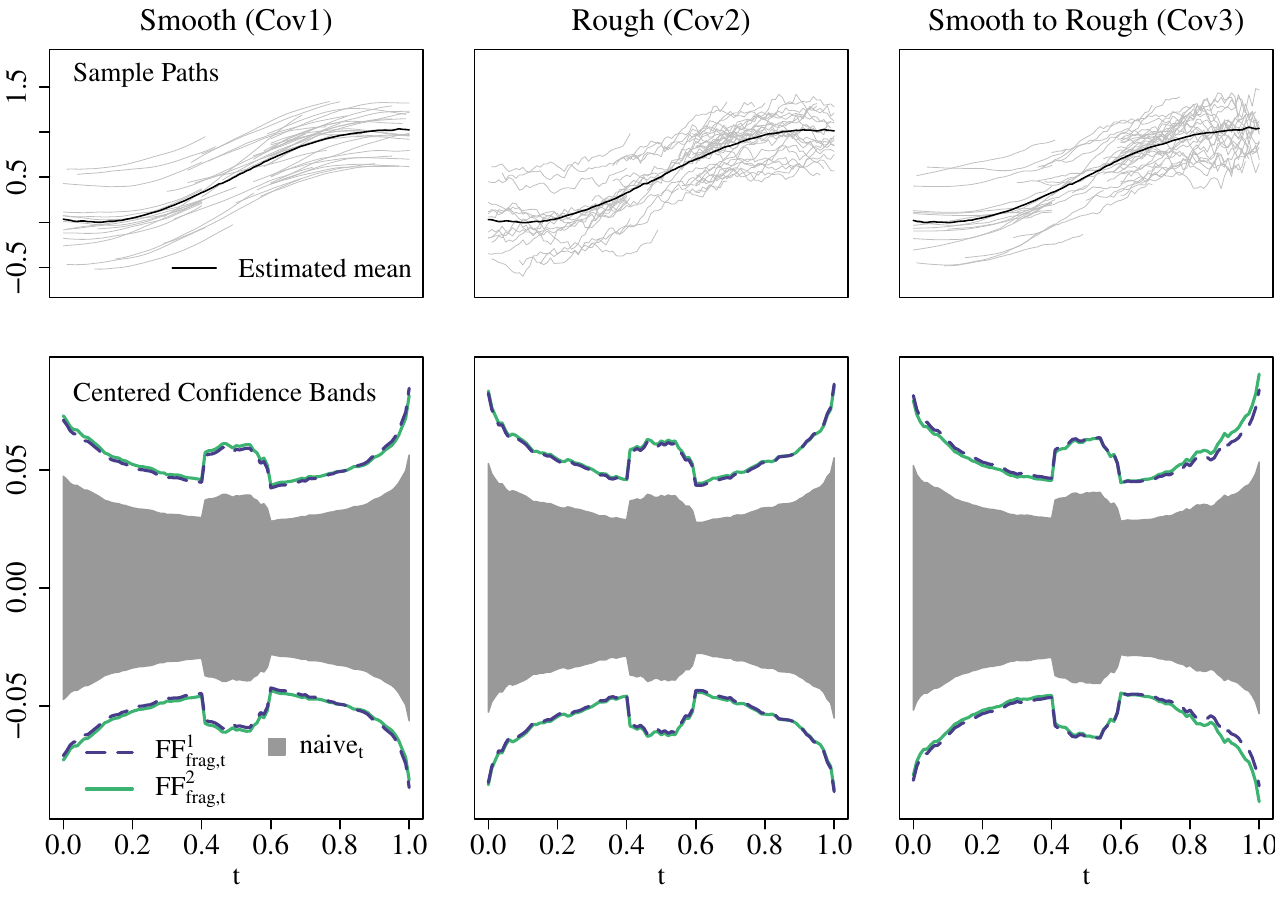}
\caption{{\sc Upper Row:} Sample paths and mean estimates,
  $\hat\theta_{\operatorname{frag}}$.  {\sc Lower Row:} Centered 95\% confidence bands, where centering by $\hat\theta_{\operatorname{frag}}$ is applied to improve the visual exposition.}
\label{FIG:SIM_3}  
\end{figure}

%%%%%%%%%%%%%%%%%%%%%%%%%%%%%%%%%%%%%%%%%%%%%%%%%%%%%%%%%%%%%
In this section, we consider fragmentary functional data generated from the mean
and covariance scenarios, Mean1-3 and Cov1-3.  In a first step, we draw random
sample functions
$\{S_i\}_{i=1}^n\overset{\operatorname{iid}}{\sim}\mathcal{N}(\theta,C_\theta)$
and evaluate them at $101$ equidistant grid points in $[0,1]$ as in
Section \ref{SEC:SIMUL_FULLFDA}.  In a second step, we fragment the curves by
declaring all grid points $t\not\in[A_i,B_i]\subset[0,1]$ as missing,
where $B_i=A_i+0.4$ and $A_i=\tilde{A}_i/100$ with
$\{\tilde{A}_i\}_{i=1}^n\overset{\operatorname{iid}}{\sim}\operatorname{BeB}(N=60,\alpha=0.3,\beta=0.3)$
and with $\operatorname{BeB}(N,\alpha,\beta)$ denoting the discrete
Beta-Binomial distribution.  This leads to a challenging case of fragmentary
functions, $S_i$, since no function covers the total domain $[0,1]$.
Consequently, the covariance function, $C_\theta$, can only be estimated over a
band along the diagonal, namely, over $\{(t,s)\in[0,1]^2:|t-s|\leq
0.4\}\subset[0,1]^2$.  We consider a relatively large sample size of $n=500$
since fragmenting the functions reduces the local sample size,
$n_t\leq n$,
considerably. In our simulation study, the local sample size varied between
$n_t=44$ and $n_t=332$.  The upper row in Figure \ref{FIG:SIM_3} shows exemplary
fragmented sample paths for each of the covariance scenarios Cov1-3.
%%%%%%%%%%%%%%%%%%%%%%
%\spacingset{1}
%\spacingset{1.00001}
%%%%%%%%%%%%%%%%%%%%%%

For estimating the mean, $\theta$, and covariance, $C_\theta$, we use the following estimators for fragmentary functional data:
\begin{align*}
  \hat\theta_{\operatorname{frag}}(t) &= \frac{I_1(t)}{n_t} \sum_{i=1}^n O_i(t)\,S_i(t),\\
  \hat{C}_{\operatorname{frag}}(t,s) &= \frac{I_2(t,s)}{n_{ts}} \sum_{i=1}^n O_i(t)O_i(s)\, \big(S_i(t) - \hat\theta_{\operatorname{frag}}(t) \big) \big(S_i(s) - \hat\theta_{\operatorname{frag}}(s) \big),
\end{align*}
with $n_t=\sum_{i=1}^n O_i(t)$ and $n_{ts}=\sum_{i=1}^n O_i(t)O_i(s)$, where $O_i(t)
= \mathbbm{1}\{S_i(t)\textrm{ is not missing}\}$. The operators $I_1(t) =
\mathbbm{1}\{\sum_{i=1}^n O_i(t) > 0\}$ and $I_2(t,s)=\mathbbm{1}\{\sum_{i=1}^n
O_i(t)O_i(s) > 0\}$  prevent divisions by zero through defining
$0/0:=\texttt{Missing}$. Recent works using these estimators are, for instance, \cite{DH2013}, \cite{K2015}, and \cite{LR2019}.

To the best of our knowledge, the following fragment versions,
$\operatorname{FF}_{\operatorname{frag}}^p$, or our FF bands are the only
simultaneous confidence bands that can be used in the challenging case of fragmentary functional data which prevents estimating the full covariance:
\begin{align*}
\operatorname{FF}_{\operatorname{frag},\text{distr}}^p(t)=[\hat\theta^\star_{\operatorname{frag},l}(t),\hat\theta^\star_{\operatorname{frag},u}(t)]&=\hat\theta_{\operatorname{frag}}(t)\pm \hat{u}_{\operatorname{frag},\alpha/2}^\star(t) \big(\hat{C}_{\operatorname{frag}}(t,t)/n_t\big)^{1/2},\quad t\in[0,1],
\end{align*}
with $\text{distr}\in\{z,t\}$, and where we focus here on $p\in\{1,2\}$.
% For
% $\operatorname{FF}_{\operatorname{frag},z}^1$ the constant threshold,
% $\hat{u}_{\operatorname{frag},\alpha}^\star$, is computed by solving
% \eqref{EQ:KacRice_Gauss} and for $\operatorname{FF}_{\operatorname{frag},t}^1$
% by solving \eqref{EQ:KacRice_t}. For
% $\operatorname{FF}_{\operatorname{frag},z}^2$, and where the adaptive threshold
% function, $\hat{u}_{\operatorname{frag},\alpha}^\star$, is computed by Algorithm
% \ref{ALGO:FAIRNESS} based on \eqref{EQ:GRF_Gauss} for
% $\operatorname{FF}_{\operatorname{frag},z}$ or \eqref{EQ:GRF_t} for
% $\operatorname{FF}_{\operatorname{frag},t}$.
As an estimator of the roughness parameter function, $\tau$, we use the following fragment version of  $\hat\tau_1$ in \eqref{EQ:tau_1}:
\begin{equation*}
\hat{\tau}_{\operatorname{frag}}(t)=\big(\partial_{12}\hat{c}_{\operatorname{frag}}(t,t)\big)^{1/2}\quad\text{with}\quad
\hat{c}_{\operatorname{frag}}(t,s)=\hat{C}_{\operatorname{frag}}(t,s)\big(\hat{C}_{\operatorname{frag}}(t,t)\hat{C}_{\operatorname{frag}}(s,s)\big)^{-1/2}
\end{equation*}
for $t,s\in[0,1]$.  This estimator is feasible even when the covariance
estimate $\hat{C}_{\operatorname{frag}}(t,s)$ can only be computed at the
diagonal and a narrow band along the diagonal. For a given
$\hat{\tau}_{\operatorname{frag}}$ estimate,
$\hat{u}_{\operatorname{frag},\alpha/2}^\star$ can be computed using Algorithm
\ref{ALGO:FAIRNESS} as in the case of fully observed functional data. As degrees of freedom of the $t$-process we use $\nu=\min_t(n_t)-1$.
%%%%
% {\color{red}We cannot recommend a corresponding adaption of $\hat\tau_2$ in \eqref{EQ:tau_2} as, to our experience, it underestimates the roughness in the case of fragmentary functional data.}
%%%%
We test the hypothesis H$_0$: $\theta(t)=\theta_0(t)$
$\forall$ $t\in[0,1]$ vs.~H$_1$: $\exists t\in[0,1]$
s.t.~$\theta(t)\neq\theta_0(t)$ at the nominal significance level $\alpha=0.05$.

The lower row of Figure \ref{FIG:SIM_3} shows examples of the $95\%$
simultaneous confidence bands $\operatorname{FF}_{\operatorname{frag},t}^1$ and $\operatorname{FF}_{\operatorname{frag},t}^2$. Note that the shapes of the bands are essentially equivalent to each other---even in the case of the non-stationary covariance scenario Cov3. The reason for this is that the missingness process, $O_i$, introduces an additional source of roughness into the covariance structure of the estimator $\hat\theta_{\operatorname{frag}}$. This additional roughness component becomes here the dominating roughness component leading to equivalent band shapes since we consider the same missingness process for all covariance scenarios Cov1-3.  The unusual shape of the bands is due to the varying local sample sizes: $n_t$ is relatively small at the boundaries and over $[0.4,0.6]$ which makes the bands wide at the boundaries of $[0,1]$ and causes the bump shape over $[0.4,0.6]$.

%%%%%%%%%%%%%%%%%%%%%%%%%%
%\spacingset{1}
\begin{table}[h!tb]
  \caption{The type-I error rates and power values of the fragment bands
    FF$_{\operatorname{frag},t}^1$ and FF$_{\operatorname{frag},t}^2$ in the non-stationary covariance scenario Cov3.}
  \label{TAB:FRAGM}
  \begin{adjustbox}{width=\columnwidth,center}
    \centering
    \begin{tabular}{ll c cccccc c}
      \toprule
      &     &H$_0$&\multicolumn{5}{c}{H$_1$}&Avg.~\\
      \cmidrule(lr){3-3}\cmidrule(lr){4-8}\\[-2ex]
      Mean & Band &$\Delta=0$ &$\Delta=0.02$ &$\Delta=0.04$ &$\Delta=0.06$ &$\Delta=0.08$ &$\Delta=0.1$&Power \\
      \midrule
      Mean1 & FF$_{\operatorname{frag},t}^1$ & 0.042 & 0.243 & 0.798 & 0.992 & 1.000 & 1.000 & 0.81 \\
      Mean1 & FF$_{\operatorname{frag},t}^2$ & 0.041 & 0.244 & 0.803 & 0.992 & 1.000 & 1.000 & 0.81 \\
      Mean2 & FF$_{\operatorname{frag},t}^1$ & 0.042 & 0.130 & 0.486 & 0.855 & 0.987 & 0.999 & 0.69 \\
      Mean2 & FF$_{\operatorname{frag},t}^2$ & 0.042 & 0.119 & 0.455 & 0.835 & 0.984 & 0.999 & 0.68 \\
      Mean3 & FF$_{\operatorname{frag},t}^1$ & 0.043 & 0.074 & 0.254 & 0.620 & 0.908 & 0.992 & 0.57 \\
      Mean3 & FF$_{\operatorname{frag},t}^2$ & 0.041 & 0.078 & 0.273 & 0.645 & 0.919 & 0.994 & 0.58 \\
      \bottomrule
      %\multicolumn{9}{l}{Nominal type-I error rate: $\alpha=0.05$}
    \end{tabular}
  \end{adjustbox}
\end{table}
%\spacingset{1.00001}
%%%%%%%%%%%%%%%%%%%%%%%%%%

Table \ref{TAB:FRAGM} summarizes the empirical type-I error rates and power
values of the fragment bands FF$_{\operatorname{frag},t}^1$ and
FF$_{\operatorname{frag},t}^2$ in the challenging non-stationary covariance
scenario Cov3; the results for all other scenarios are summarized in the tables
of Appendix \ref{APP:ADD_MATERIAL} of the supplementary paper \cite{LR_Suppl}.
They all lead to the same conclusions; namely, that the
FF$_{\operatorname{frag},t}^1$ and FF$_{\operatorname{frag},t}^2$ bands perform
very similar.  Both are able to keep the nominal $\alpha$-level at comparable
type-I error rates and both show similar power values.

% The possibility to
% perform simultaneous inference for functional parameters in the case of
% fragmentary functional data is unique in the literature.

%%%%%%%%%%%%%%%%%%%%%%%%%%% 
\section{Applications}\label{SEC:APPL}   
%%%%%%%%%%%%%%%%%%%%%%%%%%%
In this section we demonstrate the use of our fair confidence bands in two case studies.  Section \ref{SSEC:APPL_BIOM} considers an example from sports biomechanics where scientists often collect and analyze fully observed functional data.  Section \ref{SSEC:APPL_FRAGM} considers the case of fragmentary growth curves where the estimation of the total covariance function is impossible.

%%%%%%%%%%%%%%%%%%%%%%%%%%%%%%%%%%%%%%%%%%%%%%%%%%%%%
\subsection{Fully Observed Functions}\label{SSEC:APPL_BIOM}
\begin{figure}
\centering
\includegraphics[width=\textwidth]{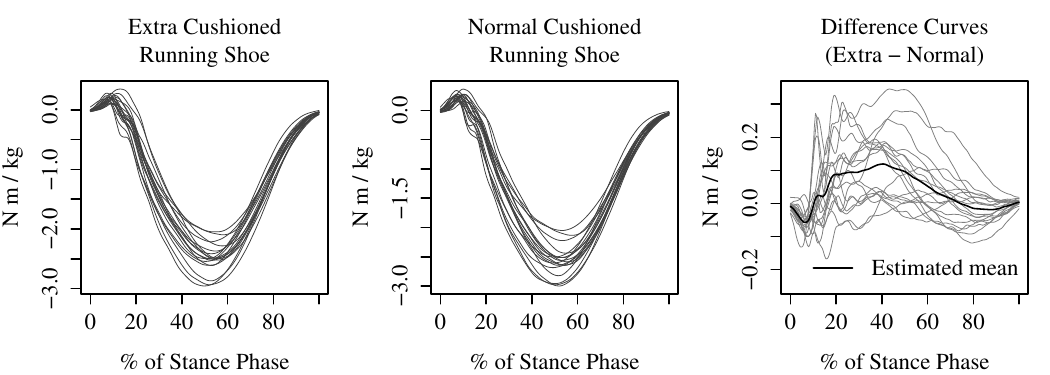}
\caption{{\sc Left and Middle:} Paired sample of $n=18$ torque curves measured at the ankle joint in sagittal pane. {\sc Right:} Sample of $n=18$ difference-curves computed from the paired sample.} \label{FIG:BIOM_1}
\end{figure}

%%%%%%%%%%%%%%%%%%%%%%%%%%%%%%%%%%%%%%%%%%%%%%%%%%%%%
Empirical research in biomechanics uses functional data for describing human (or animal) movements over time \citep{VVPR2012,HHZ2016,WCDH2019}.  The data shown in Figure \ref{FIG:BIOM_1} comes from a sports biomechanics experiment designed to assess the differences between extra cushioned vs.~normal cushioned running shoes. The experiment was conducted at the biomechanics lab of the German Sport University, Cologne.  A sample of $n=18$ recreational runners with a habitual heel strike running pattern were included into the experiment. The outcome of interest are torque curves describing the temporal torques acting at the right ankle joint in sagittal pane during the stance phase of one running stride.\footnote{The stance phase is the phase of a running stride during which the foot has ground contact. The individual stance phases are standardized to a common unit interval $[0\%,100\%]$ using simple linear affine time transformations.  This simple warping method leads to a good alignment of the data and is common in the biomechanics literature.}  The torques are measured in Newton metres (N m) standardized by the bodyweight (kg) of the participants.  At $t=0\%$ the heel strikes the ground and at $t=100\%$ the forefoot leaves the ground.  Further details on the data can be found in \cite{LWHB2014}, who consider a more exhaustive version of the data set.
%%%%%%%%%%%%%%%%%%%%%%
%\spacingset{1}
%\spacingset{1.00001}
%%%%%%%%%%%%%%%%%%%%%%

\begin{figure}
\centering
\includegraphics[width=\textwidth]{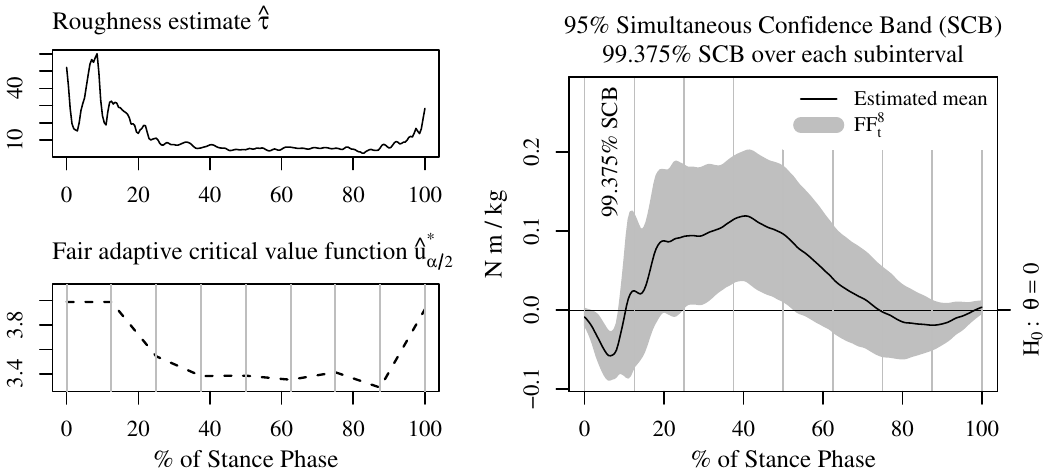}
\caption{{\sc Upper-Left:} Estimate of the roughness parameter function,
  $\tau$. {\sc Lower-Left:} Adaptive fair critical value function of the FF$_t^8$ band.  {\sc Right:} 95\% Simultaneous confidence band (SCB) over $[0,1]$ and $99.375\%$ SCBs over each $[(j-1) 12.5\%,j 12.5\%]$, $j=1,\dots,8$.}
\label{FIG:BIOM_2}
\end{figure}

For each participant there are two torque curves---one when running with a certain extra cushioned running shoe and another when running with a certain normal cushioned running shoe (left and middle plot of Figure \ref{FIG:BIOM_1}). In order to test for differences in the mean torque curves, we consider the pairwise difference-curves (extra minus normal cushioned) shown in the right plot of Figure \ref{FIG:BIOM_1} and test the hypotheses H$_0$: $\theta(t)=0$ $\forall t\in[0\%,100\%]$ vs.~H$_1$: $\exists t\in[0\%,100\%]$ s.t.~$\theta(t)\neq 0$.  To conduct this test, we use the fair $95\%$ simultaneous confidence band FF$_t^8$ with $p=8$ equidistant domain partitions (i.e., $a_0=0$, $a_1=12.5\%$, $a_2=25\%$, $a_3=37.5\%$, \dots, $a_8=100\%$) as indicated by the gray vertical lines in Figure \ref{FIG:BIOM_2}.

The upper-left plot in Figure \ref{FIG:BIOM_2} shows the estimate, $\hat\tau$, of the roughness parameter function $\tau$ which shows large roughness at the beginning and the end of the stance phase, but low roughness in the middle section. The lower-left plot demonstrates how the adaptive fair critical value function, $\hat{u}^\star_{\alpha/2}$, adapts to this roughness pattern.  
%%%%%%%%%%%%%%%%%%%%%%
%\spacingset{1}
%\spacingset{1.00001} 
%%%%%%%%%%%%%%%%%%%%%

\vspace*{-3mm}

\paragraph{Global and local interpretations.}
The simultaneous $95\%$ confidence band FF$_t^8$ in the right plot of Figure \ref{FIG:BIOM_2} shows two regions which do not contain the ``no effect'' zero parameter. Thus the global ``no effect'' null-hypothesis, H$_0$: $\theta(t)=0$ for all $t\in[0,100\%]$, can be rejected at the $\alpha=0.05$ level. Moreover, the fair $95\%$ SCB, FF$_t^8$, allows here the following two local interpretations: 
\begin{description}[wide, leftmargin=10pt, labelwidth=!, labelindent=5pt, itemsep=0ex,topsep=2pt]
  \item[The first significant region]is contained in the first subinterval $[0,12.5\%]$ over which FF$_t^8$ is a $(1-\alpha/8)\times 100\%=99.375\%$ SCB. Thus, the local ``no effect'' null-hypothesis, H$_0$: $\theta(t)=0$ for all $t\in[0,12.5\%]$, can be rejected at the $\alpha/8=0.006$ level. 
  \item[The second significant region]essentially stretches over the two neighboring subintervals $[25\%,37.5\%]$ and $[37.5\%,50\%]$ over which FF$_t^8$ is a $(1-2\alpha/8)\times 100\%=98.75\%$ SCB. Thus, the local ``no effect'' null-hypothesis, H$_0$: $\theta(t)=0$ for all $t\in[25\%,50\%]$, can be rejected at the $2(\alpha/8)=0.012$ level.
\end{description}

%%%%%%%%%%%%%%%%%%%%%%%%%%%%%%%
\subsection{Fragmentary Functional Data}\label{SSEC:APPL_FRAGM}
\begin{figure}
\centering
\includegraphics[width=.9\textwidth]{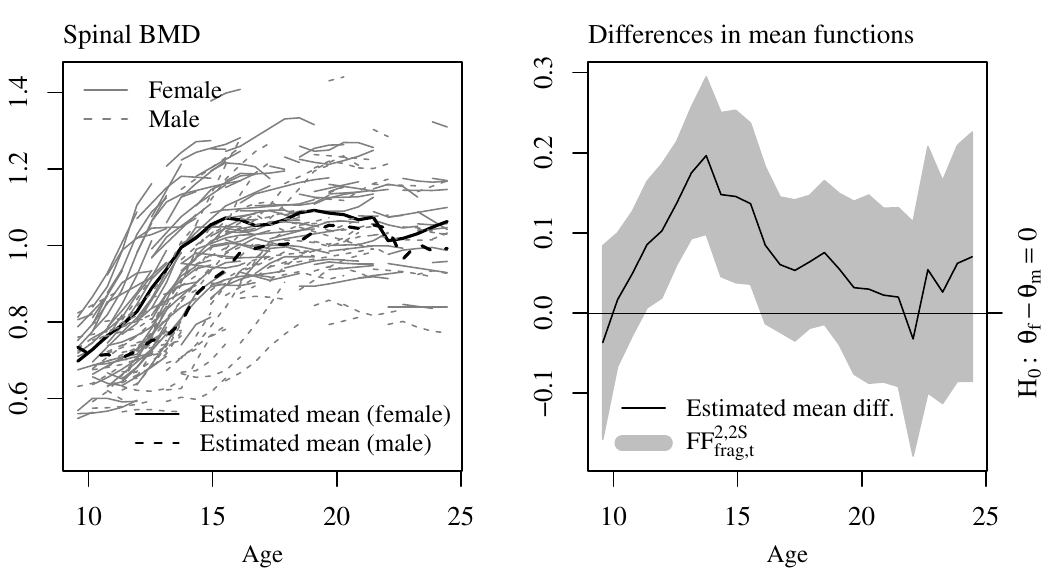}
\caption{{\sc Left Plot:} Fragmentary spinal bone mineral density aquisition curves.  {\sc Right Plot:} Simultaneous 95\% confidence band, FF$_t^2$, for the difference mean function.} \label{FIG:FRAGM}
\end{figure}

\
%%%%%%%%%%%%%%%%%%%%%%%%%%%%%%%
The bone mineral acquisition data were first described in \cite{BHWNM1999} and further analyzed by \cite{JH2001}, \cite{DH2013}, and
\cite{DH2016}.\footnote{The data is freely available at the companion website of the textbook of \cite{FHT2001_book}.} The spinal Bone Minearal Density (BMD) measurements (g/cm$^2$) were taken for each individual at two to four time points over short time intervals; therefore, linear interpolations of the measurements give good approximations of the underlying smooth mineral acquisition growth processes. The left plot in Figure \ref{FIG:FRAGM} shows the fragmentary growth curves of $n_{\operatorname{f}}=140$ female (f) and $n_{\operatorname{m}}=113$ male (m) participants after restricting the data to a common domain (9.6 to 24.5 years of age) and at least two measurements.
%%%%%%%%%%%%%%%%%%%%%%
%\spacingset{1}
%\spacingset{1.00001}
%%%%%%%%%%%%%%%%%%%%%

We test the hypotheses of equal means H$_0$: $\theta_{\operatorname{f}}(t)-\theta_{\operatorname{m}}(t)=0$ $\forall t\in[9.6\operatorname{ yr},24.5\operatorname{ yr}]$ vs.~H$_1$: $\exists t\in[9.6\operatorname{ yr},24.5\operatorname{ yr}]$ s.t.~$\theta_{\operatorname{f}}(t)\neq\theta_{\operatorname{m}}(t)$ using the following two-sample (2S) version of our FF simultaneous confidence band, 
\begin{align*}
  \operatorname{FF}^{p,\operatorname{2S}}_{\operatorname{frag},t}(t)&=\big(\hat\theta_{\operatorname{f}}(t)-\hat\theta_{\operatorname{m}}(t)\big)\pm\hat{u}^{\star\operatorname{2S}}_{\operatorname{frag}\alpha/2}(t)\big(\hat{C}^{\operatorname{2S}}_{\operatorname{frag}}(t,t)\big)^{1/2}\quad\text{with}\quad p=2\quad\text{and}\\
\hat{C}^{\operatorname{2S}}_{\operatorname{frag}}(t,s)&=\big((n^{\operatorname{f}}_{ts}-1)\hat{C}^{\operatorname{f}}_{\operatorname{frag}}(t,s)+(n^{\operatorname{m}}_{ts}-1)\hat{C}^{\operatorname{m}}_{\operatorname{frag}}(t,s)\big)/(n^{\operatorname{f}}_{ts}+n^{\operatorname{m}}_{ts}-2)
\end{align*}
denoting the weighted average of the sample covariances,
$\hat{C}^{\operatorname{f}}_{\operatorname{frag}}$ and $\hat{C}^{\operatorname{m}}_{\operatorname{frag}}$, with $n^{\operatorname{m}}_{ts}=\sum_{i=1}O^{\operatorname{m}}_i(t)O^{\operatorname{m}}_i(s)$, $n^{\operatorname{f}}_{ts}=\sum_{i=1}O^{\operatorname{f}}_i(t)O^{\operatorname{f}}_i(s)$, and $O^{\operatorname{f}}_i(t)$ and $O^{\operatorname{m}}_i(t)$ as defined in Section \ref{SEC:SIMUL_PARTFDA}.  The short fragments of the growth curves allow only an estimation of the covariances, $\hat{C}^{\operatorname{f}}_{\operatorname{frag}}$ and $\hat{C}^{\operatorname{m}}_{\operatorname{frag}}$, over narrow bands along the diagonal (see Figure \ref{FIG:FRAGM_Appendix} in Appendix \ref{APP:ADD_MATERIAL} of \cite{LR_Suppl}).  The critical value function, $\hat{u}^{\star\operatorname{2S}}_{\operatorname{frag}\alpha/2}$, is computed using Algorithm \ref{ALGO:FAIRNESS} based on the significance level $\alpha=0.05$, $p=2$ equidistant domain partitions ($a_9.6=0$, $a_1=17.05$, $a_2=24.5$), and empirical roughness parameter function, $\hat{\tau}^{\operatorname{2S}}$, computed from $\hat{C}^{\operatorname{2S}}_{\operatorname{frag}}$ using the estimator in \eqref{EQ:tau_1}. 

\vspace*{-3mm}

\paragraph{Global and local interpretations.}
The right plot in Figure \ref{FIG:FRAGM} shows a significant difference in the mean growth functions from 11.4 to 15.5 years of age, and thus the global ``no difference'' null-hypothesis, H$_0$: $\theta_{\operatorname{f}}(t)-\theta_{\operatorname{m}}(t)=0$ $\forall t\in[9.6\operatorname{ yr}, 24.5\operatorname{ yr}]$, can be rejected at the $\alpha=0.05$ level.
\begin{description}[wide, leftmargin=10pt, labelwidth=!, labelindent=5pt, itemsep=0ex,topsep=2pt]
\item[The significant region]is contained withing the first half $[9.6\operatorname{ yr}, 17.05\operatorname{ yr}]$ of the domain over which $\operatorname{FF}^{2,\operatorname{2S}}_{\operatorname{frag},t}$ is a $(1-\alpha/2)\times 100\%=97.5\%$ SCB. Thus the local ``no difference'' null-hypothesis H$_0$: $\theta_{\operatorname{f}}(t)-\theta_{\operatorname{m}}(t)=0$ $\forall t\in[9.6\operatorname{ yr},17.05\operatorname{ yr}]$ can be rejected at the $\alpha/2=0.025$ level. This confirms the adolescent gender-differences described in \cite{BHWNM1999}. 
\end{description}

%%%%%%%%%%%%%%%%%%%%%%%%%%%%%%%
\section{Discussion}\label{SEC:DISS}
%%%%%%%%%%%%%%%%%%%%%%%%%%%%%%%
In this paper, we propose a novel method for constructing simultaneous
confidence bands for (fragmentary) functional data using adaptive critical value
functions. While we focus on the construction of fair critical value functions, Theorem
\ref{TH:MAIN} can also be used to explore other selection criteria beyond fairness. For instance, a very simple approach
would be to optimize the allocation of the total false positive rate over given domain partitions such that the minimum average
width of the band is minimized. Another still unanswered question is the
extension of our approach to the case of higher dimensional function domains or
manifolds. Such extensions exist for the classic Kac-Rice formulas and
thus, in principle, should be possible for our adaptive bands as well. Finally, another useful, but highly nontrivial extension would be to further relax the
(asymptotic) distribution assumption.  In particular, estimates for models such
as scalar-on-function regression do not possess any asymptotic distribution in
the strong topology since they are not tight \citep{CMS2007}.  Instead, such
estimates satisfy the CLT only in the weak topology.  It is still unclear if
confidence bands can be constructed for such estimates. 

% Lastly, our assumptions require the underlying estimate/process to be
% continuously differentiable, but for some applications it might be useful to
% only require continuity of the process.  However, such a generalization would require serious theoretical work as non-differentiable processes can cross a smooth boundary infinitely many times over a compact domain (e.g. Brownian motion).

\bigskip

\noindent\textbf{Supplementary materials.} The supplementary paper
\cite{LR_Suppl} contains the mathematical proofs or our theoretical results,
further simulation results, and additional plots. The \textsf{R}-package
\texttt{ffscb} which implements the introduced methods is available at \url{www.dliebl.com/ffscb/}.

\bigskip

\noindent\textbf{Acknowledgments.} We thank the mathematical research institute
MATRIX in Creswick, Australia and the Simons Institute for the Theory of
Computing at the UC Berkeley where parts of this research was performed. Many
thanks go also to the four anonymous referees and the associate editor whose comments helped us to
improve our manuscript.  

%Many thanks go also to Steffen Willwacher (German Sport University Cologne) for
%providing the data of the biomechanics application in Section
%\ref{SSEC:APPL_BIOM}.

\spacingset{1}
%%%%%%%%%%%%%%%%%%%%%%%%%%%%%
% References
%%%%%%%%%%%%%%%%%%%%%%%%%%%%%
\bibliographystyle{Chicago}
\bibliography{bibfile}
%%%%%%%%%%%%%%%%%%%%%%%%%%%%%

%%%%%%%%%%%%%%%%%%%%%%%%%%%%%%%%%%%%%%%%
% Appendix-Supplement Paper
%%%%%%%%%%%%%%%%%%%%%%%%%%%%%%%%%%%%%%%%
\newpage

\setcounter{page}{1}
\thispagestyle{plain}
%\pagenumbering{Roman}

\appendix

\vspace*{.5cm}
\spacingset{1}
%%%%%%%%%%%%%%%%%%%%%%%%%%%%%%%%%%%%%%%%%%

\if0\blind
{ \begin{center}
  {\LARGE \bf Supplement to\\[1ex]
  Fast and Fair Simultaneous Confidence\\[1ex] Bands for Functional Parameters}

  \bigskip

  %%%%%%%%%%%%%%
  Dominik Liebl\\University of Bonn\\[2ex]
  %%%%%%%
  Matthew Reimherr\\Penn State University
  %%%%%%%%%%%%%%
  \end{center}
} \fi

\if1\blind
{
  \bigskip
  \bigskip
  \bigskip
  \begin{center}
    {\LARGE \bf Supplement to:\\[1ex] Fast and Fair Simultaneous Confidence\\[1ex]Bands for Functional Parameters}
\end{center}
  \medskip
} \fi

\bigskip

\appendix

%%%%%%%%%%%%%%%%%%%%%%%%%%%%%%%%%%%%%%%%%%%%%%%
\section{Proofs}% of Theorem \ref{TH:MAIN} and Corollaries \ref{CO:GAUSS}-\ref{CO:STUDt}}
%%%%%%%%%%%%%%%%%%%%%%%%%%%%%%%%%%%%%%%%%%%%%%%
The mathematical arguments are the same for any $t_0 \in [0,1]$ since the
expected number of down-crossings about $u(t)$ is equal to the expected number of up-crossings about $-u(t)$.  Thus we will fix $t_0 = 0$ in our arguments below.  We first establish all of our results conditioned on the mixing coefficient, $V$, for the elliptical distribution, which can be thought of as computing $\E[\varphi_u(X)|V]$, which is then a Gaussian calculation.  After establishing our results for the Gaussian case, we will then reintroduce $V$ and compute the final expectation $\E[\E[\varphi_u(X)|V]]$.

The proof hinges on two approximations that simplify the calculations.  The first step is to introduce a smooth approximation of the up-crossing count.  This is accomplished using a kernel, $K(x)$.  Since the kernel is merely a theoretical device, we will assume it is fairly simple in our lemmas.  Intuitively, if $K((u(t)-X(t))/h)$ is large for some small bandwidth, $h$, then $X(t)$ is very close to crossing $u(t)$.  Thus, by integrating this quantity (with appropriate normalizations) we can use it to approximate the up-crossing count.

Our second approximation consists of forming linear interpolations for both $X(t)$ and $u(t)$ on a dyadic grid.  In particular, fix an integer $k$ and an index $j=1,\dots,2^k$.  Then for any $t \in [0,1]$ such that $j - 1<  t 2^k  < j $ we form the quantities
\begin{align*}
\tX_k(t)&= X\left( \frac{j-1}{2^k} \right) \left(j - 2^k t  \right)
+ X\left( \frac{j}{2^k} \right)  \left(2^k t - j + 1  \right), \\
\tu_k(t)&= u\left( \frac{j-1}{2^k} \right) \left(j - 2^k t  \right)
+ u\left( \frac{j}{2^k} \right)  \left(2^k t - j + 1  \right), \\
\tX_k'(t)&= 2^k \left( X\left( \frac{j}{2^k} \right)  - X\left( \frac{j-1}{2^k} \right) \right), \\
\tu_k'(t)&= 2^k \left( u\left( \frac{j}{2^k} \right)  - u\left( \frac{j-1}{2^k} \right) \right).
\end{align*}
We continuously extend $\tX_k$ and $\tu_k$ at the dyadic points, $\tX_k(j/2^k) = X(j/2^k)$ and $\tu_k(j/2^k) = u_k(j/2^k)$ for $j = 0,1,\dots,2^k$.  Since the $\tX_k'$ and $\tu_k'$ are step functions, they cannot be continuously extended.  For convenience, we take $\tX'_k(j/2^k) = X'(j/2^k)$ and $\tu'_k(j/2^k) = u'_k(j/2^k)$, though for the derivatives what happens on a set of measure zero will have no impact on our calculations.  This choice also yields the convenient bounds
\begin{align*}
\sup_{0 \leq t \leq 1} |\tX_k(t)| & \leq \sup_{0 \leq t \leq 1} |X(t)|
& \sup_{0 \leq t \leq 1} |\tu_k(t)|  & \leq \sup_{0 \leq t \leq 1} |u(t)| \\
\sup_{0 \leq t \leq 1} |\tX'_k(t)| & \leq \sup_{0 \leq t \leq 1} |X'(t)|
& \sup_{0 \leq t \leq 1} |\tu'_k(t)|  & \leq \sup_{0 \leq t \leq 1} |u'(t)|.
\end{align*}

%%%%%%%%%%%%%%%%%%%%%%%%%%%%%
\subsection{Lemmas}
%%%%%%%%%%%%%%%%%%%%%%%%%%%%%

In this section we state our technical lemmas which results in a proof of Corollary \ref{CO:GAUSS} on the Gaussian case (Lemma \ref{lem:grf}).

\begin{lemma} \label{lem:intervals}
Let $f \in C[0,1]$ and suppose there only exits a finite number of zeros $\{t_1, \dots, t_k\}$, that is, $f(t) = 0$ if and only if $t =t_i$ for some $i=1,\dots,k$.  Then for any $\epsilon > 0$ there exists $h>0$ such that
\begin{align}
f^{-1}(-h,h) \subseteq \bigcup_{i=1}^k (t_i - \epsilon, t_i + \epsilon). \label{e:inv_img}
\end{align}
\end{lemma}
\noindent\textbf{Proof of Lemma \ref{lem:intervals}.}\\
%%%%%%%%%%%%%%%%%%%%%%%%%%%%%%%%%%%%%%%%%%%%%%%%%%%%%%%%%
Consider a proof by contradiction.  Suppose there exists $\epsilon > 0$ such that \eqref{e:inv_img} does not hold for any $h$.  Call the right hand side of \eqref{e:inv_img} $B_\epsilon$.
%%%%%
Then consider the sets $A_l = f^{-1}(-2^{-l},2^{-l})$ for $l = 1,2,\dots$.  By assumption, $A_l \cap B_\epsilon^c$ is nonempty for any $l$, so select an infinite sequence $\{x_l\}$ by taking $x_l \in A_l \cap B_\epsilon^c$.  By construction, $|f(x_l)| \leq 2^{-l}$, which implies that $f(x_l) \to 0$ as $l\to\infty$.  However, since $[0,1]$ is compact, there exists a convergent subsequence $x_{l'}$ such that $x_{l'} \to x$, which implies $f(x) = 0$.  However, $|x - t_i| \geq \epsilon$ for all $i=1,\dots,k$, which means $x$ cannot be one of the zeros, which is a contradiction. $\qed$

\bigskip

%%%%%%%%%%%%%%%%%%%%%%%%%%%%%%%%%%%
%
% Counting Formulas
%
%%%%%%%%%%%%%%%%%%%%%%%%%%%%%%%%%%%
\begin{lemma}{Counting Formula.}\label{lem:cf}

\begin{itemize}
%%%%%%%%%%%
\item[(a)]
%%%%%%%%%%%
Let $X$ be a Gaussian random function with $\var(X(t))>0$ for every $t\in[0,1]$ and let $u$ be a deterministic function.  If $X\in C^1[0,1]$ almost surely, $u\in C[0,1]$ and $u'$ is continuous almost everywhere, then the number of up-crossings is
\begin{align*}
N_u(X)
&=\#\{t\in[0,1]\,:\,X(t)=u(t),\,X'(t)-u'(t)>0\}\notag\\
&=\lim_{h\to 0}\int_0^1 \frac{1}{h}K\left(\frac{u(t)-X(t)}{h}\right)\big(X'(t)-u'(t)\big)\mathbbm{1}_{X'(t)-u'(t)>0}\,dt\;<\infty\;\quad a.s.,
\end{align*}
where $\mathbbm{1}$ denotes the indicator function and $K$ is a symmetric kernel function with compact support, i.e.,
$K(t) = K(-t) \geq 0$,
$K(t) = 0$ if $|t| > 1$,
$K(t)$ is continuous on $[-1,1]$,
and
$\smallint K(t) \ dt = 1$.
%%%%%%%%%%%
\item[(b)]
%%%%%%%%%%%
Let all requirements in (a) hold, %and let $\tX_k$ be a piecewise linear approximation of $X$ defined by the linear interpolation of the partition points $(j/2^k,X(j/2^k))$, $j=0,\dots,2^k$, for some fixed integer $k$. Then
then the dyadic linear interpolations satisfy
\begin{align*}
N_{\tilde{u}_k}(\tX_k)
&=\#\{t\in[0,1]\,:\,\tX_k(t)=\tu_k(t),\,\tX_k'(t)-\tu_k'(t)>0\}\\
&=\lim_{h\to 0}\int_0^1 \frac{1}{h}K\left(\frac{\tu_k(t)-\tX_k'(t)}{h}\right)\big(\tX_k'(t)-\tu_k'(t)\big)\mathbbm{1}_{\tX_k'(t)-\tu_k'(t)>0}\,dt \\
& \leq \min\{2^k, N_u(X)\},
\end{align*}
almost surely.
\end{itemize}
\end{lemma}

% %%%%%%%%%%%%%%%%%%%%%%%%%%%%%%%%%%%%%%%%%%%%%%%%%%%%%%%%%
% \noindent\textbf{Proof of Lemma \ref{lem:cf}}\\
% %%%%%%%%%%%%%%%%%%%%%%%%%%%%%%%%%%%%%%%%%%%%%%%%%%%%%%%%%
% Before we start with the proofs of parts (a) and (b) observe that the $t$ in $X(t)=u(t)$ is random and equals a discontinuity point of $u'$ with probability zero since $u'$ is continuous almost everywhere.

%%%%%%%%%%%%%%%%%%%%%%%%%%%%%%%%%%%%%%%%%%%%%%%%%%%%%%%%%
\noindent\textbf{Proof of Lemma \ref{lem:cf} Part (a).}\\
%%%%%%%%%%%%%%%%%%%%%%%%%%%%%%%%%%%%%%%%%%%%%%%%%%%%%%%%%
First consider the case of no up-crossings, $N_u(X)=0$, which can happen in only one of two ways. The first possibility is that $X(t)<u(t)$ for all $t\in[0,1]$ or $X(t) > u(t)$ for all $t \in [0,1]$.  Since $X(t) - u(t)$ is continuous and the domain is compact, there exists an $\epsilon > 0$ such that, $|X(t) - u(t)| \geq  \epsilon $ for all $t \in [0,1]$.  This implies that $K((u(t)-X(t))/h)=0$ for all $t\in[0,1]$ with $h \leq \epsilon$, thus trivially proving the claim.  The second possibility is that we have exactly one down crossing, but no up-crossing.  In this case, there is exactly one $t_d$ such that $X(t_d) = u(t_d)$, but $X'(t_d) < u'(t_d)$.  %{\color{blue}Again, since $X'(t)$ and $u'(t)$ are continuous, we can take a small enough interval such that for all $t \in (t_d -\delta, t_d+\delta)$ we have $\mathbbm{1}_{X'(t) > u'(t)} = 0$.}
{\color{black}%ALTERNATIVE:
Note that $t_d$ is random and equals a discontinuity point of $u'$ with probability zero since $u'$ is continuous almost everywhere. That is,  we can take a small enough interval $(t_d -\delta, t_d+\delta)$ such that for all $t \in (t_d -\delta, t_d+\delta)$ we have $\mathbbm{1}_{X'(t) > u'(t)} = 0$ with probability one.}  Then, $|X(t) - u(t)| \geq \epsilon > 0$ for all $t\in [0,t_d-\delta]\cup[t_d + \delta, 1]$ and we can again take $h$ small enough that the integral is exactly zero almost surely.

For the case $N_u(X)=m$ with $m>0$, define $X_u:=X-u$ and let us consider the up-crossings of $X_u$ above $0$, which is equivalent to the up-crossings of $X$ above $u$, that is,
\begin{align*}
N_u(X)
&=\#\{t\in[0,1]\,:\,X(t)-u(t)=0,\,X'(t)-u'(t)>0\}\notag\\
&=\#\{t\in[0,1]\,:\,X_u(t)=0,\,X'_u(t)>0\}\notag\\
&=\lim_{h\to 0}\int_0^1 \frac{1}{h}K\left(\frac{X_u(t)-0}{h}\right)X'_u(t)\mathbbm{1}_{X'_u(t)>0}\,dt.
\end{align*}
With probability one, no up-crossing location equals a discontinuity point of $u'$, thus In addition, with probability one, no up-crossing location equals a discontinuity point of $u'$, thus the $N_u(X)$ is a well defined quantity. $N_u(X)$ is a well defined quantity. Since $X(0)\neq u(0)$ and $X(1)\neq u(1)$ with probability 1, we can denote the (random) up-crossing locations as $0<t_1<\dots<t_m<1${\color{black}.
By Lemma \ref{lem:intervals} for any sufficiently small $h$, the inverse image of $(-h,h)$ by the function $X_u$ is contained in the union of $m$ pairwise disjoint intervals $I_1,\dots,I_m$ with $t_j\in I_j$, $j=1,\dots,m$.
Observe that $X'_u(t_j)>0$ for every up-crossing location $t_j$, $j=1,\dots,m$, and that $X'_u$ is continuous; therefore, we can take the intervals small enough that the restriction of $X_u$ to $I_j$ is almost surely a diffeomorphism for each $j=1,\dots,m$.
Since $X_u'(t) = X'(t) + u'(t) > 0$ on $I_j$ almost surely, we can perform a change of variables with $y h  = X_u(t)$ and $h dy = X'_u(t) dt$ without changing the orientation of the integral. Since $K$ is compactly supported, we can then take $h$ small enough that
\begin{align*}
 \int_{I_j} K \left( \frac{X_u(t)}{h} \right) \frac{X'_u(t)}{h} \ dt
=  \int_{-1}^1 K \left( y \right)  dy = 1, \quad\text{for every}\quad j=1,\dots,m,
\end{align*}
which implies that that
\begin{align*}
N_u(X)&=\int_0^1 \frac{1}{h}K\left(\frac{u(t)-X(t)}{h}\right)\big(X'(t)-u'(t)\big)\mathbbm{1}_{X'(t)-u'(t)>0}\,dt\\
&=\sum_{j=1}^m
\int_{I_j} \frac{1}{h} K \left( \frac{u(t) - X(t)}{h} \right) (X'(t) - u'(t))\mathbbm{1}_{X'(t) > u'(t)} \ dt
= m
\end{align*}
for all sufficiently small $h$.

Finally, the assumption that $X\in C^1([0,1])$ almost surely implies that the number of up-crossings is finite with probability 1, that is, $N_u(X)=m<\infty$ with probability 1. $\qed$
%%%%
% Proof: https://math.stackexchange.com/questions/445097/prove-that-f-has-finite-number-of-roots
%%%

\bigskip

%%%%%%%%%%%%%%%%%%%%%%%%%%%%%%%%%%%%%%%%%%%%%%%%%%%%%%%%%
\noindent\textbf{Proof of Lemma \ref{lem:cf} Part (b).}\\
%%%%%%%%%%%%%%%%%%%%%%%%%%%%%%%%%%%%%%%%%%%%%%%%%%%%%%%%%
%%%
% Remember: Countable sets such as j/2^k for j=1,...,2^k and for k\to\infty will be of measure 0 in an uncountable one such as [0,1].
%%%
Since $X(t)$ is a Gaussian process it follows that $\tX_k(j/2^k)\neq u(j/2^k)$
a.s.~for each $j=0,\dots,2^k$, meaning an up-crossing occurs only within one of
the partitioned intervals $[(j-1)/2^k,j/2^k]$, $j=1,\dots,2^k$.  Observe that,
restricted to an interval $((j-1)/2^k,j/2^k)$, $j=1,\dots,2^k$, $\tX_k$ is just
a very simple Gaussian process and $\tu_k$ is just a linear threshold -- both fulfilling the requirements of Lemma \ref{lem:cf} part (a).  Thus the result of Lemma \ref{lem:cf} part (a) holds also for each subinterval. So, almost surely, we can count the up-crossings for each partitioned interval and sum up the corresponding integrals
\begin{align*}
N_{\tilde{u}_k}(\tX_k)
&=\lim_{h\to 0}\sum_{j=1}^{2^k}\int_{(j-1)/2^k}^{j/2^k} \frac{1}{h}K\left(\frac{\tu_k(t)-\tX_k(t)}{h}\right)\big(\tX_k'(t)-\tu_k'(t)\big)\mathbbm{1}_{\tX_k'(t)-\tu_k'(t)>0}\,dt,
\end{align*}
which shows the equality statement.

Since both $\tX_k$ and $\tu_k$ are piecewise linear, their difference is also linear.  With probability one, this difference is not zero, thus by the fundamental theorem of algebra, there is at most one root/up-crossing. Therefore, the maximum number of up-crossings $N_{\tilde{u}_k}(\tX_k)$ is bounded by the number of partitioned intervals, $2^k$.  However, within an interval $[(j-1)/2^k,j/2^k]$, $j=1,\dots,2^k$, $\tX_k$ crossing $\tu_k$ implies $X$ must cross $u$ by the intermediate value theorem, but the reverse is not true since their can be a subsequent downcrossing.  Namely, it is only possible to ``miss'' up-crossings with the linear interpolation, one cannot add additional crossings, which shows the inequality statement. $\qed$

\bigskip

The result of the following lemma is equivalent to the result in Corollary \ref{CO:GAUSS} a).

%%%%%%%%%%%%%%%%%%%%%%%%%%%%%%%%%%%
%
% Generalized Gaussian Kac-Rice Formula
% for non-constant thresholds
%
%%%%%%%%%%%%%%%%%%%%%%%%%%%%%%%%%%%
\begin{lemma}{Kac-Rice Formula for Non-Constant Critical Value Functions.}\label{lem:grf}\ \\
Let $\{X(t): t \in [0,1]\}$ be a continuously differentiable Gaussian process and $u(t)$ a continuously differentiable function. Let $\tau(t)^2 = \partial_x \partial_y C(x,y) |_{x=y=t}$, where $C(x,y)$ is the covariance function of $X(t)$.  Assume $X(t)$ has a constant pointwise unit variance, $\var(X(t)) = 1$ for all $t\in[0,1]$. Then the expected value of $\varphi_{u,X}(t_0)=\mathbbm{1}_{X(0) \geq u(0)}+N_u(X)$ is
\begin{align*}
\E[\varphi_{u,X}(t_0)]
&= P(X(0) \geq u(0)) + \\
&\int_0^1 \frac{\tau(t)}{2 \pi }
\exp\left\{- \frac{1}{2} \left[u(t)^2  + \frac{u'(t)^2}{2\tau(t)^2} \right] \right\} \ dt -
\int_0^1 \frac{u'(t)}{\sqrt{2 \pi }}
\exp\left\{-  \frac{u(t)^2}{2}\right\} \Phi\left(\frac{-u'(t)}{\tau(t)}\right) \ dt,
\end{align*}
where $\Phi$ is the standard normal cdf.
\end{lemma}

%%%%%%%%%%%%%%%%%%%%%%%%%%%%%%%%%%%%%%%%%%%%%%%%%%
\noindent\textbf{Proof of Lemma \ref{lem:grf}.}\\
%%%%%%%%%%%%%%%%%%%%%%%%%%%%%%%%%%%%%%%%%%%%%%%%%%
%Let $\tX_k$ be a piecewise linear approximation of $X$ defined by the linear interpolation of the partition points $(j/2^k,X(j/2^k))$, $j=0,\dots,2^k$, for some fix $k\geq 1$.
From Lemma \ref{lem:cf} part (b) we have
\begin{align*}
\varphi_{\tu_k}(\tX_k)
% &=\mathbbm{1}_{\tX_k(0)\geq u(0)}+N_u(\tX_k)\\
&=\mathbbm{1}_{X(0)\geq u(0)}+\lim_{h\to 0}\int_0^1 \frac{1}{h}K\left(\frac{\tu_k(t)-\tX_k(t)}{h}\right)\big(\tX_k'(t)-\tu_k'(t)\big)\mathbbm{1}_{\tX_k'(t)-\tu_k'(t)>0}\,dt\quad a.s.,
\end{align*}
where we use that $\tX_k(0)=X(0)$ with probability 1.  Let $g_{\tX_k(t)\tX_k'(t)}$ denote the joint density of $(\tX_k(t),\tX_k'(t))$ and note that by Lemma \ref{lem:cf} part (b) $\smallint_0^1(1/h)K((\tu_k(t)-\tX_k'(t))/h)(\tX_k'(t)-\tu_k'(t))\mathbbm{1}_{\tX_k'(t)-\tu_k'(t)>0}\,dt\,<2^k$ (a.s.). Therefore, we can apply the dominated convergence theorem for an arbitrary but fix $k\geq 1$, such that
\begin{align*}
\E[\varphi_{\tu_k}(\tX_k)]=
&P(X(0)\geq u(0))+\\
&\lim_{h\to 0}\int_0^1 \frac{1}{h}\int_{-\infty}^\infty\int^\infty_{\tu_k'(t)}K\left(\frac{\tu_k(t)-x}{h}\right)\big(y-\tu_k'(t)\big)g_{\tX_k(t)\tX_k'(t)}(x,y)\ dy\ dx\ dt.
\end{align*}
Applying a change of variables $x = zh + \tu_k(t)$, the symmetry and compactness of $K$ yields
\begin{align*}
\E[\varphi_u(\tX_k)]=
&P(X(0)\geq u(0))+\\
&\lim_{h\to 0}\int_0^1 \int_{-1}^1\int^\infty_{\tu_k'(t)}K(z)\big(y-\tu_k'(t)\big)g_{\tX_k(t)\tX_k'(t)}(zh + \tu_k(t),y)\ dy\ dz\ dt.
\end{align*}
As $g_{\tX_k(t)\tX_k'(t)}$ consists of a finite mixture of Gaussian densities for each $t\in[0,1]$, $g_{\tX_k(t)\tX_k'(t)}(x,y)$ it bounded uniformly for all $x$ and $y$. Therefore, we can bring the limit inside the integral, which yields
\begin{align}
\E[\varphi_u(\tX_k)]=
P(X(0)\geq u(0))+
\int_0^1 \int^\infty_{\tu_k'(t)}\big(y-\tu_k'(t)\big)g_{\tX_k(t)\tX_k'(t)}(\tu_k(t),y)\ dy\ dt.\label{eq:grf1}
\end{align}

We now finally want to take the limit as $k\to\infty$. For the left hand side of \eqref{eq:grf1} we immediately have $\E[\varphi_{\tu_k}(\tX_k)]\to \E[\varphi_{u}(X)]$ as $k\to\infty$ by the monotone convergence theorem, since $N_u(\tX_k)\nearrow N_u(X)$ as $k\to\infty$. For the right hand side of \eqref{eq:grf1}, we can rewrite the integral as
\[
\int_0^1 \E[(\tX'_k(t) - \tu_k'(t))1_{\tX'_k(t) \geq \tu_k'(t)}| \tX_k(t) = \tu_k(t)] g_{\tX_k(t)}(\tu_k(t)) \ dt.
\]
To take pass the limit we need to be quite careful with our justification.  In particular, we need to pass the limit twice: under the first integral and then under the expectation.  We accomplish this in a series of steps.
\begin{enumerate}
\item %To pass under the integral, we need only find a uniform bound over $k$ and $t$, since the domain is compact.
For any $j-1 \leq t 2^k \leq j$ we can write
\[
\tX_k(t) = w_t X((j-1)/2^k) + (1-w_t) X(j/2^k),
\]
for nonrandom weights $w_t = j - 2^k t \in [0,1]$.  Since the covariance of $X$ is differentiable, it implies that $X$ is mean squared continuous, thus there exists a $k_0$ such that for any $k \geq k_0$ we have $\cov(X((j-1)/2^k, X(j/2^k)) \geq 0$, which implies that $\var(\tilde X_k(t)) \geq 1$.
Thus, for $k$ large, the density $g_{\tX_k}$ is uniformly bounded by $(2 \pi )^{-1/2}$.
\item
Next, since $(\tX_k(t),\tX'_k(t))$ are jointly mean zero Gaussian random variables (for each $t$), we have
\[
\tX_k'(t) = b_{k}(t) \tX_k(t) + \epsilon_{k}(t),
\]
where
$b_k(t)=\E[\tX_k(t)\tX_k'(t)]/\E[\tX_k(t)^2]\allowbreak=\allowbreak\cov(\tX_k(t),\tX_k'(t))/\var(\tX_k(t))$, and $\epsilon_k(t)$ and $\tX_k(t)$ are independent for each
$t$.

So then we have the bound
\begin{align*}
  &\E[(\tX'_k(t) - \tu_k'(t))1_{\tX'_k(t) \geq \tu_k'(t)}| \tX_k(t) = \tu_k(t)]\\
  & \leq \E[|\tX'_k(t) - \tu_k'(t)| \ | \tX_k(t) = \tu_k(t)] \\
  & = \E[|b_k (t)\tX_k(t) + \epsilon_k(t) - \tu_k'(t)| \ | \tX_k(t) = \tu_k(t)] \\
  %& = \E[|b_k (t) \tu_k(t) + \epsilon_k(t) - \tu_k'(t)|]
  &  \leq |b_k (t) \tu_k(t)  - \tu_k'(t)| + \E[|\epsilon_k(t)|].
\end{align*}

By the independence of $\epsilon_{k}(t)$ and $\tX_k(t)$,
\begin{align*}
  \E[\tX_k'(t)^2] = b_{k}(t)^2\E[\tX_k(t)^2] + \E[\epsilon_{k}(t)^2]\quad\Rightarrow\quad \E[\epsilon_{k}(t)^2]\leq \E[\tX_k'(t)^2].
\end{align*}

Jensen's inequality
implies that
\[
\E|\epsilon_k(t)| \leq \E[\tX_k'(t)^2]^{1/2}.% = \tau(t) \leq \sup_{0 \leq t \leq 1} \tau(t).
\]
By definition we have that $\tilde X'(t) = 2^k [X(j/2^k) - X((j-1)/2^k)]$. By
the Mean value theorem we have that $\tX_k'(t) = X'(\xi)$ for some $\xi \in
[(j-1)/2^k, j/2^k]$.  Therefore $\tX'(t)^2 \leq \left(\sup_{0\leq s \leq 1}
  X'(s) \right)^2$, which has a finite expectation by Theorem 2.9 of
\cite{AW2009_book}.

By construction $\tu_k$ and $\tu_k'$ are uniformly bounded by sup norms of $u$ and $u'$ respectively.  From arguments from step 1, for $k$ large, $\var(\tX_k(t)) \geq 1$ and so
\[
b_k(t)^2 \leq  \E(\tX_k'(t)^2)
\leq  \E\left( \left(\sup_{0 \leq s \leq 1}X'(s)\right)^2\right)  < \infty.
\]
Thus, we have a uniform bound over $t$ for all $k$ large and we can justify passing the limit as $k \to \infty$ under the integral.
\item Finally, with the limit under the integral we need only that the integrand converges for all or almost all $t$.  Clearly the density term converges
\[
g_{\tX_k(t)}(\tu_k(t))
\to \frac{1}{\sqrt{2 \pi}} \exp \left\{ - \frac{u(t)^2}{2} \right\},
\]
since $\var(\tX_k(t)) \to \var(X(t)) = 1$ and $\tu_k(t) \to u(t)$.  For the conditional expected value term, using the same arguments as before, we have that
\begin{align*}
&\E[(\tX'_k(t) - \tu_k'(t))1_{\tX'_k(t) \geq \tu_k'(t)}| \tX_k(t) = \tu_k(t)] \\
& = \E[ (b_k(t) \tu_k(t) + \epsilon_k(t) - \tu_k'(t)) 1_{b_k(t) \tu_k(t) + \epsilon_k(t) \geq \tu_k'(t)}].
\end{align*}
Clearly we have
\[
(b_k(t) \tu_k(t) + \epsilon_k(t) - \tu_k'(t)) 1_{b_k(t) \tu_k(t) + \epsilon_k(t) \geq \tu_k'(t)}
\leq |b_k(t) \tu_k(t) + \epsilon_k(t) - \tu_k'(t)|.
\]
\begin{comment}
As we already showed, $b_k(t)$, $\tu_k(t)$, and $\tu_k'(t)$ are uniformly bounded for all $k$ large (and they are deterministic).  Turning to $\epsilon_k(t)$ we have
\[
|\epsilon_k(t)|
= |\tX_k'(t) - b_k(t) \tX_k(t)|
\leq \|X'\|_1  + \|X\|_1  (\E\|X'\|_1^2)^{1/2},
\]
which has finite expectation.
\end{comment}
In the previous step we already showed that $b_k(t)$, $\tu_k(t)$, $\tu_k'(t)$, and $\E|\epsilon_k(t)|$ are uniformly bounded across $k$ and $t$.
Thus, by the DCT we can pass the limit as $k \to \infty$ to obtain
\begin{align*}
& \E[ (b_k(t) \tu_k(t) + \epsilon_k(t) - \tu_k'(t)) 1_{b_k(t) \tu_k(t) + \epsilon_k(t) \geq \tu_k'(t)}] \\
& \to \E[ (b(t) u(t) + \epsilon(t) - u'(t)) 1_{b(t) u(t) + \epsilon(t) \geq u'(t)}] \\
& = \E[(X'(t) - u'(t))1_{X'(t) \geq u'(t)} | X(t) = u(t)],
\end{align*}
as desired.
\end{enumerate}

Unraveling the conditional expectation, we have shown that
\begin{align*}
\E[\varphi_u(X)]=
P(X(0)\geq u(0))+
\int_0^1 \int^\infty_{u'(t)}\big(y-u'(t)\big)g_{X(t)X'(t)}(u(t),y)\,dydt.
\end{align*}
Since we assume that $X(t)$ has constant variance, it follows that, for each $t$, $X(t)$ and $X'(t)$ are uncorrelated and thus independent.  So we can express
\begin{align*}
\E[\varphi_u(X)]
&=P(X(0) \geq u(0)) + \\
& \int_0^1  \left[
\frac{1}{\sqrt{2 \pi }} \exp\left\{-\frac{u(t)^2}{2}\right\}
\int_{u'(t)}^\infty \big(y-u'(t)\big) \frac{1}{\sqrt{2 \pi \tau(t)^2}} \exp\left\{-\frac{y^2}{2\tau(t)^2}\right\}\ dy \right]\ dt.
\end{align*}
Turning to the second component, basic calculus implies that
\begin{align*}
&\int_{u'(t)}^\infty (y-u'(t)) \frac{1}{\sqrt{2 \pi \tau(t)^2}} \exp\left\{-\frac{y^2}{2\tau(t)^2}\right\}\ dy =
\frac{\tau(t)}{\sqrt{2 \pi}}\exp\left\{-\frac{u'(t)^2}{2 \tau(t)^2}\right\} -
u'(t) \, \Phi\left(\frac{-u'(t)}{\tau(t)}\right), %>0
\end{align*}
where $\Phi$ is the standard normal cdf.
%%%
% Wolfrahm Alpha input ( t==tau(t) and u==u'(t) ):
% integrate y * (1/(sqrt(2* pi* t^2))) * exp(-y^2/(2*t^2)) dy from u to infinity -
% integrate u * (1/(sqrt(2* pi* t^2))) * exp(-y^2/(2*t^2)) dy from u to infinity
%%%
% Wolfrahm Alpha output:
% (sqrt(t^2) e^(-u^2/(2 t^2)))/sqrt(2 π) -
% (u (1/sqrt(1/t^2) - t erf(u/(sqrt(2) t))))/(2 sqrt(t^2)) for Re(t^2)>0
%%%
% Simplifying:
% (u (1/sqrt(1/t^2) - t erf(u/(sqrt(2) t))))/(2 sqrt(t^2)) ==
% (u (1 - erf( u/(sqrt(2) t))) ) / 2                       ==
%  u (1 + erf(-(u/t)/sqrt(2)))  * (1/2)                    ==
%  u Phi(-(u/t))                                           ==
% -u Phi( (u/t))
%%%
This yields the final result
\begin{align*}
\E[\varphi_u(X)]
&= P(X(0) \geq u(0)) +\\
&\int_0^1 \frac{\tau(t)}{2 \pi }
\exp\left\{- \frac{u(t)^2}{2 }  - \frac{u'(t)^2}{2\tau(t)^2} \right\} \ dt -
\int_0^1 \frac{u'(t)}{\sqrt{2 \pi }}
\exp\left\{- \frac{u(t)^2}{2 }\right\} \Phi\left(\frac{-u'(t)}{\tau(t)}\right) \ dt. \qed
\end{align*}

This result of Lemma \ref{lem:grf} equals the result in Corollary \ref{CO:GAUSS}
a) for $t_0=0$.  The general result for any $t_0\in[0,1]$ follows now directly
from observing that (a), when moving from $t_0$ down to $0$, up-crossings as $t$
decreases are equivalent to down-crossings as $t$ increases, and (b) that the
expected number down-crossings about $u(t)$ is equal to the expected number of
up-crossings about $-u(t)$.  Corollary \ref{CO:GAUSS} b) follows directly from
part a), since $u'\equiv 0$ for constant critical values $u(t)\equiv u$,
$\tau(t)\geq 0$ for all $t\in[0,1]$ and $X(t_0)\overset{D}{=}X(0)$ for all
$t_0\in[0,1]$

Lemma \ref{lem:grf} allows us now to proof Theorem \ref{TH:MAIN}.

%%%%%%%%%%%%%%%%%%%%%%%%%%%%%
\subsection{Proof of Theorem \ref{TH:MAIN}}
%%%%%%%%%%%%%%%%%%%%%%%%%%%%%

In the following we extend the above result of Lemma \ref{lem:grf} to elliptical distributions.  Recall that every elliptical distribution can be expressed as a scalar mixture of a Gaussian and, therefore, our previous calculations were based on conditioning on the scalar mixing coefficient $V$.  %This means that for any elliptical process, $X(t)$, there exists a scalar random variable, $\lambda > 0$, and a GP, $Z(t)$, which are independent and  $X(t) = \lambda^{-1/2} Z(t)$; the reasoning behind this particular parametrization will become clear below.  Therefore handling the case where $X(t)$ is elliptical is equivalent to replacing $\sigma \to \sigma/\lambda^{1/2}$ and $\tau(t) \to \tau(t)/ \lambda^{1/2}$, and then taking the expectation over $\lambda$.
Our final expressions simplify if we use the alternative parametrization $V = \mcV^{-1/2}$ and work with $\mcV$ instead.  Our previous expectation can then be understood as being conditioned on $\mcV$.  In particular, if $\dot u(t) = \mcV^{1/2} u(t)$, then by Lemma \ref{lem:grf}
\begin{align}
\E[\varphi_{u}(X) | \mcV]
= \E[\varphi_{\dot u}(X/\mcV^{1/2}) | \mcV]
 = & P(X(0) \geq u(0)| \mcV) \notag\\
+ &\int_0^1 \frac{\tau(t)}{2 \pi }
    \exp\left\{- \mcV \left[ \frac{u(t)^2}{2}  + \frac{u'(t)^2}{2\tau(t)^2} \right]\right\} \ dt \notag\\
-&\int_0^1 \frac{u'(t) \mcV^{1/2}}{\sqrt{2 \pi}}
   \exp\left\{- \frac{ \mcV u(t)^2}{2 }\right\} \Phi\left(\frac{-u'(t) \mcV^{1/2}}{\tau(t)}\right) \ dt.\label{eq:ellip}
\end{align}
The expected value of the first component is clearly $P(X(0) \geq u(0))$.  For the second component, we can take the expected value and use Fubini's theorem. If $F_\mcV$ is the cdf of $\mcV$ then this yields
\[
\int_0^1 \frac{\tau(t)}{2 \pi}
\int_0^\infty \exp\left\{- x \left[\frac{u(t)^2}{2 }  + \frac{u'(t)^2}{2  \tau(t)^2} \right]  \right\} \ dF_\mcV(x) \ dx\ dt.
\]
Now notice that this is simply a moment generating function (mgf) calculation.  Furthermore, it is always finite since $\mcV$ is strictly positive and the argument is strictly negative.  Thus, if $M_\mcV(x)$ is the mgf of $\mcV$ then we have that the above simplifies to
\[
\int_0^1 \frac{\tau(t)}{2 \pi} M_\mcV \left(- \left[\frac{u(t)^2}{2 }  + \frac{u'(t)^2}{2  \tau(t)^2} \right]  \right) \ dt.
\]
Turning to the third component, we have
\begin{align*}
&-\int_0^1 \frac{u'(t) \mcV^{1/2}}{\sqrt{2 \pi }}
\exp\left\{- \frac{\mcV u(t)^2}{2 }\right\} \Phi\left(\frac{-u'(t) \mcV^{1/2}}{\tau(t)}\right) \ dt \\
=&-\int_0^1 \frac{u'(t) \mcV^{1/2}}{\sqrt{2 \pi }}
\exp\left\{- \frac{\mcV u(t)^2}{2 }\right\} \int_{\mcV^{1/2}u'(t)/\tau(t)}^\infty \frac{1}{ \sqrt{2 \pi}} e^{-x^2/2} \ dx \ dt.
\end{align*}
Let $y = \tau(t) x/\mcV^{1/2} - u'(t)$, then the above becomes
\begin{align*}
& \int_0^1 \frac{u'(t)\mcV^{1/2}}{\sqrt{2 \pi }}
\exp\left\{- \frac{\mcV u(t)^2}{2 }\right\} \int_{0}^\infty \frac{\mcV^{1/2}}{\sqrt{2 \pi \tau(t)^2}} e^{-\mcV (y+u'(t))^2/2 \tau(t)^2} \ dy \ dt \\
& = \int_0^1 \int_0^\infty \frac{u'(t)\mcV }{2 \pi  \tau(t)}
\exp \left\{ - \mcV \left[\frac{u(t)^2}{2 } + \frac{(y+u'(t))^2}{2 \tau(t)^2} \right]\right\} \ dy \ dt.
\end{align*}
Now we can take the expected value and use Fubini's theorem to get
\begin{align*}
-\int_0^1 \int_0^\infty \int_0^\infty \frac{ u'(t) x}{2 \pi \tau(t)}
\exp \left\{ - x \left[\frac{u(t)^2}{2 } + \frac{(y+u'(t))^2}{2 \tau(t)^2} \right]  \right\} \ dF_\mcV(x) \ dy \ dt.
\end{align*}
However as before, the inner integral can be expressed using the mgf of $\mcV$, 
though now with the derivative 
$M_{\mathcal{V}}'(\cdot)=\E(\mathcal{V}\exp(\cdot\mathcal{V}))$, resulting in
\begin{align*}
-\int_0^1 \int_0^\infty  \frac{u'(t)}{2 \pi  \tau(t)}
M_\mcV'\left( - \left[\frac{u(t)^2}{2 } + \frac{(y+u'(t))^2}{2 \tau(t)^2} \right]  \right)  \ dy \ dt.
\end{align*}
So we have the final form for elliptical distributions given as
\begin{align*}
\E[\varphi_{u}(X)]= P(X(0) \geq u(0))
&+\int_{0}^1 \frac{\tau(t)}{2 \pi } M_\mcV \left(- \left[\frac{u(t)^2}{2 }  + \frac{u'(t)^2}{2  \tau(t)^2} \right]  \right) \ dt\\
&- \int_0^1 \int_0^\infty  \frac{u'(t)}{2 \pi \tau(t)}
M_\mcV'\left( - \left[\frac{u(t)^2}{2 } + \frac{(y+u'(t))^2}{2 \tau(t)^2} \right]  \right)  \ dy \ dt
\end{align*}
which equals the result in Theorem \ref{TH:MAIN} for $t_0=0$.  The general result for any $t_0 \in [0,1]$ follows now directly from observing that (a), when moving from $t_0$ down to $0$, up-crossings as $t$ decreases are equivalent to down-crossings as $t$ increases, and (b) that the expected number down-crossings about $u(t)$ is equal to the expected number of up-crossings about $-u(t)$.  $\qed$

%%%%%%%%%%%%%%%%%%%%%%%%%%%%%
\subsection{Proof of Corollary \ref{CO:STUDt}}
%%%%%%%%%%%%%%%%%%%%%%%%%%%%%

Maybe the most useful for our purposes is the case where $X(t)$ comes from a $t$-process.  We note that pointwise standardization does not actually produce a $t$-process, but can still be used to help account for the added uncertainty of estimating the covariance.  In this case, the distribution of $\mcV$ is actually a chi-squared divided by its degrees of freedom.  So, if we write $\mcV= \chi^2_\nu/\nu$, then we have
\[
M_\mcV(x) = \left( 1 - \frac{2x}{\nu} \right)^{-\nu/2}
\qquad \text{and} \qquad M_\mcV'(x) =  \left( 1 - \frac{2x}{\nu} \right)^{-\nu/2 - 1}.
\]
Therefore, our expression becomes
\begin{align*}
\E[\varphi_{u}(X)]= F_t(-u(0);\nu)
&+\int_{0}^1 \frac{\tau(t)}{2 \pi }
\left(
1 +  \frac{u(t)^2}{ \nu }  + \frac{u'(t)^2}{  \nu \tau(t)^2}
\right)^{-\nu/2} \ dt\\
&- \int_0^1 \int_0^\infty  \frac{u'(t)}{2 \pi \tau(t)}
\left( 1+  \frac{u(t)^2}{\nu } + \frac{(y+u'(t))^2}{\nu \tau(t)^2}  \right)^{-\nu/2 -1}  \ dy \ dt.
\end{align*}
The third term can be further simplified by using the cdf of a $t$-distribution.  In particular, let $\nu' = \nu + 1$ and $a(t)^2 = \nu \tau(t)^2 (1+ u(t)^2/\nu )/\nu'$, then we have
\begin{align*}
 & \int_0^\infty
\left( 1+  \frac{u(t)^2}{\nu } + \frac{(y+u'(t))^2}{\nu \tau(t)^2}  \right)^{-\nu/2 -1}  \ dy \\
& = \left(1+ \frac{u(t)^2}{\nu  }\right)^{-\nu/2 -1} \int_0^\infty \left( 1 + \frac{(y + u'(t))^2}{\nu ' a(t)^2} \right)^{-\frac{\nu'+1}{2}} \ dy.
\end{align*}
Now use a change of variables $x = y / a(t) + u'(t)/a(t)$ to obtain
\begin{align*}
&   \left(1+ \frac{u(t)^2}{\nu }\right)^{-\nu/2 -1} a(t)\int_{u'(t) / a(t)}^\infty \left( 1 + \frac{x^2}{\nu ' } \right)^{-\frac{\nu'+1}{2}} \  dx
\\
& = \left(1+ \frac{u(t)^2}{\nu }\right)^{-\nu/2 -1} B(1/2,\nu'/2)\sqrt{\nu' }a(t) F_t(-u'(t)/a(t);\nu') \\
& = \left(1+ \frac{u(t)^2}{\nu }\right)^{-\nu/2 -1} B(1/2,(\nu+1)/2)\sqrt{\nu+1 }a(t) F_t(-u'(t)/a(t);\nu+1)\\
& = \left(1+ \frac{u(t)^2}{\nu }\right)^{-(\nu+1)/2} \frac{\Gamma((\nu+1)/2) \sqrt{(\nu+1) \pi } a(t)}{\Gamma((\nu+2)/2)} F_t(-u'(t)/a(t);\nu').
\end{align*}
where $B$ denotes the beta function, $\Gamma$ the gamma function, and $F_t(\cdot; \nu')$ denotes the distribution function of a $t$-distribution with $\nu'$ degrees of freedom.  This leads to the result in Corollary \ref{CO:STUDt} for $t_0=0$.  The general result for any $t_0 \in [0,1]$ follows now directly from observing that (a), when moving from $t_0$ down to $0$, up-crossings as $t$ decreases are equivalent to down-crossings as $t$ increases, and (b) that the expected number down-crossings about $u(t)$ is equal to the expected number of up-crossings about $-u(t)$.  $\qed$

\subsection{Proof of Proposition \ref{PRO:FAIR_BAND}}\label{APP:Prop}
%%%%%%%%%%%%%%%%%%%%%%%%%%%%%%%%%%%%%%%%%%%%%%%%%%%%%%%%%%%%%%%%%

The proof of Proposition \ref{PRO:FAIR_BAND} follows directly from Theorems
\ref{TH:MAIN}, \ref{TH:EST}, and \ref{LE:FAIRNESS}, and the classic confidence
interval derivations from inverting pivot statistics.\footnote{See, for
  instance, Ch.~5 in Panaretos, V. M. (2016). Statistics for Mathematicians: A Rigorous First Course (1. ed.).
  Compact Textbook in Mathematics. Birkhäuser/Springer. Springer.}

\bigskip

% %%%%%%%%%%%%%%%%%%%%%%%%%%%%%%%%%%%%%%%%%%%%%%%%%%
% \noindent\textbf{Proof of Propositions \ref{PRO:FAIR_BAND}.}\\
% %%%%%%%%%%%%%%%%%%%%%%%%%%%%%%%%%%%%%%%%%%%%%%%%%%
% Proposition \ref{PRO:FAIR_BAND} is a simple variation of Proposition \ref{PRO:FAIR} and follows from straight forward calculus exercises equivalent to those leading to \eqref{EQ:VALID_SCB_1} and \eqref{EQ:VALID_SCB_2}. $\qed$

%%%%%%%%%%%%%%%%%%%%%%%%%%%%%%%%%%%%%%%%%%%%%%%%
\subsection{Proof of Theorem \ref{TH:EST}}
%%%%%%%%%%%%%%%%%%%%%%%%%%%%%%%%%%%%%%%%%%%%%%%%

The arguments are the same for any elliptical distribution, so we assume that $X$ is a Gaussian process.  We recall two classic properties, one from analysis and the other from probability theory.  First, if $f(u,\tau)$ is continuous as a mapping from $C^1 \times C \to \mbR$, and $\mcU$ is assumed to be compact, then the collection of functions $\{f_u(\tau):= f(u, \tau): u \in \mcU\}$ are equicontinuous.  Second, recall that $\tau_n \to \tau$ in probability if and only if every subsequence $\tau_{n(m)}$ has a further subsequence $\tau_{n(m_l)}$ that converges almost surely.\footnote{See, for instance, Theorem 2.3.2.~in: Durrett, R. (2019). \textit{Probability: Theory and Examples}. Cambridge University Press.}  With a slight abuse of notation we will simply write $n=n(m_l)$.  So, we will assume that we are working with such a subsequence $\tau_n \to \tau$ almost surely.  We will then show that this implies that the corresponding properties (1.-3.)~of the theorem hold almost surely.  By the same logic, this implies that the main claims of the theorem hold in probability, since we will have constructed a subsequence of any subsequence where almost sure convergence holds.
\begin{enumerate}[wide, labelwidth=!, labelindent=0pt, itemsep=-1ex]
\item The first property follows from the fact that the constant functions are included, and thus a constant band with the desired coverage can always be constructed.  That the sets are closed follows from the continuity of $f(u,\tau)$.  
\item Since the $f_u$ are equicontinuous, for any $\epsilon > 0$ there exists $\delta > 0$ such that $|f(u,\tau') - f(u,\tau)| \leq \epsilon$ whenever $\|\tau - \tau'\| \leq \delta$ and for all $u\in \mcU$.  Since $\tilde \tau_n \to \tilde \tau$ almost surely, we can, with probability 1, find an $N$ large enough such that $n \geq N$ implies $\| \tilde \tau_n - \tilde \tau\| \leq \delta$.  It then follows that if $n \geq N$ then
\[
|\alpha - f(u_n, \tilde \tau)| = | f(u_n, \tilde \tau_n) - f(u_n, \tilde \tau)| \leq \epsilon,
\]
%\[
%|\alpha - f(u_n, \tau_n)| = | {\color{red}f(u,\tau)} - f(u_n, \tau_n)| \leq \epsilon,
%\]
which proves the claim.

\item Recall that the Hausdorff distance is given by
\[
d(S_n,S) = \max\left\{ \sup_{u_n \in S_n} \inf_{u \in S} \|u_n - u\|,
\sup_{u \in S} \inf_{u_n \in S_n} \|u_n - u\|
 \right\}
 = \max\left\{ d_{1,n},d_{2,n} \right\}.
\]
In the following, we examine the terms $d_{1,n}$ and $d_{2,n}$ separately.  In the first case, only compactness of $\mcU$ is needed, while in the second case, both the convexity and the inclusion of the constant functions will be crucial.
\begin{enumerate}
\item We consider a proof by contradiction, so assume
\[
 d_{1,n} =  \sup_{u_n \in S_n} \inf_{u \in S} \|u_n - u\| {\not\to}0.
\]
This implies that, with probability 1, there exists an $\epsilon > 0$ and infinitely many $n$ such that $d_{1,n} \geq \epsilon. $  Take a further subsequence such that $d_{1,n} \geq \epsilon$ for all $n$.  Since the sets are compact, this implies that there exists $u_n \in S_n$ such that $\|u_n - u\| \geq \epsilon$, for all $n$ and all $u \in S$.  Now take a convergent subsequence of $u_n$ such that $u_n \to w$.  By property 2.~we have
\[
f(u_n,\tilde \tau) \to \alpha \Longrightarrow f(w,\tilde \tau) = \alpha,
\]
which implies that $w \in S$, which is a contradiction since $\|u - w\| \geq \epsilon $ for all $u \in S$.  Thus, $d_{1,n} \to 0.$

\item We again use a proof by contradiction.    Suppose that
\[
 d_{2,n} =  \sup_{u \in S} \inf_{u_n \in S_n} \|u_n - u\| {\not\to}0.
\]
As the proof here is more complex, we first provide a high level overview.  First, we will show that if $d_{2,n}{\not \to} 0$, then we can construct a band, $w \in S$ that is isolated from all but a finite number of the other $S_n$.  We then combine the continuity of $f$ and the convexity of $\mcU$ to construct a band in $S_n$ with proper coverage, but which must be close to $w$ contradicting this isolation.

Turning to the proof, with probability 1, there exists $\epsilon>0$ such that $d_{2,n} > 2\epsilon$ for infinitely many $n$, take an appropriate subsequence such that it holds for all $n$.  Again, since the sets are compact we can find $w_n \in S$, such that $\|w_n - u_n\| > 2\epsilon$ for all $u_n \in S_n$.  Now take a further convergent subsequence such that $w_n \to w$.  Then we can find $N$ such that for all $n \geq N$ we have $\|w_n - w\| \leq \epsilon$, which implies that $w \not\in S_n$ for $n \geq N$ and even more
\[
\|u_n - w\|
=\| u_n -w_n + w_n - w\|
\geq \|u_n - w_n\| - \|w_n - w\|
\geq \epsilon
\]
for all $u_n \in S_n$.  This implies that $w$ is isolated from the $S_n$, that is, there is a ball of radius $\epsilon$ about $w$ that does not intersect with any $S_n$.
%$B_c(u',\epsilon) \cap S_n = \emptyset$, where $B_c$ denotes the close ball.

We now have to utilize the form for $f$ more than previously.  In particular, since the constant functions are in $\mcU$ and $\mcU$ is convex, this means that $w+1c \in \mcU$ ($1$ is the constant function) if $c \in \mbR$ is small (if $w$ is on the boundary then enlarge $\mcU$ slightly) and so for any $\tilde \tau$, $f(w + 1 c,\tilde \tau)$ is monotonically decreasing with $c$ (since the band raises uniformly).  If $\phi$ is the density of $X(0)$ then
\begin{align*}
| c| \phi(w(0)+|c|) & \leq |P(X(0) \geq w(0) + c) - P(X(0) \geq w(0))|
 \leq | c| \phi(w(0)-|c|),
\end{align*}
which means that we can raise/lower the band by $c$ and produce a change in $\alpha$ at least as large in magnitude as the corresponding quantity above (note the integrals in $f(w,\tilde \tau)$ will increase/decrease in the same direction). %We will consider $|c| < \epsilon/2$ so that $u'+1c \in B_c(u',\epsilon/2) \subset B_c(u',\epsilon)$.

Coming back to the problem at hand, we have that $f(w, \tilde \tau) = \alpha$,  which implies that $f(w,\tilde \tau_n) \to \alpha$.  So take $N$ such that for all $n\geq N$ we have
\[
|f(w,\tilde \tau_n) - \alpha| < \delta:= (\epsilon/2) \phi(u(0)+\epsilon/2).
\]
However, now by construction we have that there exists $u_1,u_2 \in \mcU$ with $|u_i - w| \leq \epsilon$ such that
\[
f(u_1,\tau_n) < \alpha
\qquad \text{and} \qquad f(u_2, \tau_n) > \alpha,
\]
where $u_i = w + c_i 1$, for some $|c_i| \leq \epsilon/2$.  However, $B_c(u',\epsilon)\cap \mcU$ is convex and $f(u_1 t + (1-t)u_2, \tau_n)$ is continuous in $t \in [0,1]$ thus there must exist a $u \in B_c(w,\epsilon) \cap \mcU$ such that $f(u,\tau_n) = \alpha$ implying $u\in S_n$, which is a contradiction since no element of $S_n$ is within $\epsilon$ of $w$.  Thus $d_{2,n} \to 0$.
\end{enumerate}

% \item  Both properties follow immediately from the continuity of $\Gamma$ and property 3.

\end{enumerate}

\begin{comment}
\subsection{More General Mixtures}
Suppose that
\[
X(t) = \sum_{i=1}^\infty c_i V_i Z_i(t),
\]
where $Z_i$ are GP$(0,C_i)$ with $\E\|Z_i\| \leq M < \infty$, $V_i$ are iid strictly positive random variables, and $c_i$ are positive summable weights.
\[
\sup_{0\leq t \leq 1}|X(t)|
\leq \sum_{i=1}^\infty c_i V_i \sup_{0\leq t \leq 1} |Z_i(t)|.
\]

SIDE NOTE.  Let $X_i$ be iid centered random variables and $c_i$ a sequence of numbers then the characteristic function of the weighted sum is given by
\[
\E[e^{i t \sum_{j=1}^m c_j X_j}]
= \exp\left\{\sum_{j=1}^m \log(\phi(t c_j))\right\}.
\]
However, $c_j \to 0$ and $\log(\phi(0)) = 0$, so we have for $j$ large
\[
\log(\phi(tc_j)) \leq C |t| c_j.
\]
Which means that the characteristic function converges, and thus the infinite process $\sum_{j=1}^\infty c_j X_j$ exists and is well defined.

So, as long as there exists iid $M_i$ which stochastically dominate the $\|Z_i\|$, then the above implies $X(t)$ exists as long as the $c_i$ are summable.  So, conditioned on the $V_i$, it follows that the covariance function of $X(t)$ is given by $\sum c_i^2 V_i^2 C_i(t,s).$

So, conditioned on the $V_i$, $X(t)$ is a Gaussian process, however it does not possess constant variance.  So perform the substitution
\[
\tu(t) = \mcV(t) u(t),
\]
where $\mcV(t)=\var(X(t) | \bV)^{-1}.$  Then our Gaussian formula becomes
\end{comment}

\subsection[]{Proof of Lemma \ref{LE:FAIRNESS} (Fairness of
 $ \boldsymbol{u_{\alpha/2}^\star}$)}
In equations \eqref{EQ:FAIRNESS_ALGO_1}-\eqref{EQ:FAIRNESS_ALGO_3}, Algorithm \ref{ALGO:FAIRNESS}
applies the generalized Kac-Rice formula
\eqref{EQ:GRF_Ell} interval-wise for $\{X(t):t\in[a_{j-1},a_j]\}$ by substituting $0$, $1$, and $t_0$ in \eqref{EQ:GRF_Ell} by $a_{j-1}$, $a_j$, and $a_{j}$ for odd
$j=1,3,\dots$ and by $a_{j-1}$, $a_j$, $a_{j-1}$ for even $j=2,4,\dots$. The
interval-wise critical value functions
$\{u^\star_{\alpha/2}(t):t\in[a_{j-1},a_j]\}$ are determined
by setting the interval-wise generalized Kac-Rice formulas for
$\{X(t):t\in[a_{j-1},a_j]\}$ equal to $(\alpha/2)(a_j-a_{j-1})$. Thus, by
inequality \eqref{EQ:EulerIneq_General} with corresponding substitutions
for $0$, $1$, and $t_0$
\begin{align*}
  P(\exists t\in[a_{j-1},a_j]:X(t)\geq u^\star_{\alpha/2}(t))\leq
  \frac \alpha 2 (a_j-a_{j-1})\quad\text{for each}\quad j=1,\dots,p.
\end{align*}
The result of Lemma \ref{LE:FAIRNESS} follows now by applying Boole's
inequality
\begin{align*}
  P\Big(\exists t\in\bigcup_{j\in\mathcal{J}}[a_{j-1},a_j]:X(t)\geq u^\star_{\alpha/2}(t)\Big)&\leq
                                                                                                           \sum_{j\in\mathcal{J}}P(\exists t\in[a_{j-1},a_j]:X(t)\geq u^\star_{\alpha/2}(t))\\
  &\leq
    \sum_{j\in\mathcal{J}}\frac \alpha 2 (a_j-a_{j-1})
\end{align*}
for any $\mathcal{J}\subseteq\{1,2,\dots,p\}.\hfill \qed$

\bigskip

% To guarantee that $P(\exists t\in[a_{j-1},a_j]:X(t)\geq
% u^\star_{\alpha/2}(t))\leq(\alpha/2)(a_j-a_{j-1})$ holds for every
% $j=1,\dots,p$, we apply the generalized Kac-Rice formula
% \eqref{EQ:GRF_Ell} separately for each $\{X(t):t\in[a_{j-1},a_j]\}$ by substituting $0$, $1$, and $t_0$ in \eqref{EQ:GRF_Ell} by $a_{j-1}$, $a_j$, and $a_{j}$ for odd
% $j=1,3,\dots$ and by $a_{j-1}$, $a_j$, $a_{j-1}$ for even $j=2,4,\dots$.

\subsection[]{Additional Derivations for the Discussion of the Price of
  Fairness}\label{APP:DISCUSSION}
To discuss the effect of the number of intervals $p$, let us sum up the single
components produced by Algorithm \ref{ALGO:FAIRNESS}, which yields
\begin{align*}
  &\sum_{j=1}^p
    \bigg[\mathbbm{1}_{(j \text{ is odd})}\left(P\big(X(a_j) \geq u^\star_{\alpha/2}(a_j)\big)\right)
    +\int_{a_{j-1}}^{a_{j}} \frac{\tau(t)}{2 \pi } M_\mcV \left(-\frac 1 2 \left[(u^\star_{\alpha/2}(t))^2  + \frac{u^{\star '}_{\alpha/2}(t)^2}{ \tau(t)^2} \right]  \right) \ dt \\
  & + \mathbbm{1}_{(j \text{ is odd})}\left(\int_{a_{j-1}}^{a_j} \int_0^\infty  \frac{u^{\star '}_{\alpha/2}(t)}{2 \pi\tau(t)}
    M_\mcV'\left( - \frac 1 2 \left[ (u^\star_{\alpha/2}(t))^2+ \frac{(y-u^{\star '}_{\alpha/2}(t))^2}{\tau(t)^2} \right]  \right)  \ dy \ dt\right)\\
  & - \mathbbm{1}_{(j \text{ is even})}\left( \int_{a_{j-1}}^{a_{j}} \int_0^\infty  \frac{u^{\star '}_{\alpha/2}(t)}{2 \pi\tau(t)}
    M_\mcV'\left( - \frac 1 2 \left[ (u^\star_{\alpha/2}(t))^2+ \frac{(y+u^{\star '}_{\alpha/2}(t))^2}{\tau(t)^2} \right]  \right)  \ dy \ dt\right)\bigg]=\frac{\alpha}{2}.
\end{align*}
Over intervals $[a_{j-1},a_j]$, where $j$ is odd/even, we compute the mean
number of down-/up-crossings about $u$. However, since the distribution of $X$ is symmetric, the mean of up-crossings about $u$ equals the mean of down-crossings about $-u$. This allows us to write,
\begin{align*}
  &\sum_{j=1}^p
  \mathbbm{1}_{(j \text{ is odd})}\left(P\big(X(a_j) \geq u^\star_{\alpha/2}(a_j)\big)\right)
      +\int_{0}^{1} \frac{\tau(t)}{2 \pi } M_\mcV \left(-\frac 1 2 \left[(u^\star_{\alpha/2}(t))^2  + \frac{u^{\star '}_{\alpha/2}(t)^2}{ \tau(t)^2} \right]  \right) \ dt \\
  & + \left(\int_{0}^{a_1} \int_0^\infty  \frac{u^{\star '}_{\alpha/2}(t)}{2 \pi\tau(t)}
        M_\mcV'\left( - \frac 1 2 \left[ (u^\star_{\alpha/2}(t))^2+ \frac{(y-u^{\star '}_{\alpha/2}(t))^2}{\tau(t)^2} \right]  \right)  \ dy \ dt\right)\\
  & - \left( \int_{a_{1}}^{1} \int_0^\infty  \frac{u^{\star '}_{\alpha/2}(t)}{2 \pi\tau(t)}
        M_\mcV'\left( - \frac 1 2 \left[ (u^\star_{\alpha/2}(t))^2+ \frac{(y-u^{\star '}_{\alpha/2}(t))^2}{\tau(t)^2} \right]  \right)  \ dy \ dt\right)=\frac{\alpha}{2}\\
      %%%%%%%%%%%%%%%%%%%%%%%
  \Leftrightarrow &\E[\varphi_{u^\star_{\alpha/2},X}(a_1)] + \sum_{\substack{3\leq j\leq p\\j: \textrm{odd}}}\left(P\big(X(a_j) \geq u^\star_{\alpha/2}(a_j)\big)\right) = \frac{\alpha}{2}
\end{align*}

\clearpage

%%%%%%%%%%%%%%%%%%%%%%%%%%%%%%%%%%%%%%%%%%%%
\section{Additional Simulation Results and Figures}\label{APP:ADD_MATERIAL}
%%%%%%%%%%%%%%%%%%%%%%%%%%%%%%%%%%%%%%%%%%%%
%This section contains the further simulation studies and supplementary figures.
% \subsection{Additional Simulation Results for Section
% \ref{SEC:SIMUL_FULLFDA}}\label{APP:SIMRES_1}

Table \ref{TAB:CompTimes} shows summary
statistics of the computations times for the FF bands and all comparison bands. The comparison study shown in Table \ref{TAB:CompTimes} was done using the
\textsf{R} package \texttt{microbenchmark} using 100 repeated
computations.\footnote{Mersmann, O. (2021). microbenchmark: Accurate Timing Functions. R package version
  1.4.9.} Computations were conduced using a laptop computer with AMD-Ryzen 7 pro 4750u processor.

\begin{table}[h!tb]
  \centering
  \caption{Computation times (100 repetitions) for $n=100$, Mean $\theta_0$ and Cov3.}\label{TAB:CompTimes}
  \begin{tabular}{l rrrrrr c}
    \toprule
                 &       &      &       &        &      &     &\% of \\
    Band         & Min.  & LQ   &  Mean & Median &  UQ  &  Max&MB$_t$ Median\\
    \midrule
    FF$_z^1$     & 0.025& 0.025& 0.028& 0.025& 0.028& 0.056&\phantom{10}2\\
    FF$_z^2$     & 0.027& 0.028& 0.032& 0.029& 0.032& 0.065&\phantom{10}2\\
    FF$_z^4$     & 0.032& 0.034& 0.037& 0.035& 0.037& 0.083&\phantom{10}3\\
    FF$_t^1$     & 0.026& 0.029& 0.033& 0.030& 0.032& 0.060&\phantom{10}2\\
    FF$_t^2$     & 0.030& 0.033& 0.038& 0.035& 0.038& 0.077&\phantom{10}3\\
    FF$_t^4$     & 0.039& 0.044& 0.050& 0.047& 0.051& 0.086&\phantom{10}4\\
    %\midrule
    GKF$_t$     & 0.020& 0.021& 0.027& 0.022& 0.024& 0.266&\phantom{10}2\\
    B$_{\text{EC}}$&0.030& 0.033& 0.037& 0.034& 0.038& 0.076&\phantom{10}3\\
    B$_{\text{S}}$ &0.137& 0.148& 0.163& 0.160& 0.169& 0.388&\phantom{1}13\\
    IWT         & 0.509& 0.559& 0.598& 0.589& 0.623& 0.850&\phantom{1}46\\
    MB$_t$      & 0.969& 1.220& 1.280& 1.276& 1.317& 1.571&          100\\
    \bottomrule
    \multicolumn{8}{l}{Unit: seconds (LQ: Lower Quantile; UQ: Upper Quantile)}\\
  \end{tabular}
\end{table}

%%%%%%%%%%%%%%
%\spacingset{1}
\begin{table}[h!]
\centering
\caption{Average confidence band widths and rankings (in parentheses)$^\ast$}
\label{TAB:SIM_2}
\begin{tabular}{l  ccc c ccc}
\toprule
&\multicolumn{3}{c}{$n=15$}&&\multicolumn{3}{c}{$n=100$}\\
  \cline{2-4}\cline{6-8}\\[-2.1ex]
  Band         & Cov1    & Cov2      & Cov3    && Cov1 & Cov2 & Cov3 \\
  \midrule
  FF$_t^4$     & 0.386 (6)& 0.447 (5)& 0.426 (4)&& 0.134 (6)& 0.152 (4)& 0.146 (4)\\
  FF$_t^2$     & 0.358 (5)& 0.437 (4)& 0.413 (2)&& 0.126 (4)& 0.150 (3)& 0.142 (2)\\
  FF$_t^1$     & 0.336 (2)& 0.432 (3)& 0.415 (3)&& 0.120 (2)& 0.148 (2)& 0.144 (3)\\
  B$_{\text{EC}}$& 0.324 (1)& 0.413 (1)& 0.396 (1)&& 0.131 (5)& 0.205 (6)& 0.183 (6)\\
  MB$_t$       & 0.353 (3)& 0.431 (2)& 0.426 (5)&& 0.120 (1)& 0.140 (1)& 0.139 (1)\\
  GKF$_t$      & 0.357 (4)& 0.486 (6)& 0.465 (6)&& 0.121 (3)& 0.154 (5)& 0.149 (5)\\
\bottomrule
  \multicolumn{8}{l}{$^\ast$Cannot be computed for the IWT procedure. B$_{\text{S}}$ is not shown since this}\\
  \multicolumn{8}{l}{band is too narrow and leads to invalid inference in small samples.}
\end{tabular}
\end{table}
%\spacingset{1.00001}
%%%%%%%%%%%%%%

Table \ref{TAB:SIM_2} summarizes the average widths of the bands where the
widths are averaged over the domain $[0,1]$ and over the Monte Carlo samples.
The average widths of the FF$_t^1$ band is close to that of the MB$_t$ band. The
larger $p$ the wider are the FF$_t^p$ bands. The B$_{\text{EC}}$ band is very narrow for small
samples, but becomes the widest for the rough processes (Cov2-3) and large
samples. The width rankings of the GKF$_{t}$ band are similar to those of the
FF$_t^4$ band.

\begin{table}[h!]
\caption{Size and Power for Cov1 and $n=15$. (Section \ref{SEC:SIMUL_FULLFDA} in the main paper.)}
%\begin{adjustbox}{width=.85\columnwidth,height=.38\textheight,center}
\begin{adjustbox}{width=.99\columnwidth,center}
  \begin{tabular}{ll cccc cc cccccc c}
    \toprule
    & &\multicolumn{4}{c}{4 Fair Intervals}&\multicolumn{2}{c}{2 Fair Intervals}&1 Interval&\multicolumn{5}{c}{}&\\
    & &\multicolumn{4}{c}{$\alpha/4=0.0125$}&\multicolumn{2}{c}{$\alpha/2=0.025$}&$\alpha=0.05$&\multicolumn{5}{c}{Mean perturbations $\Delta$}&Avg.\\
    \cmidrule(lr){3-6} \cmidrule(lr){7-8} \cmidrule(lr){9-9} \cmidrule(lr){10-14}\\[-1ex]
    & Band & $\left[0,\frac{1}{4}\right]$ & $\left[\frac{1}{4},\frac{2}{4}\right]$ & $\left[\frac{2}{4},\frac{3}{4}\right]$ & $\left[\frac{3}{4},1\right]$ & $\left[0,\frac{1}{2}\right]$ & $\left[\frac{1}{2},1\right]$ & $[0,1]$ & $0.05$ & $0.15$ & $0.25$ & $0.35$ & $0.45$ & Power \\
    \midrule
  Mean1  & FF$_t^4$ & 0.012 & 0.012 & 0.012 & 0.012 & 0.017 & 0.017 & 0.025 & 0.072 & 0.492 & 0.927 & 0.998 & 1.000 & 0.698 \\
  Mean1  & FF$_t^2$ & 0.019 & 0.019 & 0.018 & 0.018 & 0.026 & 0.025 & 0.037 & 0.100 & 0.572 & 0.955 & 0.999 & 1.000 & 0.725 \\
  Mean1  & FF$_t^1$ & 0.026 & 0.026 & 0.025 & 0.025 & 0.034 & 0.034 & 0.051 & 0.119 & 0.645 & 0.970 & 1.000 & 1.000 & 0.747 \\
  Mean1  & FF$_z^4$ & 0.027 & 0.028 & 0.027 & 0.027 & 0.036 & 0.036 & 0.052 & 0.122 & 0.644 & 0.968 & 1.000 & 1.000 & 0.747 \\
  Mean1  & FF$_z^2$ & 0.036 & 0.037 & 0.035 & 0.035 & 0.047 & 0.046 & 0.069 & 0.150 & 0.695 & 0.979 & 1.000 & 1.000 & 0.765 \\
  Mean1  & FF$_z^1$ & 0.045 & 0.044 & 0.044 & 0.044 & 0.058 & 0.058 & 0.085 & 0.179 & 0.732 & 0.984 & 1.000 & 1.000 & 0.779 \\
  Mean1  & B$_{\text{EC}}$ & 0.025 & 0.036 & 0.035 & 0.024 & 0.039 & 0.038 & 0.051 & 0.122 & 0.631 & 0.966 & 1.000 & 1.000 & 0.744 \\
  Mean1  & B$_{\text{S}}$ & 0.047 & 0.047 & 0.046 & 0.046 & 0.061 & 0.061 & 0.088 & 0.186 & 0.740 & 0.985 & 1.000 & 1.000 & 0.782 \\
  Mean1  & MB$_t$ & 0.020 & 0.019 & 0.019 & 0.019 & 0.026 & 0.026 & 0.039 & 0.099 & 0.593 & 0.958 & 1.000 & 1.000 & 0.730 \\
  Mean1  & GKF$_t$ & 0.018 & 0.018 & 0.018 & 0.018 & 0.025 & 0.025 & 0.037 & 0.095 & 0.581 & 0.956 & 1.000 & 1.000 & 0.726 \\
  Mean1  & IWT &  &  &  &  &  &  & 0.036 & 0.097 & 0.585 & 0.953 & 0.999 & 1.000 & 0.727 \\
  Mean2  & FF$_t^4$ & 0.012 & 0.018 & 0.016 & 0.011 & 0.022 & 0.022 & 0.029 & 0.054 & 0.325 & 0.805 & 0.984 & 1.000 & 0.633 \\
  Mean2  & FF$_t^2$ & 0.019 & 0.019 & 0.021 & 0.027 & 0.025 & 0.033 & 0.045 & 0.089 & 0.472 & 0.912 & 0.997 & 1.000 & 0.694 \\
  Mean2  & FF$_t^1$ & 0.026 & 0.026 & 0.025 & 0.025 & 0.034 & 0.034 & 0.051 & 0.091 & 0.455 & 0.902 & 0.997 & 1.000 & 0.689 \\
  Mean2  & FF$_z^4$ & 0.027 & 0.028 & 0.027 & 0.027 & 0.036 & 0.036 & 0.052 & 0.094 & 0.458 & 0.902 & 0.996 & 1.000 & 0.690 \\
  Mean2  & FF$_z^2$ & 0.036 & 0.037 & 0.035 & 0.035 & 0.047 & 0.046 & 0.069 & 0.116 & 0.515 & 0.929 & 0.998 & 1.000 & 0.712 \\
  Mean2  & FF$_z^1$ & 0.045 & 0.044 & 0.044 & 0.044 & 0.058 & 0.058 & 0.085 & 0.143 & 0.560 & 0.942 & 0.999 & 1.000 & 0.729 \\
  Mean2  & B$_{\text{EC}}$ & 0.025 & 0.036 & 0.035 & 0.024 & 0.039 & 0.038 & 0.051 & 0.088 & 0.441 & 0.898 & 0.996 & 1.000 & 0.685 \\
  Mean2  & B$_{\text{S}}$ & 0.047 & 0.047 & 0.046 & 0.046 & 0.061 & 0.061 & 0.088 & 0.149 & 0.569 & 0.944 & 0.999 & 1.000 & 0.732 \\
  Mean2  & MB$_t$ & 0.020 & 0.019 & 0.019 & 0.019 & 0.026 & 0.026 & 0.039 & 0.072 & 0.406 & 0.875 & 0.995 & 1.000 & 0.670 \\
  Mean2  & GKF$_t$ & 0.018 & 0.018 & 0.018 & 0.018 & 0.025 & 0.025 & 0.037 & 0.070 & 0.397 & 0.866 & 0.994 & 1.000 & 0.665 \\
  Mean2  & IWT &  &  &  &  &  &  & 0.036 & 0.049 & 0.218 & 0.605 & 0.923 & 0.997 & 0.558 \\
  Mean3  & FF$_t^4$ & 0.012 & 0.018 & 0.016 & 0.011 & 0.022 & 0.022 & 0.029 & 0.049 & 0.319 & 0.817 & 0.989 & 1.000 & 0.635 \\
  Mean3  & FF$_t^2$ & 0.019 & 0.019 & 0.021 & 0.027 & 0.025 & 0.033 & 0.045 & 0.074 & 0.395 & 0.873 & 0.995 & 1.000 & 0.667 \\
  Mean3  & FF$_t^1$ & 0.026 & 0.026 & 0.025 & 0.025 & 0.034 & 0.034 & 0.051 & 0.088 & 0.455 & 0.909 & 0.997 & 1.000 & 0.690 \\
  Mean3  & FF$_z^4$ & 0.027 & 0.028 & 0.027 & 0.027 & 0.036 & 0.036 & 0.052 & 0.092 & 0.460 & 0.911 & 0.997 & 1.000 & 0.692 \\
  Mean3  & FF$_z^2$ & 0.036 & 0.037 & 0.035 & 0.035 & 0.047 & 0.046 & 0.069 & 0.118 & 0.518 & 0.939 & 0.999 & 1.000 & 0.715 \\
  Mean3  & FF$_z^1$ & 0.045 & 0.044 & 0.044 & 0.044 & 0.058 & 0.058 & 0.085 & 0.142 & 0.572 & 0.955 & 0.999 & 1.000 & 0.734 \\
  Mean3  & B$_{\text{EC}}$ & 0.025 & 0.036 & 0.035 & 0.024 & 0.039 & 0.038 & 0.051 & 0.077 & 0.432 & 0.906 & 0.998 & 1.000 & 0.683 \\
  Mean3  & B$_{\text{S}}$ & 0.047 & 0.047 & 0.046 & 0.046 & 0.061 & 0.061 & 0.088 & 0.147 & 0.582 & 0.958 & 1.000 & 1.000 & 0.737 \\
  Mean3  & MB$_t$ & 0.020 & 0.019 & 0.019 & 0.019 & 0.026 & 0.026 & 0.039 & 0.068 & 0.402 & 0.877 & 0.995 & 1.000 & 0.668 \\
  Mean3  & GKF$_t$ & 0.018 & 0.018 & 0.018 & 0.018 & 0.025 & 0.025 & 0.037 & 0.065 & 0.392 & 0.871 & 0.995 & 1.000 & 0.664 \\
    Mean3 & IWT &  &  &  &  &  &  & 0.036 & 0.038 & 0.057 & 0.120 & 0.282 & 0.615 & 0.222 \\
   \bottomrule
\end{tabular}
\end{adjustbox}
\end{table}

\begin{table}[h!]
\caption{Size and Power for Cov1 and $n=100$. (Section \ref{SEC:SIMUL_FULLFDA} in the main paper.)}
%\begin{adjustbox}{width=.85\columnwidth,height=.375\textheight,center}
\begin{adjustbox}{width=.99\columnwidth,center}
    \begin{tabular}{ll cccc cc cccccc c}
  \toprule
  & &\multicolumn{4}{c}{4 Fair Intervals}&\multicolumn{2}{c}{2 Fair Intervals}&1 Interval&\multicolumn{5}{c}{}&\\
  & &\multicolumn{4}{c}{$\alpha/4=0.0125$}&\multicolumn{2}{c}{$\alpha/2=0.025$}&$\alpha=0.05$&\multicolumn{5}{c}{Mean perturbations $\Delta$}&Avg.\\
  \cmidrule(lr){3-6} \cmidrule(lr){7-8} \cmidrule(lr){9-9} \cmidrule(lr){10-14}\\[-1ex]
  & Band & $\left[0,\frac{1}{4}\right]$ & $\left[\frac{1}{4},\frac{2}{4}\right]$ & $\left[\frac{2}{4},\frac{3}{4}\right]$ & $\left[\frac{3}{4},1\right]$ & $\left[0,\frac{1}{2}\right]$ & $\left[\frac{1}{2},1\right]$ & $[0,1]$ & $0.05$ & $0.15$ & $0.25$ & $0.35$ & $0.45$ & Power \\
  \midrule
  Mean1  & FF$_t^4$ & 0.012 & 0.012 & 0.012 & 0.013 & 0.016 & 0.017 & 0.025 & 0.078 & 0.282 & 0.605 & 0.864 & 0.978 & 0.561 \\
  Mean1  & FF$_t^2$ & 0.019 & 0.019 & 0.019 & 0.019 & 0.025 & 0.025 & 0.036 & 0.104 & 0.338 & 0.671 & 0.898 & 0.984 & 0.599 \\
  Mean1  & FF$_t^1$ & 0.025 & 0.025 & 0.025 & 0.026 & 0.033 & 0.033 & 0.048 & 0.126 & 0.375 & 0.710 & 0.923 & 0.987 & 0.624 \\
  Mean1  & FF$_z^4$ & 0.014 & 0.014 & 0.015 & 0.015 & 0.019 & 0.020 & 0.028 & 0.084 & 0.293 & 0.626 & 0.880 & 0.978 & 0.572 \\
  Mean1  & FF$_z^2$ & 0.021 & 0.020 & 0.020 & 0.021 & 0.027 & 0.027 & 0.040 & 0.109 & 0.344 & 0.682 & 0.911 & 0.984 & 0.606 \\
  Mean1  & FF$_z^1$ & 0.028 & 0.028 & 0.027 & 0.028 & 0.036 & 0.036 & 0.052 & 0.134 & 0.388 & 0.723 & 0.928 & 0.988 & 0.632 \\
  Mean1  & B$_{\text{EC}}$ & 0.013 & 0.020 & 0.020 & 0.013 & 0.021 & 0.021 & 0.027 & 0.081 & 0.279 & 0.604 & 0.863 & 0.972 & 0.560 \\
  Mean1  & B$_{\text{S}}$ & 0.029 & 0.029 & 0.029 & 0.030 & 0.038 & 0.038 & 0.055 & 0.140 & 0.399 & 0.731 & 0.930 & 0.989 & 0.638 \\
  Mean1  & MB$_t$ & 0.025 & 0.025 & 0.025 & 0.026 & 0.033 & 0.033 & 0.048 & 0.126 & 0.375 & 0.711 & 0.923 & 0.987 & 0.624 \\
  Mean1  & GKF$_t$ & 0.024 & 0.024 & 0.024 & 0.025 & 0.032 & 0.032 & 0.046 & 0.123 & 0.368 & 0.703 & 0.920 & 0.986 & 0.620 \\
  Mean1  & IWT &  &  &  &  &  &  & 0.036 & 0.113 & 0.350 & 0.680 & 0.900 & 0.983 & 0.605 \\
  Mean2  & FF$_t^4$ & 0.012 & 0.013 & 0.014 & 0.014 & 0.018 & 0.019 & 0.026 & 0.054 & 0.178 & 0.442 & 0.737 & 0.924 & 0.467 \\
  Mean2  & FF$_t^2$ & 0.019 & 0.018 & 0.019 & 0.021 & 0.024 & 0.026 & 0.037 & 0.073 & 0.224 & 0.505 & 0.788 & 0.945 & 0.507 \\
  Mean2  & FF$_t^1$ & 0.025 & 0.025 & 0.025 & 0.026 & 0.033 & 0.033 & 0.048 & 0.090 & 0.254 & 0.538 & 0.812 & 0.953 & 0.530 \\
  Mean2  & FF$_z^4$ & 0.014 & 0.014 & 0.015 & 0.015 & 0.019 & 0.020 & 0.028 & 0.057 & 0.181 & 0.446 & 0.739 & 0.925 & 0.470 \\
  Mean2  & FF$_z^2$ & 0.021 & 0.020 & 0.020 & 0.021 & 0.027 & 0.027 & 0.040 & 0.077 & 0.226 & 0.507 & 0.790 & 0.946 & 0.509 \\
  Mean2  & FF$_z^1$ & 0.028 & 0.028 & 0.027 & 0.028 & 0.036 & 0.036 & 0.052 & 0.096 & 0.266 & 0.554 & 0.823 & 0.957 & 0.539 \\
  Mean2  & B$_{\text{EC}}$ & 0.013 & 0.020 & 0.020 & 0.013 & 0.021 & 0.021 & 0.027 & 0.052 & 0.159 & 0.406 & 0.700 & 0.905 & 0.444 \\
  Mean2  & B$_{\text{S}}$ & 0.029 & 0.029 & 0.029 & 0.030 & 0.038 & 0.038 & 0.055 & 0.101 & 0.273 & 0.565 & 0.831 & 0.959 & 0.546 \\
  Mean2  & MB$_t$ & 0.025 & 0.025 & 0.025 & 0.026 & 0.033 & 0.033 & 0.048 & 0.090 & 0.255 & 0.538 & 0.813 & 0.953 & 0.530 \\
  Mean2  & GKF$_t$ & 0.024 & 0.024 & 0.024 & 0.025 & 0.032 & 0.032 & 0.046 & 0.087 & 0.247 & 0.530 & 0.807 & 0.951 & 0.525 \\
  Mean2  & IWT &  &  &  &  &  &  & 0.036 & 0.053 & 0.112 & 0.224 & 0.429 & 0.672 & 0.298 \\
  Mean3  & FF$_t^4$ & 0.012 & 0.013 & 0.014 & 0.014 & 0.018 & 0.019 & 0.026 & 0.052 & 0.178 & 0.432 & 0.734 & 0.925 & 0.464 \\
  Mean3  & FF$_t^2$ & 0.019 & 0.018 & 0.019 & 0.021 & 0.024 & 0.026 & 0.037 & 0.072 & 0.224 & 0.502 & 0.794 & 0.950 & 0.508 \\
  Mean3  & FF$_t^1$ & 0.025 & 0.025 & 0.025 & 0.026 & 0.033 & 0.033 & 0.048 & 0.092 & 0.265 & 0.556 & 0.830 & 0.967 & 0.542 \\
  Mean3  & FF$_z^4$ & 0.014 & 0.014 & 0.015 & 0.015 & 0.019 & 0.020 & 0.028 & 0.057 & 0.191 & 0.455 & 0.752 & 0.933 & 0.477 \\
  Mean3  & FF$_z^2$ & 0.021 & 0.020 & 0.020 & 0.021 & 0.027 & 0.027 & 0.040 & 0.077 & 0.237 & 0.521 & 0.805 & 0.956 & 0.519 \\
  Mean3  & FF$_z^1$ & 0.028 & 0.028 & 0.027 & 0.028 & 0.036 & 0.036 & 0.052 & 0.098 & 0.278 & 0.573 & 0.842 & 0.971 & 0.553 \\
  Mean3  & B$_{\text{EC}}$ & 0.013 & 0.020 & 0.020 & 0.013 & 0.021 & 0.021 & 0.027 & 0.048 & 0.161 & 0.401 & 0.711 & 0.915 & 0.447 \\
  Mean3  & B$_{\text{S}}$ & 0.029 & 0.029 & 0.029 & 0.030 & 0.038 & 0.038 & 0.055 & 0.103 & 0.286 & 0.583 & 0.849 & 0.973 & 0.559 \\
  Mean3  & MB$_t$ & 0.025 & 0.025 & 0.025 & 0.026 & 0.033 & 0.033 & 0.048 & 0.092 & 0.266 & 0.557 & 0.831 & 0.967 & 0.542 \\
  Mean3  & GKF$_t$ & 0.024 & 0.024 & 0.024 & 0.025 & 0.032 & 0.032 & 0.046 & 0.087 & 0.259 & 0.546 & 0.825 & 0.963 & 0.536 \\
  Mean3  & IWT &  &  &  &  &  &  & 0.036 & 0.038 & 0.043 & 0.051 & 0.074 & 0.100 & 0.061 \\
	   \bottomrule
	\end{tabular}
\end{adjustbox}
\end{table}

\begin{table}[h!]
\caption{Size and Power for Cov2 and $n=15$. (Section \ref{SEC:SIMUL_FULLFDA} in the main paper.)}
%\begin{adjustbox}{width=.85\columnwidth,height=.375\textheight,center}
\begin{adjustbox}{width=.99\columnwidth,center}
  \begin{tabular}{ll cccc cc cccccc c}
    \toprule
    & &\multicolumn{4}{c}{4 Fair Intervals}&\multicolumn{2}{c}{2 Fair Intervals}&1 Interval&\multicolumn{5}{c}{}&\\
    & &\multicolumn{4}{c}{$\alpha/4=0.0125$}&\multicolumn{2}{c}{$\alpha/2=0.025$}&$\alpha=0.05$&\multicolumn{5}{c}{Mean perturbations $\Delta$}&Avg.\\
    \cmidrule(lr){3-6} \cmidrule(lr){7-8} \cmidrule(lr){9-9} \cmidrule(lr){10-14}\\[-1ex]
    & Band & $\left[0,\frac{1}{4}\right]$ & $\left[\frac{1}{4},\frac{2}{4}\right]$ & $\left[\frac{2}{4},\frac{3}{4}\right]$ & $\left[\frac{3}{4},1\right]$ & $\left[0,\frac{1}{2}\right]$ & $\left[\frac{1}{2},1\right]$ & $[0,1]$ & $0.05$ & $0.15$ & $0.25$ & $0.35$ & $0.45$ & Power \\
    \midrule
    Mean1  & FF$_t^4$ & 0.011 & 0.012 & 0.011 & 0.011 & 0.018 & 0.018 & 0.032 & 0.079 & 0.542 & 0.947 & 0.999 & 1.000 & 0.713 \\
    Mean1  & FF$_t^2$ & 0.013 & 0.013 & 0.013 & 0.013 & 0.021 & 0.021 & 0.036 & 0.088 & 0.572 & 0.955 & 1.000 & 1.000 & 0.723 \\
    Mean1  & FF$_t^1$ & 0.014 & 0.013 & 0.014 & 0.014 & 0.022 & 0.023 & 0.038 & 0.098 & 0.592 & 0.960 & 1.000 & 1.000 & 0.730 \\
    Mean1  & FF$_z^4$ & 0.029 & 0.030 & 0.030 & 0.031 & 0.046 & 0.047 & 0.078 & 0.169 & 0.725 & 0.985 & 1.000 & 1.000 & 0.776 \\
    Mean1  & FF$_z^2$ & 0.032 & 0.032 & 0.033 & 0.033 & 0.051 & 0.052 & 0.084 & 0.177 & 0.740 & 0.987 & 1.000 & 1.000 & 0.781 \\
    Mean1  & FF$_z^1$ & 0.034 & 0.034 & 0.035 & 0.036 & 0.053 & 0.054 & 0.089 & 0.186 & 0.749 & 0.988 & 1.000 & 1.000 & 0.785 \\
    Mean1  & B$_{\text{EC}}$ & 0.012 & 0.015 & 0.016 & 0.012 & 0.021 & 0.022 & 0.036 & 0.095 & 0.623 & 0.971 & 1.000 & 1.000 & 0.738 \\
    Mean1  & B$_{\text{S}}$ & 0.049 & 0.048 & 0.048 & 0.050 & 0.074 & 0.075 & 0.122 & 0.237 & 0.808 & 0.993 & 1.000 & 1.000 & 0.808 \\
    Mean1  & MB$_t$ & 0.013 & 0.013 & 0.013 & 0.013 & 0.021 & 0.021 & 0.036 & 0.095 & 0.596 & 0.963 & 1.000 & 1.000 & 0.731 \\
    Mean1  & GKF$_t$ & 0.006 & 0.006 & 0.006 & 0.006 & 0.010 & 0.010 & 0.018 & 0.055 & 0.448 & 0.915 & 0.998 & 1.000 & 0.683 \\
    Mean1  & IWT &  &  &  &  &  &  & 0.028 & 0.100 & 0.632 & 0.980 & 1.000 & 1.000 & 0.742 \\
    Mean2  & FF$_t^4$ & 0.011 & 0.012 & 0.013 & 0.013 & 0.019 & 0.021 & 0.034 & 0.061 & 0.369 & 0.842 & 0.991 & 1.000 & 0.653 \\
    Mean2  & FF$_t^2$ & 0.012 & 0.012 & 0.014 & 0.014 & 0.020 & 0.022 & 0.036 & 0.066 & 0.387 & 0.857 & 0.992 & 1.000 & 0.661 \\
    Mean2  & FF$_t^1$ & 0.014 & 0.013 & 0.014 & 0.014 & 0.022 & 0.023 & 0.038 & 0.066 & 0.374 & 0.840 & 0.991 & 1.000 & 0.654 \\
  Mean2  & FF$_z^4$ & 0.029 & 0.030 & 0.030 & 0.031 & 0.046 & 0.047 & 0.078 & 0.122 & 0.513 & 0.917 & 0.997 & 1.000 & 0.710 \\
  Mean2  & FF$_z^2$ & 0.032 & 0.032 & 0.033 & 0.033 & 0.051 & 0.052 & 0.084 & 0.132 & 0.531 & 0.926 & 0.998 & 1.000 & 0.717 \\
  Mean2  & FF$_z^1$ & 0.034 & 0.034 & 0.035 & 0.036 & 0.053 & 0.054 & 0.089 & 0.138 & 0.538 & 0.926 & 0.998 & 1.000 & 0.720 \\
  Mean2  & B$_{\text{EC}}$ & 0.012 & 0.015 & 0.016 & 0.012 & 0.021 & 0.022 & 0.036 & 0.066 & 0.396 & 0.871 & 0.995 & 1.000 & 0.666 \\
  Mean2  & B$_{\text{S}}$ & 0.049 & 0.048 & 0.048 & 0.050 & 0.074 & 0.075 & 0.122 & 0.182 & 0.608 & 0.952 & 0.999 & 1.000 & 0.748 \\
  Mean2  & MB$_t$ & 0.013 & 0.013 & 0.013 & 0.013 & 0.021 & 0.021 & 0.036 & 0.064 & 0.373 & 0.845 & 0.992 & 1.000 & 0.655 \\
  Mean2  & GKF$_t$ & 0.006 & 0.006 & 0.006 & 0.006 & 0.010 & 0.010 & 0.018 & 0.036 & 0.254 & 0.738 & 0.973 & 0.999 & 0.600 \\
  Mean2  & IWT &  &  &  &  &  &  & 0.028 & 0.045 & 0.225 & 0.695 & 0.961 & 0.998 & 0.585 \\
  Mean3  & FF$_t^4$ & 0.011 & 0.012 & 0.013 & 0.013 & 0.019 & 0.021 & 0.034 & 0.049 & 0.275 & 0.771 & 0.983 & 1.000 & 0.616 \\
    Mean3  & FF$_t^2$ & 0.012 & 0.012 & 0.014 & 0.014 & 0.020 & 0.022 & 0.036 & 0.054 & 0.297 & 0.795 & 0.986 & 1.000 & 0.627 \\
    Mean3  & FF$_t^1$ & 0.014 & 0.013 & 0.014 & 0.014 & 0.022 & 0.023 & 0.038 & 0.058 & 0.310 & 0.804 & 0.987 & 1.000 & 0.632 \\
  Mean3  & FF$_z^4$ & 0.029 & 0.030 & 0.030 & 0.031 & 0.046 & 0.047 & 0.078 & 0.115 & 0.454 & 0.903 & 0.998 & 1.000 & 0.694 \\
  Mean3  & FF$_z^2$ & 0.032 & 0.032 & 0.033 & 0.033 & 0.051 & 0.052 & 0.084 & 0.124 & 0.474 & 0.913 & 0.998 & 1.000 & 0.702 \\
  Mean3  & FF$_z^1$ & 0.034 & 0.034 & 0.035 & 0.036 & 0.053 & 0.054 & 0.089 & 0.131 & 0.486 & 0.915 & 0.998 & 1.000 & 0.706 \\
  Mean3  & B$_{\text{EC}}$ & 0.012 & 0.015 & 0.016 & 0.012 & 0.021 & 0.022 & 0.036 & 0.053 & 0.306 & 0.830 & 0.994 & 1.000 & 0.637 \\
  Mean3  & B$_{\text{S}}$ & 0.049 & 0.048 & 0.048 & 0.050 & 0.074 & 0.075 & 0.122 & 0.175 & 0.572 & 0.950 & 1.000 & 1.000 & 0.739 \\
  Mean3  & MB$_t$ & 0.013 & 0.013 & 0.013 & 0.013 & 0.021 & 0.021 & 0.036 & 0.056 & 0.309 & 0.810 & 0.988 & 1.000 & 0.633 \\
  Mean3  & GKF$_t$ & 0.006 & 0.006 & 0.006 & 0.006 & 0.010 & 0.010 & 0.018 & 0.029 & 0.199 & 0.679 & 0.964 & 0.999 & 0.574 \\
  Mean3  & IWT &  &  &  &  &  &  & 0.028 & 0.030 & 0.056 & 0.136 & 0.376 & 0.722 & 0.264 \\
	   \bottomrule
	\end{tabular}
\end{adjustbox}
\end{table}

\begin{table}[h!]
\caption{Size and Power for Cov2 and $n=100$. (Section \ref{SEC:SIMUL_FULLFDA} in the main paper.)}
%\begin{adjustbox}{width=.85\columnwidth,height=.375\textheight,center}
\begin{adjustbox}{width=.99\columnwidth,center}
  \begin{tabular}{ll cccc cc cccccc c}
    \toprule
    & &\multicolumn{4}{c}{4 Fair Intervals}&\multicolumn{2}{c}{2 Fair Intervals}&1 Interval&\multicolumn{5}{c}{}&\\
    & &\multicolumn{4}{c}{$\alpha/4=0.0125$}&\multicolumn{2}{c}{$\alpha/2=0.025$}&$\alpha=0.05$&\multicolumn{5}{c}{Mean perturbations $\Delta$}&Avg.\\
    \cmidrule(lr){3-6} \cmidrule(lr){7-8} \cmidrule(lr){9-9} \cmidrule(lr){10-14}\\[-1ex]
    & Band & $\left[0,\frac{1}{4}\right]$ & $\left[\frac{1}{4},\frac{2}{4}\right]$ & $\left[\frac{2}{4},\frac{3}{4}\right]$ & $\left[\frac{3}{4},1\right]$ & $\left[0,\frac{1}{2}\right]$ & $\left[\frac{1}{2},1\right]$ & $[0,1]$ & $0.05$ & $0.15$ & $0.25$ & $0.35$ & $0.45$ & Power \\
    \midrule
    Mean1  & FF$_t^4$ & 0.009 & 0.010 & 0.010 & 0.010 & 0.015 & 0.016 & 0.025 & 0.085 & 0.306 & 0.648 & 0.906 & 0.987 & 0.586 \\
    Mean1  & FF$_t^2$ & 0.010 & 0.011 & 0.011 & 0.011 & 0.017 & 0.017 & 0.029 & 0.092 & 0.321 & 0.665 & 0.916 & 0.988 & 0.596 \\
    Mean1  & FF$_t^1$ & 0.012 & 0.012 & 0.013 & 0.013 & 0.019 & 0.020 & 0.033 & 0.102 & 0.337 & 0.684 & 0.916 & 0.988 & 0.605 \\
  Mean1  & FF$_z^4$ & 0.012 & 0.013 & 0.013 & 0.013 & 0.019 & 0.020 & 0.033 & 0.101 & 0.336 & 0.683 & 0.914 & 0.988 & 0.604 \\
  Mean1  & FF$_z^2$ & 0.013 & 0.014 & 0.014 & 0.015 & 0.021 & 0.022 & 0.036 & 0.111 & 0.353 & 0.699 & 0.923 & 0.990 & 0.615 \\
  Mean1  & FF$_z^1$ & 0.014 & 0.015 & 0.015 & 0.016 & 0.023 & 0.023 & 0.039 & 0.115 & 0.364 & 0.707 & 0.927 & 0.990 & 0.621 \\
  Mean1  & B$_{\text{EC}}$ & 0.000 & 0.000 & 0.000 & 0.000 & 0.000 & 0.001 & 0.001 & 0.007 & 0.048 & 0.220 & 0.536 & 0.833 & 0.329 \\
  Mean1  & B$_{\text{S}}$ & 0.023 & 0.025 & 0.024 & 0.025 & 0.037 & 0.038 & 0.061 & 0.161 & 0.448 & 0.775 & 0.952 & 0.995 & 0.666 \\
  Mean1  & MB$_t$ & 0.018 & 0.020 & 0.019 & 0.020 & 0.029 & 0.031 & 0.050 & 0.138 & 0.407 & 0.745 & 0.943 & 0.993 & 0.645 \\
  Mean1  & GKF$_t$ & 0.008 & 0.009 & 0.009 & 0.010 & 0.014 & 0.014 & 0.024 & 0.081 & 0.293 & 0.638 & 0.893 & 0.983 & 0.578 \\
  Mean1  & IWT &  &  &  &  &  &  & 0.029 & 0.111 & 0.376 & 0.717 & 0.932 & 0.993 & 0.626 \\
    Mean2  & FF$_t^4$ & 0.009 & 0.011 & 0.011 & 0.011 & 0.016 & 0.017 & 0.028 & 0.053 & 0.189 & 0.452 & 0.731 & 0.925 & 0.470 \\
    Mean2  & FF$_t^2$ & 0.011 & 0.011 & 0.012 & 0.013 & 0.018 & 0.019 & 0.031 & 0.058 & 0.200 & 0.470 & 0.746 & 0.932 & 0.481 \\
    Mean2  & FF$_t^1$ & 0.012 & 0.012 & 0.013 & 0.013 & 0.019 & 0.020 & 0.033 & 0.061 & 0.203 & 0.473 & 0.747 & 0.933 & 0.483 \\
  Mean2  & FF$_z^4$ & 0.012 & 0.013 & 0.013 & 0.013 & 0.019 & 0.020 & 0.033 & 0.060 & 0.204 & 0.473 & 0.748 & 0.933 & 0.484 \\
  Mean2  & FF$_z^2$ & 0.013 & 0.014 & 0.014 & 0.015 & 0.021 & 0.022 & 0.036 & 0.067 & 0.216 & 0.493 & 0.762 & 0.940 & 0.496 \\
  Mean2  & FF$_z^1$ & 0.014 & 0.015 & 0.015 & 0.016 & 0.023 & 0.023 & 0.039 & 0.069 & 0.224 & 0.500 & 0.768 & 0.942 & 0.501 \\
  Mean2  & B$_{\text{EC}}$ & 0.000 & 0.000 & 0.000 & 0.000 & 0.000 & 0.001 & 0.001 & 0.003 & 0.017 & 0.100 & 0.292 & 0.601 & 0.203 \\
  Mean2  & B$_{\text{S}}$ & 0.023 & 0.025 & 0.024 & 0.025 & 0.037 & 0.038 & 0.061 & 0.100 & 0.284 & 0.580 & 0.828 & 0.964 & 0.551 \\
  Mean2  & MB$_t$ & 0.018 & 0.020 & 0.019 & 0.020 & 0.029 & 0.031 & 0.050 & 0.084 & 0.257 & 0.545 & 0.803 & 0.955 & 0.529 \\
  Mean2  & GKF$_t$ & 0.008 & 0.009 & 0.009 & 0.010 & 0.014 & 0.014 & 0.024 & 0.046 & 0.171 & 0.424 & 0.707 & 0.915 & 0.453 \\
  Mean2  & IWT &  &  &  &  &  &  & 0.029 & 0.052 & 0.122 & 0.268 & 0.493 & 0.762 & 0.339 \\
  Mean3  & FF$_t^4$ & 0.009 & 0.011 & 0.011 & 0.011 & 0.016 & 0.017 & 0.028 & 0.044 & 0.142 & 0.380 & 0.676 & 0.902 & 0.429 \\
    Mean3  & FF$_t^2$ & 0.011 & 0.011 & 0.012 & 0.013 & 0.018 & 0.019 & 0.031 & 0.050 & 0.155 & 0.401 & 0.696 & 0.911 & 0.443 \\
    Mean3  & FF$_t^1$ & 0.012 & 0.012 & 0.013 & 0.013 & 0.019 & 0.020 & 0.033 & 0.054 & 0.162 & 0.412 & 0.706 & 0.915 & 0.450 \\
  Mean3  & FF$_z^4$ & 0.012 & 0.013 & 0.013 & 0.013 & 0.019 & 0.020 & 0.033 & 0.052 & 0.159 & 0.409 & 0.703 & 0.914 & 0.448 \\
  Mean3  & FF$_z^2$ & 0.013 & 0.014 & 0.014 & 0.015 & 0.021 & 0.022 & 0.036 & 0.059 & 0.171 & 0.426 & 0.723 & 0.924 & 0.461 \\
  Mean3  & FF$_z^1$ & 0.014 & 0.015 & 0.015 & 0.016 & 0.023 & 0.023 & 0.039 & 0.063 & 0.180 & 0.437 & 0.731 & 0.928 & 0.468 \\
  Mean3  & B$_{\text{EC}}$ & 0.000 & 0.000 & 0.000 & 0.000 & 0.000 & 0.001 & 0.001 & 0.002 & 0.011 & 0.068 & 0.236 & 0.534 & 0.170 \\
  Mean3  & B$_{\text{S}}$ & 0.023 & 0.025 & 0.024 & 0.025 & 0.037 & 0.038 & 0.061 & 0.095 & 0.242 & 0.525 & 0.799 & 0.959 & 0.524 \\
  Mean3  & MB$_t$ & 0.018 & 0.020 & 0.019 & 0.020 & 0.029 & 0.031 & 0.050 & 0.078 & 0.211 & 0.487 & 0.768 & 0.948 & 0.498 \\
  Mean3  & GKF$_t$ & 0.008 & 0.009 & 0.009 & 0.010 & 0.014 & 0.014 & 0.024 & 0.040 & 0.133 & 0.368 & 0.662 & 0.894 & 0.419 \\
  Mean3 & IWT &  &  &  &  &  &  & 0.029 & 0.030 & 0.042 & 0.059 & 0.080 & 0.116 & 0.065 \\
\bottomrule
\end{tabular}
\end{adjustbox}
\end{table}

\begin{table}[h!]
\caption{Size and Power for Cov3 and $n=15$. (Section \ref{SEC:SIMUL_FULLFDA} in the main paper.)}
%\begin{adjustbox}{width=.85\columnwidth,height=.375\textheight,center}
\begin{adjustbox}{width=.99\columnwidth,center}
  \begin{tabular}{ll cccc cc cccccc c}
    \toprule
    & &\multicolumn{4}{c}{4 Fair Intervals}&\multicolumn{2}{c}{2 Fair Intervals}&1 Interval&\multicolumn{5}{c}{}&\\
    & &\multicolumn{4}{c}{$\alpha/4=0.0125$}&\multicolumn{2}{c}{$\alpha/2=0.025$}&$\alpha=0.05$&\multicolumn{5}{c}{Mean perturbations $\Delta$}&Avg.\\
    \cmidrule(lr){3-6} \cmidrule(lr){7-8} \cmidrule(lr){9-9} \cmidrule(lr){10-14}\\[-1ex]
    & Band & $\left[0,\frac{1}{4}\right]$ & $\left[\frac{1}{4},\frac{2}{4}\right]$ & $\left[\frac{2}{4},\frac{3}{4}\right]$ & $\left[\frac{3}{4},1\right]$ & $\left[0,\frac{1}{2}\right]$ & $\left[\frac{1}{2},1\right]$ & $[0,1]$ & $0.05$ & $0.15$ & $0.25$ & $0.35$ & $0.45$ & Power \\
    \midrule
    Mean1  & FF$_t^4$ & 0.013 & 0.012 & 0.011 & 0.012 & 0.017 & 0.019 & 0.031 & 0.084 & 0.561 & 0.956 & 1.000 & 1.000 & 0.720 \\
    Mean1  & FF$_t^2$ & 0.016 & 0.019 & 0.017 & 0.010 & 0.025 & 0.023 & 0.038 & 0.094 & 0.587 & 0.961 & 1.000 & 1.000 & 0.728 \\
    Mean1  & FF$_t^1$ & 0.008 & 0.010 & 0.015 & 0.030 & 0.013 & 0.037 & 0.044 & 0.116 & 0.641 & 0.976 & 1.000 & 1.000 & 0.747 \\
    Mean1  & FF$_z^4$ & 0.027 & 0.027 & 0.029 & 0.040 & 0.037 & 0.055 & 0.077 & 0.171 & 0.738 & 0.988 & 1.000 & 1.000 & 0.779 \\
  Mean1  & FF$_z^2$ & 0.030 & 0.037 & 0.037 & 0.034 & 0.047 & 0.058 & 0.083 & 0.180 & 0.746 & 0.989 & 1.000 & 1.000 & 0.783 \\
  Mean1  & FF$_z^1$ & 0.019 & 0.024 & 0.034 & 0.067 & 0.031 & 0.082 & 0.097 & 0.205 & 0.780 & 0.993 & 1.000 & 1.000 & 0.796 \\
  Mean1  & B$_{\text{EC}}$ & 0.015 & 0.022 & 0.017 & 0.008 & 0.025 & 0.021 & 0.035 & 0.097 & 0.601 & 0.970 & 1.000 & 1.000 & 0.734 \\
  Mean1  & B$_{\text{S}}$ & 0.025 & 0.031 & 0.043 & 0.082 & 0.039 & 0.101 & 0.120 & 0.238 & 0.818 & 0.995 & 1.000 & 1.000 & 0.810 \\
  Mean1  & MB$_t$ & 0.006 & 0.008 & 0.012 & 0.024 & 0.010 & 0.030 & 0.036 & 0.101 & 0.614 & 0.971 & 1.000 & 1.000 & 0.737 \\
  Mean1  & GKF$_t$ & 0.004 & 0.004 & 0.007 & 0.016 & 0.006 & 0.019 & 0.023 & 0.066 & 0.510 & 0.945 & 0.999 & 1.000 & 0.704 \\
  Mean1  & IWT &  &  &  &  &  &  & 0.027 & 0.094 & 0.625 & 0.976 & 1.000 & 1.000 & 0.739 \\
  Mean2  & FF$_t^4$ & 0.011 & 0.017 & 0.012 & 0.014 & 0.020 & 0.023 & 0.034 & 0.067 & 0.389 & 0.866 & 0.992 & 1.000 & 0.663 \\
    Mean2  & FF$_t^2$ & 0.014 & 0.018 & 0.018 & 0.016 & 0.023 & 0.029 & 0.041 & 0.079 & 0.418 & 0.881 & 0.995 & 1.000 & 0.675 \\
    Mean2  & FF$_t^1$ & 0.008 & 0.010 & 0.015 & 0.030 & 0.013 & 0.037 & 0.044 & 0.098 & 0.521 & 0.931 & 0.998 & 1.000 & 0.710 \\
  Mean2  & FF$_z^4$ & 0.027 & 0.027 & 0.029 & 0.040 & 0.037 & 0.055 & 0.077 & 0.140 & 0.577 & 0.946 & 0.999 & 1.000 & 0.732 \\
  Mean2  & FF$_z^2$ & 0.030 & 0.037 & 0.037 & 0.034 & 0.047 & 0.058 & 0.083 & 0.142 & 0.561 & 0.940 & 0.998 & 1.000 & 0.728 \\
  Mean2  & FF$_z^1$ & 0.019 & 0.024 & 0.034 & 0.067 & 0.031 & 0.082 & 0.097 & 0.182 & 0.675 & 0.973 & 1.000 & 1.000 & 0.766 \\
  Mean2  & B$_{\text{EC}}$ & 0.015 & 0.022 & 0.017 & 0.008 & 0.025 & 0.021 & 0.035 & 0.062 & 0.353 & 0.853 & 0.993 & 1.000 & 0.652 \\
  Mean2  & B$_{\text{S}}$ & 0.025 & 0.031 & 0.043 & 0.082 & 0.039 & 0.101 & 0.120 & 0.217 & 0.721 & 0.980 & 1.000 & 1.000 & 0.784 \\
  Mean2  & MB$_t$ & 0.006 & 0.008 & 0.012 & 0.024 & 0.010 & 0.030 & 0.036 & 0.085 & 0.491 & 0.923 & 0.998 & 1.000 & 0.699 \\
  Mean2  & GKF$_t$ & 0.004 & 0.004 & 0.007 & 0.016 & 0.006 & 0.019 & 0.023 & 0.055 & 0.392 & 0.869 & 0.994 & 1.000 & 0.662 \\
  Mean2  & IWT &  &  &  &  &  &  & 0.027 & 0.048 & 0.231 & 0.681 & 0.958 & 0.997 & 0.583 \\
  Mean3  & FF$_t^4$ & 0.011 & 0.017 & 0.012 & 0.014 & 0.020 & 0.023 & 0.034 & 0.056 & 0.317 & 0.810 & 0.990 & 1.000 & 0.634 \\
    Mean3  & FF$_t^2$ & 0.014 & 0.018 & 0.018 & 0.016 & 0.023 & 0.029 & 0.041 & 0.067 & 0.355 & 0.843 & 0.994 & 1.000 & 0.652 \\
    Mean3  & FF$_t^1$ & 0.008 & 0.010 & 0.015 & 0.030 & 0.013 & 0.037 & 0.044 & 0.058 & 0.271 & 0.756 & 0.980 & 1.000 & 0.613 \\
  Mean3  & FF$_z^4$ & 0.027 & 0.027 & 0.029 & 0.040 & 0.037 & 0.055 & 0.077 & 0.115 & 0.469 & 0.909 & 0.998 & 1.000 & 0.698 \\
  Mean3  & FF$_z^2$ & 0.030 & 0.037 & 0.037 & 0.034 & 0.047 & 0.058 & 0.083 & 0.125 & 0.498 & 0.924 & 0.999 & 1.000 & 0.709 \\
  Mean3  & FF$_z^1$ & 0.019 & 0.024 & 0.034 & 0.067 & 0.031 & 0.082 & 0.097 & 0.124 & 0.429 & 0.879 & 0.996 & 1.000 & 0.686 \\
  Mean3  & B$_{\text{EC}}$ & 0.015 & 0.022 & 0.017 & 0.008 & 0.025 & 0.021 & 0.035 & 0.060 & 0.367 & 0.867 & 0.997 & 1.000 & 0.658 \\
  Mean3  & B$_{\text{S}}$ & 0.025 & 0.031 & 0.043 & 0.082 & 0.039 & 0.101 & 0.120 & 0.154 & 0.485 & 0.910 & 0.998 & 1.000 & 0.710 \\
  Mean3  & MB$_t$ & 0.006 & 0.008 & 0.012 & 0.024 & 0.010 & 0.030 & 0.036 & 0.049 & 0.244 & 0.732 & 0.975 & 0.999 & 0.600 \\
  Mean3  & GKF$_t$ & 0.004 & 0.004 & 0.007 & 0.016 & 0.006 & 0.019 & 0.023 & 0.030 & 0.177 & 0.636 & 0.947 & 0.998 & 0.558 \\
  Mean3  & IWT &  &  &  &  &  &  & 0.027 & 0.033 & 0.053 & 0.137 & 0.363 & 0.714 & 0.260 \\
	   \bottomrule
	\end{tabular}
\end{adjustbox}
\end{table}

\begin{table}[h!]
\caption{Size and Power for Cov3 and $n=100$. (Section \ref{SEC:SIMUL_FULLFDA} in the main paper.)}
%\begin{adjustbox}{width=.85\columnwidth,height=.375\textheight,center}
\begin{adjustbox}{width=.99\columnwidth,center}
  \begin{tabular}{ll cccc cc cccccc c}
    \toprule
    & &\multicolumn{4}{c}{4 Fair Intervals}&\multicolumn{2}{c}{2 Fair Intervals}&1 Interval&\multicolumn{5}{c}{}&\\
    & &\multicolumn{4}{c}{$\alpha/4=0.0125$}&\multicolumn{2}{c}{$\alpha/2=0.025$}&$\alpha=0.05$&\multicolumn{5}{c}{Mean perturbations $\Delta$}&Avg.\\
    \cmidrule(lr){3-6} \cmidrule(lr){7-8} \cmidrule(lr){9-9} \cmidrule(lr){10-14}\\[-1ex]
    & Band & $\left[0,\frac{1}{4}\right]$ & $\left[\frac{1}{4},\frac{2}{4}\right]$ & $\left[\frac{2}{4},\frac{3}{4}\right]$ & $\left[\frac{3}{4},1\right]$ & $\left[0,\frac{1}{2}\right]$ & $\left[\frac{1}{2},1\right]$ & $[0,1]$ & $0.05$ & $0.15$ & $0.25$ & $0.35$ & $0.45$ & Power \\
\midrule
    Mean1  & FF$_t^4$ & 0.013 & 0.012 & 0.011 & 0.011 & 0.017 & 0.018 & 0.029 & 0.092 & 0.330 & 0.660 & 0.918 & 0.990 & 0.598 \\
    Mean1  & FF$_t^2$ & 0.015 & 0.019 & 0.016 & 0.009 & 0.024 & 0.021 & 0.034 & 0.102 & 0.346 & 0.674 & 0.921 & 0.991 & 0.607 \\
    Mean1  & FF$_t^1$ & 0.007 & 0.009 & 0.013 & 0.025 & 0.012 & 0.031 & 0.038 & 0.118 & 0.379 & 0.730 & 0.933 & 0.993 & 0.631 \\
    Mean1  & FF$_z^4$ & 0.014 & 0.014 & 0.012 & 0.013 & 0.019 & 0.021 & 0.033 & 0.104 & 0.356 & 0.706 & 0.927 & 0.992 & 0.617 \\
    Mean1  & FF$_z^2$ & 0.018 & 0.020 & 0.018 & 0.010 & 0.027 & 0.024 & 0.038 & 0.112 & 0.370 & 0.717 & 0.927 & 0.993 & 0.624 \\
    Mean1  & FF$_z^1$ & 0.009 & 0.011 & 0.015 & 0.029 & 0.014 & 0.036 & 0.044 & 0.132 & 0.404 & 0.754 & 0.943 & 0.994 & 0.645 \\
    Mean1  & B$_{\text{EC}}$ & 0.004 & 0.004 & 0.001 & 0.000 & 0.005 & 0.001 & 0.006 & 0.024 & 0.134 & 0.388 & 0.706 & 0.922 & 0.435 \\
    Mean1  & B$_{\text{S}}$ & 0.013 & 0.015 & 0.021 & 0.039 & 0.020 & 0.049 & 0.059 & 0.161 & 0.462 & 0.796 & 0.957 & 0.996 & 0.674 \\
    Mean1  & MB$_t$ & 0.010 & 0.012 & 0.017 & 0.032 & 0.016 & 0.040 & 0.048 & 0.141 & 0.422 & 0.768 & 0.947 & 0.995 & 0.655 \\
    Mean1  & GKF$_t$ & 0.005 & 0.007 & 0.010 & 0.019 & 0.009 & 0.024 & 0.029 & 0.096 & 0.338 & 0.690 & 0.917 & 0.991 & 0.606 \\
    Mean1  & IWT &  &  &  &  &  &  & 0.028 & 0.110 & 0.379 & 0.722 & 0.929 & 0.994 & 0.627 \\
    Mean2  & FF$_t^4$ & 0.012 & 0.013 & 0.011 & 0.011 & 0.017 & 0.018 & 0.029 & 0.059 & 0.209 & 0.482 & 0.783 & 0.947 & 0.496 \\
    Mean2  & FF$_t^2$ & 0.015 & 0.018 & 0.016 & 0.009 & 0.024 & 0.021 & 0.034 & 0.062 & 0.198 & 0.460 & 0.763 & 0.937 & 0.484 \\
    Mean2  & FF$_t^1$ & 0.007 & 0.009 & 0.013 & 0.025 & 0.012 & 0.031 & 0.038 & 0.091 & 0.296 & 0.603 & 0.866 & 0.975 & 0.566 \\
    Mean2  & FF$_z^4$ & 0.014 & 0.014 & 0.012 & 0.013 & 0.019 & 0.021 & 0.033 & 0.067 & 0.225 & 0.505 & 0.799 & 0.954 & 0.510 \\
    Mean2  & FF$_z^2$ & 0.018 & 0.020 & 0.018 & 0.010 & 0.027 & 0.024 & 0.038 & 0.070 & 0.214 & 0.481 & 0.779 & 0.944 & 0.497 \\
    Mean2  & FF$_z^1$ & 0.009 & 0.011 & 0.015 & 0.029 & 0.014 & 0.036 & 0.044 & 0.104 & 0.317 & 0.626 & 0.877 & 0.978 & 0.580 \\
    Mean2  & B$_{\text{EC}}$ & 0.004 & 0.004 & 0.001 & 0.000 & 0.005 & 0.001 & 0.006 & 0.007 & 0.022 & 0.089 & 0.258 & 0.537 & 0.183 \\
    Mean2  & B$_{\text{S}}$ & 0.013 & 0.015 & 0.021 & 0.039 & 0.020 & 0.049 & 0.059 & 0.127 & 0.365 & 0.672 & 0.899 & 0.984 & 0.610 \\
    Mean2  & MB$_t$ & 0.010 & 0.012 & 0.017 & 0.032 & 0.016 & 0.040 & 0.048 & 0.110 & 0.330 & 0.639 & 0.885 & 0.980 & 0.589 \\
    Mean2  & GKF$_t$ & 0.005 & 0.007 & 0.010 & 0.019 & 0.009 & 0.024 & 0.029 & 0.074 & 0.260 & 0.559 & 0.840 & 0.968 & 0.540 \\
    Mean2  & IWT &  &  &  &  &  &  & 0.028 & 0.052 & 0.113 & 0.242 & 0.494 & 0.748 & 0.330 \\
    Mean3  & FF$_t^4$ & 0.012 & 0.013 & 0.011 & 0.011 & 0.017 & 0.018 & 0.029 & 0.054 & 0.176 & 0.425 & 0.731 & 0.928 & 0.463 \\
    Mean3  & FF$_t^2$ & 0.015 & 0.018 & 0.016 & 0.009 & 0.024 & 0.021 & 0.034 & 0.061 & 0.196 & 0.457 & 0.758 & 0.938 & 0.482 \\
    Mean3  & FF$_t^1$ & 0.007 & 0.009 & 0.013 & 0.025 & 0.012 & 0.031 & 0.038 & 0.054 & 0.149 & 0.367 & 0.673 & 0.895 & 0.428 \\
    Mean3  & FF$_z^4$ & 0.014 & 0.014 & 0.012 & 0.013 & 0.019 & 0.021 & 0.033 & 0.060 & 0.190 & 0.448 & 0.750 & 0.935 & 0.476 \\
    Mean3  & FF$_z^2$ & 0.018 & 0.020 & 0.018 & 0.010 & 0.027 & 0.024 & 0.038 & 0.067 & 0.212 & 0.477 & 0.774 & 0.945 & 0.495 \\
    Mean3  & FF$_z^1$ & 0.009 & 0.011 & 0.015 & 0.029 & 0.014 & 0.036 & 0.044 & 0.063 & 0.164 & 0.390 & 0.696 & 0.909 & 0.444 \\
    Mean3  & B$_{\text{EC}}$ & 0.004 & 0.004 & 0.001 & 0.000 & 0.005 & 0.001 & 0.006 & 0.017 & 0.086 & 0.275 & 0.580 & 0.847 & 0.361 \\
    Mean3  & B$_{\text{S}}$ & 0.013 & 0.015 & 0.021 & 0.039 & 0.020 & 0.049 & 0.059 & 0.081 & 0.200 & 0.444 & 0.742 & 0.933 & 0.480 \\
    Mean3  & MB$_t$ & 0.010 & 0.012 & 0.017 & 0.032 & 0.016 & 0.040 & 0.048 & 0.067 & 0.175 & 0.406 & 0.712 & 0.917 & 0.455 \\
    Mean3  & GKF$_t$ & 0.005 & 0.007 & 0.010 & 0.019 & 0.009 & 0.024 & 0.029 & 0.042 & 0.123 & 0.327 & 0.627 & 0.872 & 0.398 \\
    Mean3  & IWT &  &  &  &  &  &  & 0.028 & 0.038 & 0.037 & 0.058 & 0.075 & 0.121 & 0.066 \\
    \bottomrule
  \end{tabular}
\end{adjustbox}
\end{table}

\newpage

%%%%%%%%%%%%%%%%%%%%%%%%%%
%\spacingset{1}
\begin{table}[h!tb]
	\caption{The type-I error rates and power values of the fragment bands
    FF$_{\operatorname{frag},t}^p$ and FF$_{\operatorname{frag},z}^p$,
    $p\in\{1,2\}$, in the stationary covariance scenario Cov1. (Section \ref{SEC:SIMUL_PARTFDA} in the main paper.)}
  \begin{adjustbox}{width=\columnwidth,center}
    \centering
    \begin{tabular}{ll c cccccc c}
      \toprule
      &     &H$_0$&\multicolumn{5}{c}{H$_1$}&Avg.~\\
      \cmidrule(lr){3-3}\cmidrule(lr){4-8}\\[-2ex]
      DGP & Band &$\Delta=0$ &$\Delta=0.02$ &$\Delta=0.04$ &$\Delta=0.06$ &$\Delta=0.08$ &$\Delta=0.1$&Power \\
      \midrule
      Mean1 Cov1 & FF$_{\operatorname{frag},t}^2$ & 0.042 & 0.241 & 0.774 & 0.991 & 1.000 & 1.000 & 0.80 \\
      Mean1 Cov1 & FF$_{\operatorname{frag},t}^1$ & 0.044 & 0.247 & 0.781 & 0.991 & 1.000 & 1.000 & 0.80 \\
      Mean1 Cov1 & FF$_{\operatorname{frag},z}^2$ & 0.055 & 0.282 & 0.811 & 0.994 & 1.000 & 1.000 & 0.82 \\
      Mean1 Cov1 & FF$_{\operatorname{frag},z}^1$ & 0.058 & 0.290 & 0.820 & 0.994 & 1.000 & 1.000 & 0.82 \\
      Mean2 Cov1 & FF$_{\operatorname{frag},t}^2$ & 0.042 & 0.117 & 0.410 & 0.800 & 0.975 & 0.999 & 0.66 \\
      Mean2 Cov1 & FF$_{\operatorname{frag},t}^1$ & 0.044 & 0.118 & 0.412 & 0.802 & 0.975 & 1.000 & 0.66 \\
      Mean2 Cov1 & FF$_{\operatorname{frag},z}^2$ & 0.055 & 0.140 & 0.454 & 0.828 & 0.981 & 1.000 & 0.68 \\
      Mean2 Cov1 & FF$_{\operatorname{frag},z}^1$ & 0.058 & 0.147 & 0.460 & 0.832 & 0.983 & 1.000 & 0.68 \\
      Mean3 Cov1 & FF$_{\operatorname{frag},t}^2$ & 0.042 & 0.080 & 0.288 & 0.640 & 0.928 & 0.993 & 0.59 \\
      Mean3 Cov1 & FF$_{\operatorname{frag},t}^1$ & 0.044 & 0.085 & 0.297 & 0.651 & 0.931 & 0.994 & 0.59 \\
      Mean3 Cov1 & FF$_{\operatorname{frag},z}^2$ & 0.055 & 0.100 & 0.330 & 0.683 & 0.944 & 0.996 & 0.61 \\
      Mean3 Cov1 & FF$_{\operatorname{frag},z}^1$ & 0.058 & 0.104 & 0.338 & 0.692 & 0.947 & 0.996 & 0.62 \\
      \bottomrule
    \end{tabular}
  \end{adjustbox}
\end{table}
%\spacingset{1.00001}
%%%%%%%%%%%%%%%%%%%%%%%%%%

%%%%%%%%%%%%%%%%%%%%%%%%%%
%\spacingset{1}
\begin{table}[h!tb]
	\caption{The type-I error rates and power values of the fragment bands
    FF$_{\operatorname{frag},t}^p$ and FF$_{\operatorname{frag},z}^p$,
    $p\in\{1,2\}$, in the stationary covariance scenario Cov2. (Section \ref{SEC:SIMUL_PARTFDA} in the main paper.)}
\begin{adjustbox}{width=\columnwidth,center}
	\centering
	\begin{tabular}{ll c cccccc c}
	\toprule
		&     &H$_0$&\multicolumn{5}{c}{H$_1$}&Avg.~\\
	\cmidrule(lr){3-3}\cmidrule(lr){4-8}\\[-2ex]
	DGP & Band &$\Delta=0$ &$\Delta=0.02$ &$\Delta=0.04$ &$\Delta=0.06$ &$\Delta=0.08$ &$\Delta=0.1$&Power \\
    \midrule
    Mean1 Cov2 & FF$_{\operatorname{frag},t}^2$ & 0.042 & 0.249 & 0.806 & 0.994 & 1.000 & 1.000 & 0.81 \\
    Mean1 Cov2 & FF$_{\operatorname{frag},t}^1$ & 0.044 & 0.254 & 0.812 & 0.994 & 1.000 & 1.000 & 0.81 \\
    Mean1 Cov2 & FF$_{\operatorname{frag},z}^2$ & 0.056 & 0.292 & 0.847 & 0.996 & 1.000 & 1.000 & 0.83 \\
    Mean1 Cov2 & FF$_{\operatorname{frag},z}^1$ & 0.058 & 0.299 & 0.852 & 0.997 & 1.000 & 1.000 & 0.83 \\
    Mean2 Cov2 & FF$_{\operatorname{frag},t}^2$ & 0.042 & 0.118 & 0.458 & 0.832 & 0.983 & 0.999 & 0.68 \\
    Mean2 Cov2 & FF$_{\operatorname{frag},t}^1$ & 0.044 & 0.120 & 0.459 & 0.832 & 0.983 & 0.999 & 0.68 \\
    Mean2 Cov2 & FF$_{\operatorname{frag},z}^2$ & 0.056 & 0.147 & 0.503 & 0.862 & 0.986 & 0.999 & 0.70 \\
    Mean2 Cov2 & FF$_{\operatorname{frag},z}^1$ & 0.058 & 0.151 & 0.510 & 0.864 & 0.987 & 0.999 & 0.70 \\
    Mean3 Cov2 & FF$_{\operatorname{frag},t}^2$ & 0.042 & 0.077 & 0.275 & 0.650 & 0.920 & 0.994 & 0.58 \\
    Mean3 Cov2 & FF$_{\operatorname{frag},t}^1$ & 0.044 & 0.081 & 0.283 & 0.656 & 0.923 & 0.994 & 0.59 \\
    Mean3 Cov2 & FF$_{\operatorname{frag},z}^2$ & 0.056 & 0.103 & 0.320 & 0.695 & 0.937 & 0.996 & 0.61 \\
    Mean3 Cov2 & FF$_{\operatorname{frag},z}^1$ & 0.058 & 0.107 & 0.326 & 0.701 & 0.940 & 0.996 & 0.61 \\
	\bottomrule
	\end{tabular}
\end{adjustbox}
	\end{table}
	%\spacingset{1.00001}
	%%%%%%%%%%%%%%%%%%%%%%%%%%

%%%%%%%%%%%%%%%%%%%%%%%%%%
%\spacingset{1}
\begin{table}[h!tb]
	\caption{The type-I error rates and power values of the fragment bands
    FF$_{\operatorname{frag},t}^p$ and FF$_{\operatorname{frag},z}^p$,
    $p\in\{1,2\}$, in the non-stationary covariance scenario Cov3. (Section \ref{SEC:SIMUL_PARTFDA} in the main paper.)}
\begin{adjustbox}{width=\columnwidth,center}
	\centering
	\begin{tabular}{ll c cccccc c}
	\toprule
		&     &H$_0$&\multicolumn{5}{c}{H$_1$}&Avg.~\\
	\cmidrule(lr){3-3}\cmidrule(lr){4-8}\\[-2ex]
	DGP & Band &$\Delta=0$ &$\Delta=0.02$ &$\Delta=0.04$ &$\Delta=0.06$ &$\Delta=0.08$ &$\Delta=0.1$&Power \\
    \midrule
    Mean1 Cov3 & FF$_{\operatorname{frag},t}^2$ & 0.042 & 0.244 & 0.803 & 0.992 & 1.000 & 1.000 & 0.81 \\
    Mean1 Cov3 & FF$_{\operatorname{frag},t}^1$ & 0.042 & 0.243 & 0.798 & 0.992 & 1.000 & 1.000 & 0.81 \\
    Mean1 Cov3 & FF$_{\operatorname{frag},z}^2$ & 0.055 & 0.286 & 0.835 & 0.994 & 1.000 & 1.000 & 0.82 \\
    Mean1 Cov3 & FF$_{\operatorname{frag},z}^1$ & 0.056 & 0.286 & 0.836 & 0.994 & 1.000 & 1.000 & 0.82 \\
    Mean2 Cov3 & FF$_{\operatorname{frag},t}^2$ & 0.042 & 0.119 & 0.455 & 0.835 & 0.984 & 0.999 & 0.68 \\
    Mean2 Cov3 & FF$_{\operatorname{frag},t}^1$ & 0.042 & 0.130 & 0.486 & 0.855 & 0.987 & 0.999 & 0.69 \\
    Mean2 Cov3 & FF$_{\operatorname{frag},z}^2$ & 0.055 & 0.149 & 0.504 & 0.865 & 0.988 & 0.999 & 0.70 \\
    Mean2 Cov3 & FF$_{\operatorname{frag},z}^1$ & 0.056 & 0.160 & 0.538 & 0.884 & 0.990 & 0.999 & 0.71 \\
    Mean3 Cov3 & FF$_{\operatorname{frag},t}^2$ & 0.042 & 0.078 & 0.273 & 0.645 & 0.919 & 0.994 & 0.58 \\
    Mean3 Cov3 & FF$_{\operatorname{frag},t}^1$ & 0.042 & 0.074 & 0.254 & 0.620 & 0.908 & 0.992 & 0.57 \\
    Mean3 Cov3 & FF$_{\operatorname{frag},z}^2$ & 0.055 & 0.102 & 0.318 & 0.688 & 0.936 & 0.997 & 0.61 \\
    Mean3 Cov3 & FF$_{\operatorname{frag},z}^1$ & 0.056 & 0.096 & 0.298 & 0.666 & 0.927 & 0.995 & 0.60 \\
	\bottomrule
	\end{tabular}
\end{adjustbox}
	\end{table}
	%\spacingset{1.00001}
	%%%%%%%%%%%%%%%%%%%%%%%%%%

% %%%%%%%%%%%%%%%%%%%%%%
% \begin{figure}[h!]
% \vspace*{4cm}
% \centering
% \includegraphics[width=.9\textwidth]{Fig_Appl_Biom_Appendix.pdf}
% \caption{The adjusted $p$-value function  (left plot) of the IWT procedure of \cite{PV2017_IWT} and the B$_{\text{EC}}$ band (right plot) of \cite{CR2018} applied to the Biomechanics data application of Section \ref{SSEC:APPL_BIOM}.}
% \label{FIG:BIOM_Appendix}
% \end{figure}
% %%%%%%%%%%%%%%%%%%%%%

%%%%%%%%%%%%%%%%%%%%%%
%\spacingset{1}
\begin{figure}[h!]
\centering
\includegraphics[width=.9\textwidth]{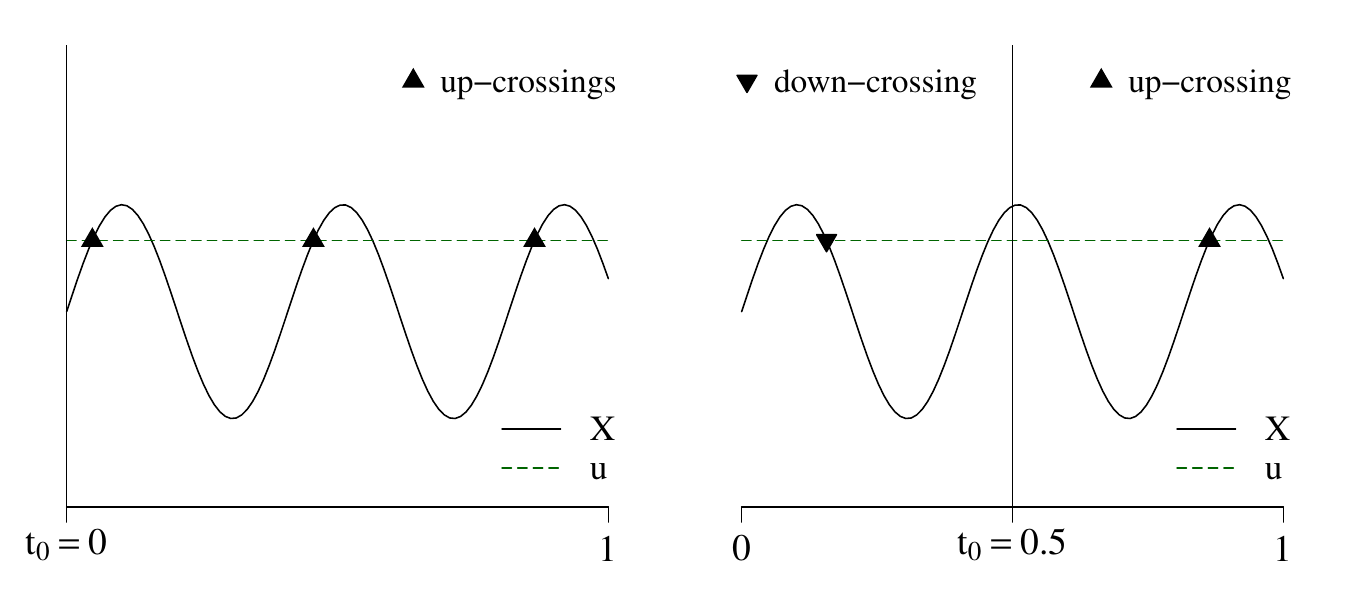}
\caption{{\sc Left Plot:} Three exceedance events counted as three up-crossing events to the right of $t_0=0$. {\sc Right Plot:} The same three exceedance events, but counted as one down-crossing event to the left of $t_0=0.5$, one exceedance event at $t_0=0.5$, and one up-crossing event to the right of $t_0=0.5$.}
\label{FIG:APP_CROSS}
\end{figure}
%\spacingset{1.00001}
%%%%%%%%%%%%%%%%%%%%%%

%%%%%%%%%%%%%%%%%%%%%%
\begin{figure}[h]
\centering
\includegraphics[width=.9\textwidth]{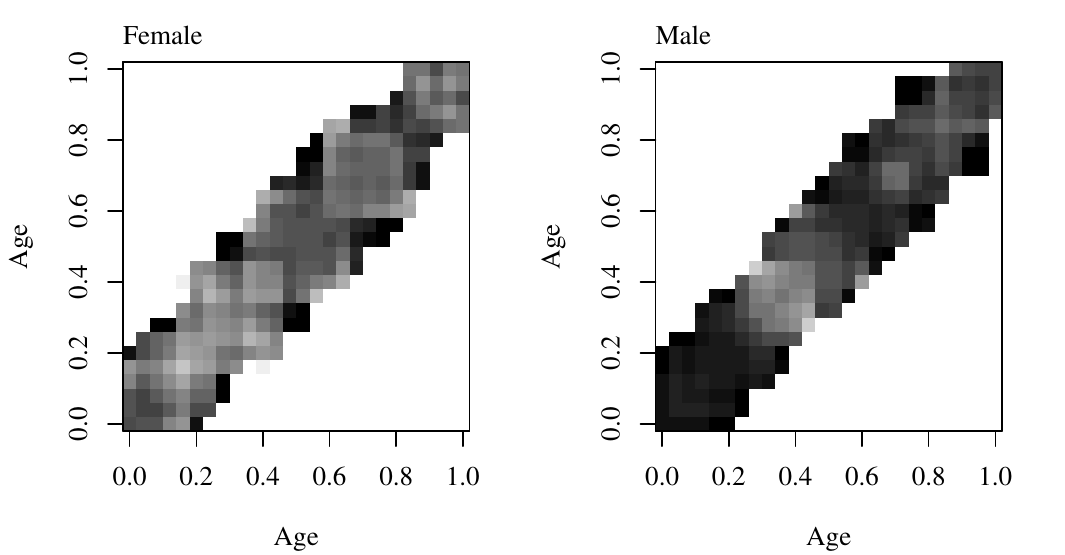}
\caption{The estimable parts of the sample covariances, $\hat{C}^{\operatorname{f}}_{\operatorname{frag}}$ and $\hat{C}^{\operatorname{m}}_{\operatorname{frag}}$, of the bone mineral acquisition data application in Section \ref{SSEC:APPL_FRAGM}.}
\label{FIG:FRAGM_Appendix}
\end{figure}
%%%%%%%%%%%%%%%%%%%%%

\end{document}